\documentclass[10pt,journal,comsoc]{IEEEtran}
\usepackage{titlesec}
\titleclass{\subsubsubsection}{straight}[
  \subsubsection]
\newcounter{subsubsubsection}[subsubsection]
\renewcommand\thesubsubsubsection{\thesubsubsection.
\arabic{subsubsubsection}}
\titleformat{\subsubsubsection}[runin]
  {\normalfont\normalsize\bfseries}{\thesubsubsubsection}{1em}{}
\titlespacing*{\subsubsubsection}{0pt}{3.25ex plus 1ex minus .2ex}{1em}
\ifCLASSOPTIONcompsoc
\usepackage[nocompress]{cite}
\else
  \usepackage{cite}
\fi
\usepackage{multirow}
\usepackage{booktabs}
\newtheorem{lemma}{Lemma}
\ifCLASSINFOpdf
   \usepackage[pdftex]{graphicx}
\else
\fi
\usepackage{amsmath}
\usepackage{algorithmic}
\usepackage{array}
\usepackage{mdwmath}
\usepackage{mdwtab}
\usepackage[T1]{fontenc}
\usepackage{graphicx}%
\usepackage[caption=false,font=footnotesize,labelfont=sf,textfont=sf]{subfig}
\begin{document}
\title{Locally Fair PageRank: Mean-Field Approximation and One-Step Refinement}
\author{Mukesh Kumar, Gaurav Dixit, Akrati Saxena%
\IEEEcompsocitemizethanks{
\IEEEcompsocthanksitem M. Kumar and G. Dixit are with the Mehta Family
School of Data Science and Artificial Intelligence, Indian Institute of
Technology Roorkee, Roorkee, Uttarakhand 247667, India.
E-mail: mukesh\_k@mfs.iitr.ac.in, gaurav.dixit@mfs.iitr.ac.in.

\IEEEcompsocthanksitem A. Saxena is with LIACS, Leiden University,
Leiden, The Netherlands.
E-mail: a.saxena@liacs.leidenuniv.nl.}} \IEEEtitleabstractindextext{%
\begin{abstract}
Graph-based ranking methods such as PageRank can amplify structural disparities in networks, motivating fairness-aware ranking mechanisms for sensitive groups. Locally Fair PageRank (LFPR) enforces fairness through local propagation, but exact computation requires repeated iterations until convergence, limiting scalability on large graphs. We develop a scalable analytical framework for approximating Neighborhood Locally Fair PageRank and Uniform Locally Fair PageRank. By introducing a group-aware heterogeneous mean-field representation, the framework aggregates structurally similar nodes into degree classes and derives closed-form approximations of stationary LFPR scores, avoiding repeated propagation over the fairness-aware transition matrix. We develop a One-Step Refinement (ORF) mechanism that applies the fairness-aware propagation operator once to the mean-field estimate, incorporating node-specific neighborhood information without iterative convergence. The fluctuation analysis characterizes degree-dependent variability around the mean-field solution and shows that the coefficient of variation decreases with increasing in-degree. The mean-field approximation reduces the computational cost of exact LFPR from iterative graph-scale propagation to linear-time node-level estimation, while ORF requires one graph traversal. Experiments on six real-world networks show strong agreement with exact LFPR scores and rankings, preservation of group-level fairness, and substantial runtime reductions. The mean-field approximation reduces complexity to $\mathcal{O}(n)$, while ORF improves accuracy with $\mathcal{O}(m+n)$.
\end{abstract}

}

\maketitle

\IEEEdisplaynontitleabstractindextext
\IEEEpeerreviewmaketitle

\ifCLASSOPTIONcompsoc
\IEEEraisesectionheading{\section{Introduction}\label{sec:introduction}}
\else
\section{Introduction}
\label{sec:introduction}
\fi
\IEEEPARstart{G}{raph}-based ranking algorithms play a fundamental role in the analysis of complex networks and automated decision-making processes, with applications spanning web search, recommendation systems, social network analysis, information diffusion, and influence-aware ranking  \cite{li2024recent, saxena2022nodesim, solanki2025survey, gao2026importance}. In many real-world networks, however, structural inequalities arising from connectivity disparities, homophily, and group imbalance are embedded within the underlying graph topology, often affecting the visibility, exposure, influence, and accessibility of individuals or sensitive groups \cite{saxena2024fairsna, wang2023survey, razaghi2022group, saxena2025homophily}. Such structural biases can significantly influence the outputs of network-based ranking and link-analysis algorithms, potentially amplifying existing disparities through iterative propagation dynamics \cite{karimi2018homophily, liu2024promoting}. Consequently, developing fairness-aware network algorithms for ranking, recommendation, and influence propagation has become a fundamental challenge in trustworthy and responsible network analytics \cite{meena2025achieving, gao2020fair, has2026fairness}.

Among graph-based ranking methods, PageRank \cite{brin1998anatomy} is one of the most widely used approaches for quantifying node importance through random-walk dynamics over network structures. Owing to its ability to propagate influence through graph connectivity, PageRank has been extensively applied in web ranking, recommendation systems, centrality analysis, and influence estimation \cite{avella2018centrality, wayama2025generalized, saxena2020centrality, li2018influence, gleich2015pagerank}. However, the ranking scores produced by PageRank-based algorithms are inherently shaped by the network topology and may therefore reflect or amplify structural disparities across sensitive groups \cite{espin2022inequality, tsioutsiouliklis2021fairness}. To address this limitation, recent studies have proposed fairness-aware variants of PageRank that explicitly incorporate group-level fairness constraints into the transition dynamics of the random walk \cite{tsioutsiouliklis2021fairness, wang2026fairness, kariotakis2026fairrari}. In particular, the Neighborhood Locally Fair PageRank (LFPR$_N$) and residual-based Uniform Locally Fair PageRank (LFPR$_U$) frameworks enforce fairness by modifying the local transition probabilities governing the propagation of PageRank mass across sensitive groups. Both retain the probabilistic interpretation of PageRank while enforcing fairness during local propagation.

Despite their effectiveness in promoting group-level fairness, LFPR methods require repeated propagation over fairness-aware random-walk dynamics to estimate the stationary distribution, resulting in substantial computational overhead on large networks \cite{tsioutsiouliklis2021fairness}. In these methods, fairness is embedded directly into the local transition probabilities, requiring iterative propagation until convergence to the stationary distribution. As graphs grow, the iterative computation incurs substantial computational overhead, limiting scalability on networks with millions of nodes and edges. Prior studies have explored efficient estimation of node rankings from structural information without computing the centrality of all nodes \cite{saxena2017global}. Consequently, there is a need for scalable approximation methods that can efficiently estimate locally fair PageRank scores while preserving the underlying fairness objectives.

Motivated by these limitations, we propose a scalable approximation framework for the LFPR$_N$ and LFPR$_U$ methods. The proposed approach reformulates the fairness-aware PageRank dynamics through a group-aware mean-field approximation, in which nodes are aggregated according to their structural and group-specific degree characteristics. This coarse-grained representation enables estimation of the stationary fairness-aware ranking scores without explicitly constructing or repeatedly propagating over fairness-aware transition matrices. To capture local structural variations beyond the mean-field approximation, we introduce a lightweight one-step local refinement mechanism based on the original fairness-aware update dynamics. Together, these components provide a scalable approximation of locally fair PageRank while retaining local structural information.

The contribution of this work is summarized as follows:
\begin{itemize}
\item We propose a scalable group-aware mean-field approximation framework for the Locally Fair PageRank methods LFPR$_N$ and LFPR$_U$, enabling efficient estimation of fairness-aware ranking scores through closed-form stationary approximations without repeated propagation to convergence.

\item We establish analytical expressions for the variance and coefficient of variation of LFPR scores, showing that fluctuations diminish with increasing node in-degree.

\item We introduce a lightweight One-Step Refinement (ORF) mechanism that incorporates local structural information and improves mean-field approximation accuracy while retaining computational efficiency.

\item Extensive experiments on multiple real-world networks show that the proposed methods closely match exact LFPR scores, preserve fairness characteristics, and substantially reduce computational cost and runtime.
\end{itemize}

\section{Related Work}
Algorithmic fairness has become an important research direction due to the increasing use of automated ranking and recommendation systems across networked applications \cite{li2023fairness, christoforou2021ranking}, including personalized recommendations \cite{patro2020fairrec}, citation analysis \cite{radicchi2012testing}, and information retrieval \cite{xu2026economic, bernard2025systematic}. Accordingly, recent research has increasingly focused on fairness-aware network analysis across graph-mining tasks, including fair node ranking \cite{gao2026importance}, link prediction \cite{saxena2022hm}, influence maximization \cite{saxena2026dq4fairim}, influence blocking \cite{saxena2023fairness}, opinion dynamics \cite{stkepien2026fairness}, and community detection \cite{de2024group, corriera2026individual}. In particular, fairness-aware PageRank methods mitigate structural biases in graph-based ranking by incorporating fairness-aware propagation mechanisms, transition reweighting strategies, and group-level fairness constraints \cite{tsioutsiouliklis2021fairness, wang2026fairness, kariotakis2026fairrari}.

Tsioutsiouliklis et al. \cite{tsioutsiouliklis2021fairness} introduced the Locally Fair PageRank (LFPR) framework, comprising the Neighborhood Locally Fair PageRank and Uniform Locally Fair PageRank algorithms, to extend PageRank with locally fair ranking mechanisms for mitigating group-level ranking disparities. These methods provide group-level fairness guarantees while preserving the stochastic interpretation of PageRank-based ranking processes. More recently, alternative fairness-aware PageRank frameworks, including edge-reweighting approaches \cite{wang2026fairness} and plug-and-play fairness-aware ranking methods such as FairRARI \cite{kariotakis2026fairrari}, have been proposed to improve fairness in graph-based ranking while preserving important structural characteristics of the network. However, these methods primarily focus on fairness-aware ranking mechanisms, while computation of the corresponding PageRank solutions generally relies on solving the underlying ranking process.

A large body of research has focused on improving the scalability of PageRank and related random-walk-based algorithms through efficient approximation and computation techniques, including local and personalized PageRank estimation \cite{bar2008local, liao2023efficient}, distributed iterative frameworks \cite{das2013fast}, scalable approximation algorithms \cite{wu2024efficient}, and mean-field approximation method \cite{fortunato2006approximating}. While these approaches improve the computational efficiency of PageRank, they do not explicitly address fairness-aware local propagation. 
Efficient rank estimation for other centrality measures has also been investigated using heuristic approaches and local structural information \cite{saxena2016estimating,saxena2019heuristic,saxena2017fast, saxena2017degree}. Approximation techniques have also been developed for fairness-aware PageRank models. In particular, Fairness-Sensitive PageRank (FSPR) \cite{kumar2026fairness} employs degree-based mean-field approximations for PageRank models in which fairness is enforced through optimized teleportation dynamics. In FSPR, fairness is enforced by modifying the teleportation (jump) vector while largely preserving the underlying random-walk transition structure. Consequently, the corresponding approximation addresses teleportation-based fairness rather than fairness embedded directly in local transition dynamics.
\begin{table*}[!t]
\centering
\caption{Comparison of related PageRank and fairness-aware ranking approaches with the proposed LFPR approximation framework.}
\label{tab:related_work_comparison}

\renewcommand{\arraystretch}{1.15}
\setlength{\tabcolsep}{4pt}
\footnotesize

\begin{tabular*}{\textwidth}{@{\extracolsep{\fill}}lccccc}
\toprule

\textbf{Method} &
\textbf{Fairness} &
\textbf{Approximation} &
\textbf{Iteration-Free} &
\textbf{Refinement} &
\textbf{Fluctuation} \\

\midrule

LFPR \cite{tsioutsiouliklis2021fairness}
&
Local propagation
&
$\times$
&
$\times$
&
$\times$
&
$\times$
\\

Edge-Reweighted Fair PageRank \cite{wang2026fairness}
&
Edge-reweighting
&
$\times$
&
$\times$
&
$\times$
&
$\times$
\\

FairRARI \cite{kariotakis2026fairrari}
&
Fairness constraints 
&
$\times$
&
--
&
$\times$
&
$\times$
\\

PageRank approximation \cite{bar2008local,liao2023efficient,
das2013fast,wu2024efficient,fortunato2006approximating}
&
No fairness constraint
&
$\checkmark$
&
Varies
&
Varies
&
$\times$
\\

FSPR approximation \cite{kumar2026fairness}
&
Teleportation
&
$\checkmark$
&
$\checkmark$
&
$\times$
&
$\checkmark$
\\

\textbf{Proposed LFPR framework}
&
\textbf{Local propagation}
&
$\boldsymbol{\checkmark}$
&
$\boldsymbol{\checkmark}$
&
\textbf{ORF}
&
$\boldsymbol{\checkmark}$
\\

\bottomrule
\end{tabular*}

\end{table*}
In contrast, LFPR$_N$ and LFPR$_U$ embed fairness directly into the local transition dynamics. LFPR$_N$ employs group-aware transition probabilities, whereas LFPR$_U$ uses residual-based redistribution to satisfy the prescribed local fairness constraint.

To address this computational challenge, we develop a group-aware mean-field approximation directly for the fairness-aware transition dynamics of LFPR$_N$ and LFPR$_U$. The proposed framework derives closed-form analytical estimates that avoid repeated propagation to convergence, characterizes degree-dependent fluctuations around the mean-field solution, and introduces a One-Step Refinement (ORF) mechanism that incorporates node-specific local information through a single fairness-aware propagation step. Table~\ref{tab:related_work_comparison} summarizes the key methodological differences between representative PageRank approximation and fairness-aware PageRank methods and the proposed LFPR framework.

\section{Preliminaries}
In this section, we review the background on the classical PageRank algorithm  \cite{brin1998anatomy} and the  Locally Fair PageRank (LFPR) framework proposed in \cite{tsioutsiouliklis2021fairness}.
\subsection{The PageRank Algorithm}
The PageRank algorithm, introduced by Brin and Page \cite{brin1998anatomy}, ranks nodes in a graph by propagating importance through the link structure. Given a graph $G = (V, E)$ with $|V| = n$ nodes and $|E| = m$ edges, the PageRank algorithm computes a stationary probability distribution over the nodes, representing their relative importance. At each step, the walker follows an outgoing edge with probability $1 - \gamma$ and jumps to a node sampled from the teleportation vector $\mu$, which is typically initialized as a uniform distribution over all nodes, with probability $\gamma$, where $\gamma$ is the damping factor (typically $0.15$). 

Moreover, let $A \in \{0,1\}^{n \times n}$ denote the adjacency matrix of the graph, where $A[i,j] = 1$ if an edge exists from node $i$ to node $j$, and $A[i,j]=0$ otherwise. The transition matrix $P$ is obtained by normalizing the rows of $A$, with appropriate handling of sink nodes. The stationary distribution ( PageRank vector $p$) satisfies the equation:
\begin{equation*}
    p^{T} = (1-\gamma)p^{T}P + \gamma \mu^{T}
\end{equation*}
Equivalently, its closed-form solution can be expressed as:
\begin{equation*}
    p^{T} = \gamma \mu^{T}[I-(1-\gamma)P]^{-1}
\end{equation*}
where $I\in \mathcal{R}^{n\times n}$ is the identity matrix. 
 
\subsection{Locally Fair PageRank Algorithms}
\label{sec:subsec2}
Tsioutsiouliklis et al.~\cite{tsioutsiouliklis2021fairness} proposed a fairness-aware extension of the PageRank algorithm under a local expectation framework, based on group fairness. Given a graph $G$, the nodes are partitioned according to a sensitive attribute (e.g., race or gender). For simplicity, a binary partition is considered, consisting of two disjoint groups: Red ($R$) and Blue ($B$), such that $R, B \subseteq V$, $R \cap B = \emptyset$, and $R \cup B = V$. To promote fairness across groups, the authors introduced two locally fair variants: LFPR$_N$ and the residual-based LFPR$_U$. These methods modify the underlying random walk to ensure that each node distributes its PageRank mass across groups according to the prescribed fairness constraint.

The central idea of LFPR$_N$ is to enforce group-level fairness during random-walk transitions. Specifically, a node is $\phi$-fair if it allocates a fraction $\phi$ of its outgoing probability mass to nodes in the Red group and the remaining fraction $1 - \phi$ to nodes in the Blue group, where $\phi \in (0,1)$ denotes the desired fairness parameter. Typically, $\phi$ is chosen proportional to the fraction of Red nodes in the graph. Accordingly, the teleportation vector $\mu$ satisfies $\mu(v) = \phi/|R|$ for $v \in R$, and $\mu(v) = (1 - \phi)/|B|$ for $v \in B$.

Unlike LFPR$_N$, which enforces fairness through group-dependent transition probabilities, LFPR$_U$ adopts a residual-based redistribution mechanism. Instead of assigning different transition probabilities to neighbors based on their sensitive group, LFPR$_U$ distributes a fraction of the PageRank mass uniformly across the outgoing neighborhood, while the remaining probability mass, referred to as the residual, is redistributed to the locally underrepresented group to satisfy the prescribed fairness constraint. Formally, the nodes are partitioned into two subsets, $L_R$ and $L_B$, where $L_R$ ($L_B$) contains nodes whose neighborhoods underrepresent Red (Blue) nodes relative to the fairness parameter $\phi$. Consequently, nodes in $L_R$ redistribute the residual probability mass to Red nodes, while nodes in $L_B$ redistribute it to Blue nodes. The corresponding residual probability masses are denoted by $\delta_R(i)$ for nodes in $L_R$ and $\delta_B(i)$ for nodes in $L_B$, respectively. The residual mass is then distributed uniformly over the associated target group.

\section{Problem Formulation}
Although the LFPR$_N$ and LFPR$_U$ achieve group-level fairness by modifying the random-walk dynamics, their exact computation requires fairness-aware transition matrices and iterative stationary-distribution estimation, resulting in substantial computational overhead on large networks.

In this work, we address the problem of efficiently approximating the stationary distributions of LFPR$_N$ and LFPR$_U$ without explicitly constructing group-specific transition matrices. Given a graph $G$, a binary group partition $(R, B)$ and a fairness parameter $\phi \in (0,1)$, the goal is to estimate the corresponding LFPR scores with substantially lower computational cost. Specifically, our objective is to develop a scalable approximation framework that avoids the explicit construction and aggregation of group-specific transition matrices while maintaining the group-level fairness specified by the target parameter $\phi$. At the same time, the approximated scores should remain close to the exact LFPR stationary distributions while requiring substantially lower computational complexity. The primary challenge is therefore to balance computational efficiency, approximation accuracy, and fidelity to the fairness-aware ranking dynamics, since overly coarse approximations may distort the distribution of PageRank mass across groups.

\subsection{Approximation of LFPR$_N$}
\label{subsec4.1}
To facilitate approximation, we reformulate the matrix-based LFPR$_N$ model into an equivalent node-level iterative representation. The stationary PageRank vector satisfies

\begin{equation*}
p_N^{T} = \gamma v_N^{T} + (1 - \gamma)p_N^{T}P_N.
\end{equation*}

where $P_N = \phi P_R + (1 - \phi)P_B$ is the fairness-aware transition matrix, $P_R$ and $P_B$ denote the group-specific transition matrices for the Red ($R$) and Blue ($B$) groups, respectively, and $v_N$ represent the $\phi$-fair teleportation distribution introduced in Section \ref{sec:subsec2}, while $\phi \in (0,1)$ specifies the desired fairness proportion toward the Red group.

The stationary equation admits the following equivalent iterative formulation. For each node $i\in V$, the LFPR$_N$ score at iteration $t$ is given by

\begin{equation*}
p_N^{(t)}(i) = \gamma v_N(i) + (1 - \gamma)\sum_{j \in V} p_N^{(t-1)}(j)\, P_N[j,i].
\end{equation*}

The iterative formulation can be expressed in a unified group-wise form. For a target node $i\in g$, where $g\in\{R,B\}$, the LFPR$_N$ score at iteration $t$ is given by
\begin{equation}
\begin{aligned}
p_N^{(t)}(i) & = \gamma\frac{\alpha_g}{|g|} + (1-\gamma)\alpha_g \left[ \sum_{j:\operatorname{out}_g(j)\neq 0} \frac{a_{ji}}{\operatorname{out}_g(j)} p_N^{(t-1)}(j) \right. \\ & \qquad \left. + \sum_{j:\operatorname{out}_g(j)=0} \frac{1}{|g|} p_N^{(t-1)}(j) \right], \qquad i\in g, 
\label{eq:lfprn_unified}
\end{aligned}
\end{equation}
where $\alpha_R=\phi$ and $\alpha_B=1-\phi$, and $\operatorname{out}_g(j)$ denotes the number of outgoing edges from node $j$ to nodes belonging to group $g$. For $g=R$, Eq. (\ref{eq:lfprn_unified}) recovers the Red-group LFPR$_N$ update, whereas for $g=B$, it yields the Blue-group update.

The second summation term in Eq. (\ref{eq:lfprn_unified}) accounts for contributions from nodes with zero group-specific out-degree. The associated probability mass is redistributed uniformly among the nodes of the corresponding target group, resulting in an identical additive contribution to every node within that group. As established in Lemma \ref{lem:vanishing}, this contribution is independent of the target node's degree class and therefore does not introduce degree-dependent variation within the group. Accordingly, this group-wise uniform term is not retained in the reduced degree-based approximation. The remaining degree-dependent terms are then used to derive the mean-field approximation. The accuracy of the resulting approximation is evaluated empirically against the exact LFPR$_N$ solution.

\noindent\textbf{Mean-field Approximation:}
\label{subsec:subsec4.1.1}
Rather than analyzing the LFPR$_N$ scores at the individual-node level, nodes are aggregated into group-aware degree classes according to their structural characteristics, with separate partitions defined for the Red and Blue groups. Let

\begin{equation*}
\mathcal{C}_R = \{C_R^i : i = 1, 2, \dots, K_R\}
\end{equation*}
\begin{equation*}
\mathcal{C}_B = \{C_B^j : j = 1, 2, \dots, K_B\}
\end{equation*}
denote the collections of degree classes for the Red and Blue groups, respectively, where $K_R$ and $K_B$ are the numbers of distinct degree classes in the corresponding groups. Each class is characterized by the group-specific in-degree and out-degree structure of its constituent nodes. In particular, a Red-group class $C_R^i$ is represented by the tuple 

\begin{equation*}
C_R^i = (d^{R}_{\text{in}},\, d^{B}_{\text{in}},\, d^{R}_{\text{out}},\, d^{B}_{\text{out}},\, R),
\end{equation*}
while a Blue-group degree class $C_B^j$ is represented by

\begin{equation*}
C_B^j = (d^{R}_{\text{in}},\, d^{B}_{\text{in}},\, d^{R}_{\text{out}},\, d^{B}_{\text{out}},\, B).
\end{equation*}
Here, $d^{R}_{\text{in}}$ and $d^{B}_{\text{in}}$ denoted by $\text{in}_R$ and $\text{in}_B$, respectively, represent the numbers of incoming edges from Red and Blue nodes. Similarly, $d^{R}_{\text{out}}$ and $d^{B}_{\text{out}}$ denoted by $\text{out}_R$ and $\text{out}_B$, respectively, represent the numbers of outgoing edges to Red and Blue nodes. These classes capture group-specific interaction patterns and provide a coarse-grained representation of the network that retains the structural information relevant to the proposed fairness-aware approximation.

We first derive the approximation for the Red group, while the corresponding Blue-group formulation follows analogously. For a Red-group degree class $k_R$, we define the average LFPR$_N$ score as

\begin{equation}
\label{eq:MFAeq1}
\overline{p}^{(t)}_{N}(k_{R}) \equiv \frac{1}{n\,P(k_{R})} \sum_{i \in k_{R}} p^{(t)}_N(i),
\end{equation}

where $P(k_R)$ denotes the fraction of nodes belonging to degree class $k_R$, and $nP(k_R)$ denotes the number of nodes in that class. The resulting quantity represents the average LFPR$_N$ score within the class and provides the basis for the class-level mean-field approximation. Setting $g=R$ and $\alpha_R = \phi$ in  Eq. (\ref{eq:lfprn_unified}), and averaging the resulting node-level update over all nodes in $k_R$, we obtain

\begin{equation}
\begin{aligned}
  \frac{1}{n\,P(k_{R})} \sum_{i \in k_{R}} p^{(t)}_{N}(i) & \approx \frac{\gamma \phi}{|R|} +  \frac{(1 - \gamma)\phi}{n\,P(k_{R})} \sum_{i \in k_{R}} \\& \sum_{j:\,\text{out}_R(j)\neq 0} \frac{a_{ji}}{\text{out}_R(j)} p^{(t-1)}_{N}(j).
  \label{eq:MFAeq2}
\end{aligned}
\end{equation}

The left-hand side of Eq. (\ref{eq:MFAeq2}) corresponds to $\overline{p}_{N}(k_{R})$, as defined in Eq. (\ref{eq:MFAeq1}). To simplify the right-hand side, we decompose the summation over predecessor nodes into two levels: an outer summation over all degree classes $k^{'}$, including both Red and Blue groups, and an inner summation over nodes within each class. This yields

\begin{equation}
\begin{aligned}
  \overline{p}^{(t)}_{N}(k_{R}) & \approx \frac{\gamma \phi}{|R|} + \frac{(1 - \gamma)\phi}{n\,P(k_R)} \\ & \sum_{{k^{'} :\mathrm{out}_{R}(k^{'})>0}} \frac{1}{\text{out}_R(k^{'})} \sum_{i \in k_{R}} \sum_{j \in k^{'}} a_{ji} \, p^{(t-1)}_{N}(j).
  \label{eq:eq5}
\end{aligned}
\end{equation}

Applying the mean-field approximation, the influence of predecessor nodes is represented by the average LFPR$_N$ score of their corresponding degree classes.

\begin{equation}
\label{eq:eq6}
  \sum_{i \in k_{R}} \sum_{j \in k^{'}} a_{ji} \, p^{(t-1)}_{N}(j)\;\approx\;\overline{p}^{(t-1)}_{N}(k^{'}) \sum_{i \in k_{R}} \sum_{j \in k'} a_{ji}.
\end{equation}

Let $E_{k^{'} \rightarrow k_{R}}$ denote the total number of edges directed from nodes belonging to degree class $k^{'}$ to nodes in the Red degree class $k_{R}$. Formally, this quantity is defined as

\begin{equation*}
  E_{k^{'} \rightarrow k_{R}} = \sum_{i \in k_{R}} \sum_{j \in k^{'}} a_{ji}
  \label{eq:eq7}
\end{equation*}

where $a_{ji}$ is the $(j,i)$-th entry of the adjacency matrix $A$. The quantity $E_{k^{'} \rightarrow k_{R}}$ can be expressed as

\begin{align}
E_{k' \rightarrow k_{R}}  &= k_{R,\text{in}} \, n \, P(k_{R}) \cdot \frac{E_{k' \rightarrow k_{R}}}{k_{R,\text{in}} \, n \, P(k_{R})} \notag \\ &= k_{R,\text{in}} \, n \, P(k_{R}) \, P_{\text{in}}(k' \mid k_{R})
\label{eq:eq8}
\end{align}

where $P_{\mathrm{in}}(k^{'} \mid k_{R})$ denotes the probability that a randomly selected predecessor of a node in the Red degree class $k_{R}$ belongs to degree class $k^{'}$, and $k_{R,\text{in}}$ denotes the total in-degree of nodes in degree class $k_R$.

In general, for networks exhibiting degree-degree correlations, the solution of the resulting class-level equation becomes non-trivial, and the LFPR$_N$ values may exhibit a complex dependence on degree structure. However, under the commonly adopted assumption of uncorrelated networks, the transition probability $P_{\text{in}}(k^{'} \mid k_{R})$ becomes independent of the target degree class $k_R$ and admits a simplified form given by

\begin{equation}
\label{eq: eq9}
    P_{\textbf{in}}(k^{'}\mid k_{R}) = \frac{\text{out}_{R}(k^{'})P(k^{'})}{\langle\text{out}_{R}\rangle}
\end{equation}

where $\langle \text{out}_{R} \rangle$ denotes the average number of outgoing edges toward Red nodes in the network. 
Substituting Eqs. (\ref{eq:eq6}) and (\ref{eq:eq8}) into Eq. (\ref{eq:eq5}), we obtain the following:
\begin{equation}
\begin{aligned}
\overline{p}^{(t)}_N(k_{R}) \approx & \frac{\gamma \phi}{|R|} + (1 - \gamma)\,\phi\,k_{R, \text{in}} \\ &\sum_{{k^{'} :\mathrm{out}_{R}(k^{'})>0}} \frac{P_{\text{in}}(k^{'}\mid k_{R})}{\text{out}_{R}(k^{'})} \overline{p}^{(t-1)}_{N}(k^{'}).
    \label{eq:eq10}
\end{aligned}
\end{equation}

Now, under the condition of a degree-degree uncorrelated network, using Eqs. (\ref{eq: eq9}) and (\ref{eq:eq10}), we get
\begin{equation}
\overline{p}^{(t)}_{N}(k_{R}) \approx \frac{\gamma \phi}{|R|} + (1 - \gamma)\,\phi\,\frac{k_{R,\text{in}}}{\langle \text{out}_{R} \rangle} \sum_{{k^{'} :\mathrm{out}_{R}(k^{'})>0}} P(k^{'}) \, \overline{p}^{(t-1)}_N(k^{'}).
\label{eq:eq11}
\end{equation}

Since $\langle \text{out}_R\rangle$ and $\langle d_{\text{in}}\rangle_{R}$ both quantify the average number of edges directed toward Red nodes, they are equal, i.e., $\langle \text{out}_R\rangle$ = $\langle d_{\text{in}}\rangle_{R}$. Furthermore, $\sum_{k^{'}} P(k^{'})\, \overline{p}_N^{(t-1)}(k^{'})$ represents the expected LFPR$_N$ value over all degree classes. Using the definition of $\overline{p}_N(k^{'})$ and the normalization condition $\sum_{i \in V} \overline{p}_N^{(t-1)}(i) = 1$, this term simplifies to $1/n$. 
Since the propagation term in Eq. (\ref{eq:eq11}) involves only source degree classes satisfying $\text{out}_R(k^{'}) > 0$, we adopt the following reduced mean-field closure
\begin{equation*}
\sum_{k^{'} :\mathrm{out}_{R}(k^{'})>0}P(k^{'})\,\overline{p}_{N}^{(t-1)}(k^{'})  \approx \sum_{k^{'}} P(k^{'})\,\overline{p}_{N}^{(t-1)}(k^{'}) = \frac{1}{n}.
\label{eq:restricted_mass_approx}
\end{equation*}
This closure replaces the class-weighted LFPR$_N$ mass over the contributing source degree classes by its normalized network-wide average, thereby yielding a tractable closed-form approximation. Consequently, the stationary limit as $t \to \infty$, Eq. (\ref{eq:eq11}) reduces to
\begin{equation}
\overline{p}_N(k_{R}) \approx \gamma \frac{\phi}{|R|} + (1 -\gamma)\,\phi\,\frac{k_{R,\text{in}}}{\langle d_{\text{in}} \rangle_{R}}\cdot \frac{1}{n}
\label{eq:eq12}
\end{equation}

Following an analogous derivation for the Blue-group degree classes, in the stationary limit $t \to \infty$, we obtain
\begin{equation}
\overline{p}_N(k_{B}) \approx \gamma \frac{(1 - \phi)}{|B|} + (1 -\gamma)\,(1 - \phi)\,\frac{k_{B,\text{in}}}{\langle d_{\text{in}} \rangle_{B}}\cdot \frac{1}{n}.
\end{equation}

The stationary LFPR$_N$ approximation for a node $u \in V$ can be written in the unified form
\begin{equation}
\label{eq:eq14}
  \hat{p}_N(u) \approx \gamma \frac{\alpha_{g}}{|g|} + (1-\gamma)\alpha_{g} \frac{k_{u,\text{in}}}{\langle d_{\text{in}}\rangle_{g}} \cdot\frac{1}{n}
\end{equation}

where $g \in \{R,B\}$ denotes the sensitive group to which node $u$ belongs, with $\alpha_R=\phi$ and $\alpha_B=1-\phi$. Here, $k_{u,\mathrm{in}}$ represents the number of incoming edges to node $u$, $\langle d_{\text{in}}\rangle_{g}$ denotes the total in-degree of nodes belonging to group $g$ normalized by $n$, and $n = |V|$ denotes the total number of nodes. 
Since the stationary mean-field approximation is derived independently for each node (or degree class), the resulting LFPR$_N$ scores are normalized to satisfy $\sum_{u\in V}\hat{p}_N(u)=1$, thereby preserving their probabilistic interpretation while maintaining the relative ranking.

\begin{lemma}[Uniform Group-Dangling Contribution]
\label{lem:vanishing}
Let $g\in\{R,B\}$ denote a sensitive group, and define the
set of nodes having zero group-specific out-degree as
\begin{equation*}
D_g^{-} = \left\{u\in V: \operatorname{out}_g(u)=0 \right\}.
\end{equation*}
Let $\alpha_R=\phi$ and $\alpha_B=1-\phi$. For every target
node $i\in g$, the contribution of nodes in $D_g^{-}$ to the
LFPR$_N$ update at iteration $t$ is
\begin{equation*}
c_g^{(t)} = (1-\gamma) \frac{\alpha_g}{|g|} \sum_{u\in D_g^{-}} p_N^{(t-1)}(u).
\label{eq:group_dangling_contribution}
\end{equation*}
The quantity $c_g^{(t)}$ satisfies

\begin{equation*}
0 \leq c_g^{(t)} \leq (1-\gamma) \frac{\alpha_g}{|g|}.
\label{eq:group_dangling_bound_general}
\end{equation*}
Since $c^{(t)}_g$ contains no quantity depending on the target node $i$ or its degree class, it is identical for all target nodes belonging to group $g$. Hence, the group-dangling contribution constitutes a group-wise uniform additive correction and does not introduce degree-dependent variation within the group. Accordingly, the reduced mean-field approximation omits this uniform correction and retains only the degree-dependent propagation component.
\end{lemma}
\subsection{Approximation of LFPR$_U$}
\label{subsec4.2}
We next derive a mean-field approximation for (LFPR$_U$). Similar to LFPR$_N$, we first express the matrix-based stationary equation in an equivalent node-level form and then aggregate structurally similar nodes into group-aware degree classes. 
The stationary LFPR$_U$ vector satisfies
\begin{equation}
    p^T_U =  \gamma v^T_N + (1-\gamma)p^T_U(P_L + \delta_R x^T + \delta_B y^T)
\label{eq:eqCloseFormU}.
\end{equation}
where $P_L$ denotes the locally fair neighborhood transition matrix, $\delta_R$ and $\delta_B$ denote the residual redistribution masses associated with groups $R$ and $B$, and $x$ and $y$ define the corresponding uniform redistribution vector. Specifically, $x(i) = 1/|R|$ for $i\in R$ and $y(j) = 1/|B|$ for $j\in B$. The vector $v_N$ denotes the $\phi$-fair teleportation distribution introduced in Section \ref{sec:subsec2}, where $\phi\in (0,1)$ specifies the desired fairness proportion assigned to the Red group.

The dangling-node set is defined as
\begin{equation*}
    D = \{u\in V \mid \text{out}(u) = 0\}
\end{equation*}
The non-dangling nodes are partitioned into two locally deficient sets according to the sensitive-group composition of their outgoing neighborhoods. The Red-deficient set is
\begin{equation*}
L_R=\left\{u\in V\setminus D:\frac{\operatorname{out}_R(u)}{\operatorname{out}(u)}<\phi\right\}
\end{equation*}
whereas the Blue-deficient set is
\begin{equation*}
L_B=\left\{u\in V\setminus D:\frac{\operatorname{out}_R(u)}{\operatorname{out}(u)}\geq\phi\right\}
\end{equation*}
Thus, $L_R$ contains nodes whose outgoing neighborhoods contain fewer Red neighbors than required by the target proportion $\phi$. Similarly, $L_B$ contains nodes whose outgoing neighborhoods are deficient in Blue neighbors. For a source node $u\in L_R$, the locally fair neighborhood propagation coefficient and the corresponding Red residual mass are given by
\begin{equation}
\rho_R(u)=\frac{1-\phi}{\operatorname{out}_B(u)}, \qquad \delta_R(u)=\phi-(1-\phi)\frac{\operatorname{out}_R(u)}{\operatorname{out}_B(u)}.
\label{eq:ueq1}
\end{equation}
Similarly, for a source node $u\in L_B$, the corresponding neighborhood propagation coefficient and Blue residual mass are given by
\begin{equation}
\rho_B(u)=\frac{\phi}{\operatorname{out}_R(u)}, \qquad \delta_B(u)=(1-\phi)-\phi\frac{\operatorname{out}_B(u)}{\operatorname{out}_R(u)}.
\label{eq:ueq2}
\end{equation}
By the definitions of $L_R$ and $L_B$, the residual values in Eqs. (\ref{eq:ueq1}) and (\ref{eq:ueq2}) are non-negative. For a dangling node $u\in D$, neighborhood propagation is not possible. Therefore, its probability mass is redistributed according to the target proportions, with fraction $\phi$ assigned to the Red group and fraction $1-\phi$ assigned to the Blue group.

The stationary formulation in Eq. (\ref{eq:eqCloseFormU}) induces the following iterative update for each node $i\in V$:
\begin{equation}
\begin{aligned}
p_U^{(t)}(i) = \gamma v_N(i) & + (1-\gamma) \sum_{u\in V} p_U^{(t-1)}(u) \left[P_L[u,i] \right. \\ & \left. + \delta_R(u)\,x(i) + \delta_B(u)\,y(i)\right].
\label{eq:ueq3}
\end{aligned}
\end{equation}
The first term in Eq. (\ref{eq:ueq3}) corresponds to teleportation, and the remaining terms account for locally fair neighborhood propagation and the uniform redistribution of residual probability mass. The node-level LFPR$_U$ update can be expressed in a unified group-wise form. For a target node $i\in g$, where
$g\in\{R,B\}$,
\begin{equation}
\begin{aligned}
p_U^{(t)}(i)
&\;= \gamma\frac{\alpha_g}{|g|} +(1-\gamma) \Bigg[ \sum_{u\in L_R} \frac{(1-\phi)a_{ui}}{\operatorname{out}_B(u)} p_U^{(t-1)}(u) \\ &+ \sum_{u\in L_B} \frac{\phi a_{ui}}{\operatorname{out}_R(u)} p_U^{(t-1)}(u) + \frac{1}{|g|} \sum_{u\in L_g} \delta_g(u)p_U^{(t-1)}(u) \\ &+ \frac{\alpha_g}{|g|} \sum_{u\in D} p_U^{(t-1)}(u) \Bigg], \qquad i\in g .
\end{aligned}
\label{eq:lfpru_unified}
\end{equation}
where $a_{ui}=1$ if $(u,i)\in E$, and $a_{ui}=0$ otherwise. Here, $\alpha_R=\phi$ and $\alpha_B=1-\phi$, while $L_g=L_R$ and $\delta_g=\delta_R$ for $g=R$, and $L_g=L_B$ and $\delta_g=\delta_B$ for $g=B$. Thus, setting $g=R$ or $g=B$ in Eq. (\ref{eq:lfpru_unified}) recovers the corresponding Red and Blue group LFPR$_U$ update rules, respectively.

Unlike LFPR$_N$, which enforces local fairness through group-normalized transition probabilities, LFPR$_U$ achieves local fairness by uniformly redistributing residual propagation mass to the locally underrepresented group. The average LFPR$_U$ score of degree class $k_g$ at iteration $t$ is defined as
\begin{equation}
p_U^{(t)}(k_g) \equiv \frac{1}{nP(k_g)} \sum_{i\in k_g} p_U^{(t)}(i).
\label{eq:ueq4}
\end{equation}
We first derive the mean-field approximation for a Red target degree class $k_R$. Averaging Eq. (\ref{eq:ueq4}) over all nodes
\begin{equation}
\begin{aligned}
& \overline{p}_U^{(t)}(k_R) =\; \frac{\gamma\phi}{|R|} + \frac{1-\gamma}{nP(k_R)} \sum_{i\in k_R} \Bigg[\sum_{u\in L_R} \frac{(1-\phi)a_{ui}}{\operatorname{out}_B(u)} \,p_U^{(t-1)}(u) \\ & \qquad + \sum_{u\in L_B} \frac{\phi a_{ui}}{\operatorname{out}_R(u)} \,p_U^{(t-1)}(u)\Bigg] + \frac{(1-\gamma)S_R^{(t-1)}}{|R|},
\label{eq:ueq5}
\end{aligned}
\end{equation}
where
\begin{equation*}
S_R^{(t-1)} = \sum_{u\in L_R} \delta_R(u)\, p_U^{(t-1)}(u) + \phi \sum_{u\in D} p_U^{(t-1)}(u).
\end{equation*}
denotes the total residual mass redistributed to the Red group.

The residual term $S^{(t-1)}_R$ depends only on the source-side probability mass and is independent of the target node $i$. Hence, averaging over the target degree class leaves this term unchanged, yielding a class-independent additive contribution in Eq. (\ref{eq:ueq5}). To simplify the predecessor propagation terms, we partition the source nodes by degree class. Eq. (\ref{eq:ueq5}) can then be written as
\begin{equation}
\begin{aligned}
& \overline{p}_U^{(t)}(k_R) = \frac{\gamma\phi}{|R|} + \frac{1-\gamma}{nP(k_R)} \sum_{k^{'}} \Bigg[\frac{1-\phi}{\operatorname{out}_B(k')} \sum_{i\in k_R} \sum_{u\in k^{'}\cap L_R} a_{ui}\, \\ & p_U^{(t-1)}(u) + \frac{\phi}{\operatorname{out}_R(k^{'})} \sum_{i\in k_R} \sum_{u\in k^{'}\cap L_B} a_{ui}\, p_U^{(t-1)}(u)\Bigg] + \frac{1-\gamma}{|R|} S_R^{(t-1)}.
\end{aligned}
\label{eq:ueq7}
\end{equation}
Under the degree-class mean-field approximation, nodes belonging to the same source degree class are assumed to have approximately equal LFPR$_U$ scores. Therefore,
\begin{equation}
p_U^{(t-1)}(u) \approx \overline{p}_U^{(t-1)}(k'), \qquad u\in k'.
\label{eq:ueq6}
\end{equation}
Using Eq. (\ref{eq:ueq6}), the first predecessor term is approximated as
\begin{equation}
\sum_{i\in k_R} \sum_{u\in k'\cap L_R} a_{ui}\, p_U^{(t-1)}(u) \approx \overline{p}_U^{(t-1)}(k') \sum_{i\in k_R} \sum_{u\in k'\cap L_R} a_{ui}.
\label{eq:ueq8}
\end{equation}
and similarly,
\begin{equation}
\sum_{i\in k_R} \sum_{u\in k'\cap L_B} a_{ui}\, p_U^{(t-1)}(u) \approx \overline{p}_U^{(t-1)}(k') \sum_{i\in k_R} \sum_{u\in k'\cap L_B} a_{ui}.
\label{eq:ueq9}
\end{equation}
Define
\begin{equation}
E_{k^{'} \cap L_R \rightarrow k_R} = \sum_{i\in k_R} \sum_{u\in k^{'}\cap L_R} a_{ui}.
\label{eq:ueq10}
\end{equation}
as the total number of directed edges from source nodes belonging to degree class $k^{'}$ and deficient set $L_R$ to nodes in the target degree class $k_R$. Similarly, define
\begin{equation}
E_{k^{'} \cap L_B \rightarrow k_R} = \sum_{i\in k_R} \sum_{u\in k^{'}\cap L_B} a_{ui}.
\label{eq:ueq11}
\end{equation}
Let $k_{R,\text{in}}$ denote the in-degree associated with the target degree class $k_R$. The corresponding edge counts satisfy
\begin{equation}
E_{k^{'} \cap L_R \rightarrow k_R} = k_{R,\mathrm{in}}\, nP(k_R)\, P_{\mathrm{in}}\!\left(k^{'} \cap L_R \mid k_R\right).
\label{eq:ueq12}
\end{equation}
and 
\begin{equation}
E_{k^{'} \cap L_B \rightarrow k_R} = k_{R,\mathrm{in}}\, nP(k_R)\, P_{\mathrm{in}}\!\left(k^{'} \cap L_B \mid k_R\right).
\label{eq:ueq13}
\end{equation}
Here, $P_{\mathrm{in}}(k^{'} \cap L_R \mid k_R)$ denotes the probability that a predecessor of a node in the target degree class $k_R$ belongs simultaneously to the source degree class $k'$ and the deficient set $L_R$. The quantity $P_{\mathrm{in}}(k^{'} \cap L_B \mid k_R)$ is defined analogously. Substituting Eqs. (\ref{eq:ueq6})-(\ref{eq:ueq13}) into Eq. (\ref{eq:ueq7}) yields
\begin{equation}
\label{eq:ueq14}
\begin{aligned}
& \overline{p}_U^{(t)}(k_R) \approx\; \frac{\gamma\phi}{|R|} + (1-\gamma)\, k_{R,\mathrm{in}} \sum_{k^{'}\subseteq L_R} \frac{(1-\phi)\, P_{\mathrm{in}}(k^{'}\cap L_R \mid k_R)} {\operatorname{out}_B(k^{'})} \,\\ &\overline{p}_U^{(t-1)}(k^{'})  + (1-\gamma)\, k_{R,\mathrm{in}} \sum_{k^{'}\subseteq L_B} \frac{\phi\, P_{\mathrm{in}}(k^{'}\cap L_B \mid k_R)} {\operatorname{out}_R(k^{'})} \,\overline{p}_U^{(t-1)}(k^{'}) \\ & \qquad \qquad \qquad \qquad + \frac{1-\gamma}{|R|} S_R^{(t-1)}.
\end{aligned}
\end{equation}
Under the degree-degree uncorrelated network assumption, the degree class of a predecessor becomes independent of the target degree class. Consequently,
\begin{equation}
P_{\mathrm{in}}(k^{'} \mid k_R) =  \frac{\operatorname{out}(k^{'})\,P(k^{'})} {\langle d_{\mathrm{in}}\rangle}.
\label{eq:ueq15}
\end{equation}
where $\langle d_{in}\rangle$ denotes the average in-degree of the network.

Because membership in $L_R$ and $L_B$ is determined by the group-specific outgoing degrees contained in degree class $k^{'}$, we define the indicators
\begin{equation}
I_R(k^{'}) = \mathbf{1}_{\{k^{'} \in L_R\}}, \qquad I_B(k^{'}) = \mathbf{1}_{\{k^{'} \in L_B\}}.
\label{eq:ueq16}
\end{equation}
It follows that
\begin{equation}
P_{\mathrm{in}}(k^{'} \cap L_R \mid k_R) = \frac{\operatorname{out}(k^{'})\,P(k^{'})\,I_R(k^{'})} {\langle d_{\mathrm{in}}\rangle}.
\label{eq:ueq17}
\end{equation}
and 
\begin{equation}
P_{\mathrm{in}}(k^{'} \cap L_B \mid k_R) = \frac{\operatorname{out}(k^{'})\,P(k^{'})\,I_B(k^{'})} {\langle d_{\mathrm{in}}\rangle}.
\label{eq:ueq18}
\end{equation}
Substituting Eqs. (\ref{eq:ueq17}) and (\ref{eq:ueq18}) into Eq. (\ref{eq:ueq14}) and restricting the sums to the corresponding deficient degree classes yields
\begin{equation}
\label{eq:ueq19}
\begin{aligned}
&\overline{p}_U^{(t)}(k_R) \approx\; \frac{\gamma\phi}{|R|} + \frac{(1-\gamma)\,k_{R,\mathrm{in}}(1-\phi)} {\langle d_{\mathrm{in}}\rangle} \sum_{k^{'}\subseteq L_R} \frac{\operatorname{out}(k^{'})} {\operatorname{out}_B(k^{'})} \\ & P(k^{'})\, \overline{p}_U^{(t-1)}(k^{'}) + \frac{(1- \gamma)\,k_{R,\mathrm{in}}\phi} {\langle d_{\mathrm{in}}\rangle} \sum_{k^{'}\subseteq L_B} \frac{\operatorname{out}(k^{'})} {\operatorname{out}_R(k^{'})} \\ & P(k^{'})\, \overline{p}_U^{(t-1)}(k') + \frac{1-\gamma}{|R|} S_R^{(t-1)}.
\end{aligned}
\end{equation}
To obtain a tractable closed-form expression, we introduce conditional edge-mixing coefficients over the deficient sets. For $L_R$, define
\begin{equation*}
\beta_{R}^{L_R} = \frac{ \displaystyle \sum_{k^{'}\subseteq L_R} P(k^{'})\,\operatorname{out}_R(k^{'})}{ \displaystyle \sum_{k^{'}\subseteq L_R} P(k^{'})\,\operatorname{out}(k^{'})}, \beta_{B}^{L_R} = \frac{\displaystyle \sum_{k^{'}\subseteq L_R} P(k^{'})\,\operatorname{out}_B(k^{'})}{\displaystyle \sum_{k^{'}\subseteq L_R} P(k^{'})\,\operatorname{out}(k^{'})}.
\end{equation*}
Similarly, for $L_B$, define
\begin{equation*}
\beta_{R}^{L_B} = \frac{ \displaystyle \sum_{k^{'}\subseteq L_B} P(k^{'})\,\operatorname{out}_R(k^{'})}{ \displaystyle \sum_{k^{'}\subseteq L_B} P(k^{'})\,\operatorname{out}(k^{'})}, \beta_{B}^{L_B} = \frac{\displaystyle \sum_{k^{'}\subseteq L_B} P(k^{'})\,\operatorname{out}_B(k^{'})}{\displaystyle \sum_{k^{'}\subseteq L_B} P(k^{'})\,\operatorname{out}(k^{'})}.
\end{equation*}
These quantities satisfy
\begin{equation*}
\label{eq:beta_sum}
\beta_{R}^{L_R}+\beta_{B}^{L_R}=1,
\qquad
\beta_{R}^{L_B}+\beta_{B}^{L_B}=1.
\end{equation*}
Under the conditional edge-mixing closure, the group-specific outgoing degrees of a source class in $L_R$ are approximated using the corresponding aggregate edge-mixing proportions of the Red-deficient set:

\begin{equation}
\left. 
\begin{aligned} \operatorname{out}_R(k^{'}) &\approx \beta_R^{L_R}\operatorname{out}(k^{'})\\ \operatorname{out}_B(k^{'}) &\approx \beta_B^{L_R}\operatorname{out}(k^{'})
\end{aligned}
\right\} \qquad k^{'}\subseteq L_R .
\label{eq:outB_LR}
\end{equation}
Similarly, for a source class in $L_B$, the group-specific outgoing degrees are approximated using the corresponding aggregate edge-mixing proportions of the Blue-deficient set:

\begin{equation}
\left.
\begin{aligned} 
\operatorname{out}_R(k^{'}) &\approx \beta_R^{L_B}\,\operatorname{out}(k^{'})\\ \operatorname{out}_B(k^{'}) &\approx \beta_B^{L_B}\,\operatorname{out}(k^{'})
\end{aligned}
\right\} \qquad k^{'} \subseteq L_B .
\label{eq:outR_LB}
\end{equation}
Using Eqs. (\ref{eq:outB_LR}) and (\ref{eq:outR_LB}), Eq. (\ref{eq:ueq19}) becomes
\begin{equation}
\begin{aligned}
&\overline{p}_{U}^{(t)}(k_R) \approx  \frac{\gamma\phi}{|R|} + \frac{(1-\gamma)k_{R,\mathrm{in}}}{\langle d_{\mathrm{in}}\rangle} \Bigg[ \frac{1-\phi}{\beta_{B}^{L_R}} \sum_{k^{'} \subseteq L_R} P(k^{'})\,\overline{p}_{U}^{(t-1)}(k^{'}) \\ &\qquad + \frac{\phi}{\beta_{R}^{L_B}} \sum_{k^{'} \subseteq L_B} P(k^{'})\,\overline{p}_{U}^{(t-1)}(k^{'}) \Bigg] + \frac{1-\gamma}{|R|} S_{R}^{(t-1)} .
\end{aligned}
\label{eq:LFPRU_reduced_47}
\end{equation}

Because the LFPR$_U$ score vector is normalized, the degree-class averages satisfy the exact identity
\begin{equation*}
\label{eq:normalization} \sum_{k^{'}} P(k^{'})\,\overline{p}_U^{(t-1)}(k^{'}) = \frac{1}{n}.
\end{equation*}
To obtain a closed-form approximation for the restricted deficient-set sums, we introduce a deficient-set stationary-mass closure. Specifically, we assume that the average LFPR$_U$ score per node within each deficient set is approximately equal to the network-wide average score $1/n$. Accordingly,
\begin{equation}
\label{eq:LR_mass} \sum_{k^{'} \subseteq L_R} P(k^{'})\,\overline{p}_U^{(t-1)}(k^{'}) \approx \frac{P(L_R)}{n},
\end{equation}
and 
\begin{equation}
\label{eq:LB_mass} \sum_{k^{'} \subseteq L_B} P(k^{'})\,\overline{p}_U^{(t-1)}(k^{'}) \approx \frac{P(L_B)}{n},
\end{equation}
Substituting Eqs. (\ref{eq:LR_mass}) and (\ref{eq:LB_mass}) into Eq. (\ref{eq:LFPRU_reduced_47}) gives
\begin{equation}
\label{eq:ueq_final_R}
\begin{aligned}
\overline{p}_U^{(t)}(k_R) \approx\;& \frac{\gamma\phi}{|R|} + \frac{(1-\gamma)\,k_{R,\mathrm{in}}} {n\langle d_{\mathrm{in}}\rangle} \left(\frac{(1-\phi)\,P(L_R)}{\beta_{B}^{L_R}} + \frac{\phi\,P(L_B)}{\beta_{R}^{L_B}} \right) \\ & + \frac{1-\gamma}{|R|} S_R^{(t-1)}.
\end{aligned}
\end{equation}
Approximation of the residual redistribution mass. The residual mass associated with the Red-deficient set is approximated as
\begin{equation}
\label{eq:SR}
S_R^{(t-1)} = \sum_{u\in L_R} \delta_R(u)\, p_U^{(t-1)}(u) + \phi \sum_{u\in D} p_U^{(t-1)}(u).
\end{equation}
Partitioning the source nodes according to their degree classes gives
\begin{equation}
\label{eq:SR_degree}S_R^{(t-1)} = \sum_{k^{'} \subseteq L_R} \sum_{u \in k^{'}} \delta_R(u)\, p_U^{(t-1)}(u) + \phi \sum_{k^{'} \subseteq D} \sum_{u \in k^{'}} p_U^{(t-1)}(u).
\end{equation}
Within each source degree class, we apply
\begin{equation}
p_U^{(t-1)}(u) \approx \overline{p}_U^{(t-1)}(k^{'}), \qquad \delta_R(u) \approx \overline{\delta}_R(k^{'}), \qquad u\in k^{'}.
\end{equation}
Because degree class $k^{'}$ contains $nP(k^{'})$ nodes, Eq.(\ref{eq:SR_degree}) becomes
\begin{equation}
\label{eq:SR_mean_field_55}
\begin{aligned}
S_R^{(t-1)} \approx\;& n \sum_{k^{'} \subseteq L_R} P(k^{'})\, \overline{\delta}_R(k^{'})\, \overline{p}_U^{(t-1)}(k^{'}) \\ & + \phi\, n \sum_{k^{'} \subseteq D} P(k^{'})\, \overline{p}_U^{(t-1)}(k^{'}).
\end{aligned}
\end{equation}
Under the same conditional edge-mixing approximation, the residual coefficient for classes in $L_R$ is approximated by a deficient-set-level representative residual coefficient.
\begin{equation}
\label{eq:deltaR_mean}
\delta_{R}(k^{'}) \approx \overline{\delta}_{R} = \phi - (1-\phi) \frac{\beta_{R}^{L_R}}
{\beta_{B}^{L_R}}.
\end{equation}
Similarly, for classes in $L_B$,
\begin{equation}
\label{eq:deltaB_mean}
\delta_{B}(k^{'}) \approx \overline{\delta}_{B} = (1-\phi) - \phi \frac{\beta_{B}^{L_B}}
{\beta_{R}^{L_B}}.
\end{equation}
Using Eq. (\ref{eq:deltaR_mean}), the Red residual mass in Eq. (\ref{eq:SR_mean_field_55}) can be simplified. For the dangling-node classes, we adopt the analogous mass closure
\begin{equation}
\label{eq:dangling_mass}
\sum_{k^{'} \subseteq D} P(k^{'})\,\overline{p}_U^{(t-1)}(k^{'}) \approx \frac{P(D)}{n}.
\end{equation}
Substituting Eqs. (\ref{eq:LR_mass}), (\ref{eq:deltaR_mean}), and (\ref{eq:dangling_mass}) into Eq. (\ref{eq:SR_mean_field_55}) yields
\begin{equation}
\label{eq:SR_final}
S_R^{(t-1)} \approx P(L_R) \left(\phi - (1-\phi) \frac{\beta_{R}^{L_R}} {\beta_{B}^{L_R}}\right) + \phi\,P(D) = \hat{S}_R
\end{equation}
By an analogous derivation, the residual redistribution mass toward the Blue group is
\begin{equation}
\label{eq:SB_final}
S_B^{(t-1)} \approx P(L_B) \left((1-\phi) - \phi \frac{\beta_{B}^{L_B}} {\beta_{R}^{L_B}} \right) + (1-\phi)\,P(D) = \hat{S}_B
\end{equation}
substituting Eq. (\ref{eq:SR_final}) into Eq. (\ref{eq:ueq_final_R}) and considering the stationary limit gives the following approximation for a Red degree class:
\begin{equation}
\label{eq:pU_final_R}
\begin{aligned}
& \overline{p}_U(k_R) \approx \; \frac{\gamma\phi}{|R|} + \frac{(1-\gamma)\,k_{R,\mathrm{in}}} {n\langle d_{\mathrm{in}}\rangle} \left[\frac{(1-\phi)\,P(L_R)}{\beta_{B}^{L_R}} + \frac{\phi\,P(L_B)}{\beta_{R}^{L_B}} \right] \\ & \qquad + \frac{1-\gamma}{|R|} \left[P(L_R) \left(\phi - (1-\phi) \frac{\beta_{R}^{L_R}}{\beta_{B}^{L_R}} \right) + \phi\,P(D)\right].
\end{aligned}
\end{equation}
Similarly, for a Blue target degree class, we obtain 
\begin{equation}
\label{eq:pU_final_B}
\begin{aligned}
& \overline{p}_U(k_B) \approx\frac{\gamma(1-\phi)}{|B|} + \frac{(1-\gamma)\,k_{B,\mathrm{in}}} {n\langle d_{\mathrm{in}}\rangle} \left[\frac{(1-\phi)\,P(L_R)}{\beta_{B}^{L_R}} + \phi \right.\\ & \left. \frac{\,P(L_B)}{\beta_{R}^{L_B}} \right]  + \frac{1-\gamma}{|B|} \left[P(L_B) \left((1-\phi) - \frac{\phi\beta_{B}^{L_B}}{\beta_{R}^{L_B}} \right) + (1-\phi)\,P(D) \right].
\end{aligned}
\end{equation}
Define the common propagation coefficient
\begin{equation}
\label{eq:ThetaU}
\Theta_U = \frac{(1-\phi)\,P(L_R)}{\beta_{B}^{L_R}} + \frac{\phi\,P(L_B)}{\beta_{R}^{L_B}}.
\end{equation}
The stationary LFPR$_U$ approximation for a node u belonging to sensitive group $g\in \{R,B\}$ can be written compactly as
\begin{equation}
\label{eq:LFPRU_final}
\hat{p}_U(u) \approx \gamma\frac{\alpha_g}{|g|} + (1-\gamma) \frac{k_{u,\mathrm{in}}} {n\langle d_{\mathrm{in}}\rangle} \Theta_U + (1-\gamma) \frac{\hat{S}_g}{|g|},
\end{equation}
where $\alpha_R = \phi$ and $\alpha_B = 1-\phi$. 
Similarly, the stationary LFPR$_U$ approximation is normalized after computation such that $\sum_{u\in V}\hat{p}_U(u)=1$. This normalization ensures that the approximate scores form a valid probability distribution while preserving the relative ordering produced by the mean-field approximation.

\section{Fluctuation Analysis of LFPR Mean-Field Approximations}\label{sec:Fluct}
The mean-field approximations derived in Sections \ref{subsec4.1} and \ref{subsec4.2} characterize the average stationary behavior of locally fair propagation at the degree-class level. By replacing node-specific neighborhood structure with class-averaged interaction patterns, the framework assigns the same approximate score to nodes within a given structural class. Consequently, the approximation does not capture the full node-level variability arising from differences in local neighborhood composition and fairness-aware propagation pathways.

To quantify these deviations, we next analyze the fluctuation behavior of the proposed LFPR mean-field approximations. In particular, we examine the second-order behavior of stationary LFPR scores around their degree-class means and derive analytical expressions for the corresponding degree-class variance under locally fair propagation. The analysis further characterizes the asymptotic scaling of these fluctuations in the large-degree regime, providing theoretical insight into the accuracy of the heterogeneous mean-field approximations.

\subsection{Fluctuation Analysis of LFPR$_N$}\label{sec:FluctN}
We first analyze the fluctuation behavior of LFPR$_N$ for Red group nodes by squaring the node-level LFPR$_N$ update in Eq. (\ref{eq:lfprn_unified}). Lemma \ref{lem:vanishing} shows that the group-dangling contribution is a uniform group-wise additive term, independent of the target node’s degree class. Following the reduced mean-field approximation adopted in Section \ref{subsec4.1}, we omit this uniform contribution and analyze the fluctuations induced by the degree-dependent propagation term. Accordingly, the second-order LFPR$_N$ dynamics are approximated as
\begin{equation}
\begin{aligned}
& (p^{(t)}_N)^2(i) = \; \gamma^2 \frac{\phi^2}{|R|^2} + 2\gamma(1 - \gamma)\frac{\phi^2}{|R|} \sum_{j:\mathrm{out}_R(j)>0} \frac{a_{ji}}{\mathrm{out}_R(j)} p^{(t-1)}_{N}(j) \\ & + (1-\gamma)^2\phi^2 \sum_{j:\mathrm{out}_R(j)>0} \frac{a_{ji}}{out_R^2(j)} (p^{(t-1)}_{N}(j))^2 + (1-\gamma)^2\phi^2 \\ & \sum_{{\substack{j\neq j'\\
\operatorname{\mathrm{out}}_R(j)>0,\;
\operatorname{\mathrm{out}}_R(j')>0}}} \frac{a_{ji}a_{j^{'}i}} {\mathrm{out}_R(j)\,\mathrm{out}_R(j^{'})} p^{(t-1)}_{N}(j)p^{(t-1)}_{N}(j^{'}). 
\end{aligned}
\label{eq:eqF1}
\end{equation}
The leading term represents the teleportation contribution, whereas the remaining terms capture second-order locally fair propagation effects. In particular, the final term accounts for interactions between distinct incoming predecessor contributions. To characterize these fluctuations at the degree-class level, we average the second-order dynamics over the same Red-group degree class $k_R$. Accordingly, the degree-class second-order moment is defined as
\begin{equation}
\overline{p^2}_N^{(t)}(k_R) = \frac{1}{n P(k_R)} \sum_{i \in k_R} (p_N^{(t)}(i))^2
\label{eq:eqF2}
\end{equation}
Substituting Eq. (\ref{eq:eqF1}) into Eq. (\ref{eq:eqF2}) and applying the mean-field approximation yields the following degree-class second-order dynamics at the degree-class level:
\begin{equation}
\begin{aligned}
& \overline{p^2}_N^{(t)}(k_R) =\; \frac{\gamma^2 \phi^2}{|R|^2} + \frac{2\gamma(1-\gamma)\phi^2}{|R|} k_{R,\mathrm{in}} \sum_{k^{'}:\mathrm{out}_R(k^{'})>0} \frac{P_{\mathrm{in}}(k^{'}\mid k_R)} {\mathrm{out}_R(k^{'})} \\ & \overline{p}_N^{(t-1)}(k^{'}) + (1-\gamma)^2\phi^2 k_{R,\mathrm{in}} \sum_{k^{'}:\mathrm{out}_R(k^{'})>0} \frac{P_{\mathrm{in}}(k^{'}\mid k_R)}{\mathrm{out}_R^2(k^{'})} \\ & \overline{p^2}_N^{(t-1)}(k^{'})  + (1 - \gamma)^2 \phi^2  k_{R,\mathrm{in}} (k_{R,\mathrm{in}}-1) \sum_{k^{'}:\mathrm{out}_R(k^{'})>0}\\&\sum_{k^{''}:\mathrm{out}_R(k^{''})>0} \frac{ P_{\mathrm{in}}(k^{'},k^{''}\mid k_R)} { \mathrm{out}_R(k^{'})\,\mathrm{out}_R(k^{''})} \overline{p}_N^{(t-1)}(k^{'})\overline{p}_N^{(t-1)}(k^{''})
\end{aligned}
\label{eq:eqF3}
\end{equation}

Here, $P_{\mathrm{in}}(k^{'},k^{''}\mid k_R)$ denotes the joint conditional probability that two distinct incoming predecessors of a node in the Red-group degree class $k_R$ belong to degree classes $k^{'}$ and $k^{''}$, respectively. Under the degree-degree uncorrelated mean-field assumption, higher-order neighboring degree dependencies are neglected, allowing the joint conditional probability to factorize as
\begin{equation*}
P_{\mathrm{in}}(k^{'},k^{''}\mid k_R) \approx P_{\mathrm{in}}(k^{'}\mid k_R)\, P_{\mathrm{in}}(k^{''}\mid k_R)
\end{equation*}
Accordingly, incoming locally fair propagation contributions from distinct predecessor degree classes are treated as statistically independent at the degree-class level. The variance of the Red-group degree class $k_R$ is then defined as
\begin{equation}
\sigma^2_{(t)}(k_R) = \overline{p^2}_N^{(t)}(k_R) - (\overline{p}_N^{(t)})^2(k_R)
\label{eq:eqF4}
\end{equation}
where $\overline{p}_N^{(t)}(k_R)$ denotes the corresponding degree-class mean-field approximation derived in Eq. (\ref{eq:eq10}). Combining Eqs. (\ref{eq:eqF3}) and (\ref{eq:eqF4}), yields the following recursive relation for the variance of the Red-group degree class $k_R$.
\begin{equation}
\begin{aligned}
 & \frac{\sigma_{(t)}^2(k_R)}{(1-\gamma)^2 \phi^2} = \; k_{R,\mathrm{in}} \sum_{k^{'}:\mathrm{out}_R(k^{'})>0} \frac{P_{\mathrm{in}}(k^{'} \mid k_R)} {\mathrm{out}_R^2(k^{'})} \sigma_{(t-1)}^2(k^{'}) \\ & + k_{R,\mathrm{in}} \sum_{k^{'}:\mathrm{out}_R(k^{'})>0} \frac{P_{\mathrm{in}}(k^{'} \mid k_R)} {\mathrm{out}_R^2(k^{'})} (\overline{p}_N^{(t-1)})^2(k^{'})  \\& - k_{R,\mathrm{in}} \left[\sum_{k^{'}:\mathrm{out}_R(k^{'})>0} \frac{P_{\mathrm{in}}(k^{'} \mid k_R)} {\mathrm{out}_R(k^{'})} \overline{p}_N^{(t-1)}(k^{'}) \right]^2 
\end{aligned}
\label{eq:eqF5}
\end{equation}
In the stationary limit $t \rightarrow \infty$, and under the degree-degree uncorrelated networks, Eq. (\ref{eq:eqF5}) reduces to the following analytical expression for the degree-class variance.

\begin{equation}
\label{eq:eqF6}
\begin{aligned}
& \sigma^2(k_R) = \frac{(1-\gamma)^2\phi^2 k_{R,\mathrm{in}}} {1- \dfrac{(1-\gamma)^2\phi^2} {\langle \mathrm{out}_R\rangle} \left\langle \frac{k_{\mathrm{in}}} {\mathrm{out}_R(k)} \right\rangle_{\mathrm{out}_R(k)>0} } \\[2ex] &\quad\times \left[\frac{1}{\langle \mathrm{out}_R\rangle} \left\langle \frac{(\overline{p}_N(k))^2} {\mathrm{out}_R(k)} \right\rangle_{\mathrm{out}_R(k)>0} - \frac{1} {n^2\langle \mathrm{out}_R\rangle^2} \right].
\end{aligned}
\end{equation}
In large networks, the term $n^{-2}\langle \mathrm{out}_R\rangle^{-2}$ is assumed to be negligible compared with the leading term in Eq. (\ref{eq:eqF6}). Moreover, under the mean-field assumption, the denominator correction term is assumed to remain sufficiently small. Consequently, the variance can be approximated by retaining the leading contribution.

Furthermore, in fluctuation analysis, the degree-dependent locally fair propagation component of the LFPR$_N$ approximation in Eq. (\ref{eq:eq12}) determines the variation across degree classes. The teleportation contribution $\gamma \phi/|R|$ remains uniform across all nodes in the Red group and therefore does not contribute to intra-class fluctuations. Accordingly, the degree-dependent part of the LFPR$_N$ approximation can be written as
\begin{equation*}
\overline{p}_N(k_R) \approx (1-\gamma)\phi\frac{k_{R,\mathrm{in}}}{n\, \langle \mathrm{out}_R\rangle }
\end{equation*}
Accordingly, under these assumptions, Eq. (\ref{eq:eqF6}) simplifies to
\begin{equation*}
\sigma^2(k_R) \approx \frac{(1-\gamma)^4\phi^{4}k_{R, \mathrm{in}}}{n^2\langle \mathrm{out}_R\rangle^{3}}\left\langle\ \frac{k^{2}_{\mathrm{in}}}{\mathrm{out}_R(k)} \right\rangle_{\text{out}_R(k) > 0}
\label{eq:eqF7}
\end{equation*}
Consequently, the coefficient of variation for the Red-group degree class $k_R$ is given by 
\begin{equation}
\frac{\sigma{(k_R)}}{\overline{p}_N(k_R)} \simeq (1-\gamma)\phi\left[\left\langle\ \frac{k^{2}_{\mathrm{in}}}{\mathrm{out}_R(k)} \right\rangle_{\mathrm{out}_R(k)>0} \frac{1}{\langle \mathrm{out}_R \rangle k_{R,\mathrm{in}}} \right]^{\frac{1}{2}}
\label{eq:eqF8}
\end{equation}
Following an analogous derivation for the Blue-group degree classes $k_B$, the corresponding coefficient of variation is given by
\begin{equation}
\frac{\sigma(k_B)}{\overline{p}_N(k_B)} \simeq (1-\gamma)(1-\phi)\left[\left\langle\ \frac{k^{2}_{\mathrm{in}}}{\mathrm{out}_B(k)} \right\rangle_{\mathrm{out}_B(k)>0} \frac{1}{\langle \mathrm{out}_B \rangle k_{B,\mathrm{in}}} \right]^{\frac{1}{2}}
\label{eq:eqF691}
\end{equation}
For sufficiently large in-degree, Eqs. (\ref{eq:eqF8}) and (\ref{eq:eqF691}) exhibit the following asymptotic scaling behavior:
\begin{equation}
\frac{\sigma(k_R)}{\overline{p}_N(k_R)} \sim (k_{R,\mathrm{in}})^{-1/2}, \qquad \frac{\sigma(k_B)}{\overline{p}_N(k_B)} \sim (k_{B,\mathrm{in}})^{-1/2}
\label{eq:eqF9}
\end{equation}
Equation (\ref{eq:eqF9}) shows that the relative fluctuations decay according to an inverse-square-root scaling law with increasing in-degree. Consequently, the proposed LFPR$_N$ mean-field approximation is expected to become increasingly accurate for high-in-degree nodes, as their relative fluctuations around the degree-class mean decrease with increasing in-degree. In contrast, low-degree nodes may exhibit comparatively larger fluctuations due to increased sensitivity to local neighborhood heterogeneity and fairness-conditioned transition effects.

\subsection{Fluctuation Analysis of LFPR$_U$}\label{FluctU}
The fluctuation behavior of LFPR$_U$ is analyzed within the same mean-field framework developed for LFPR$_N$ in Section \ref{sec:FluctN}, while incorporating the residual-aware locally fair propagation dynamics introduced in Section \ref{subsec4.2}. Unlike LFPR$_N$, LFPR$_U$ depends on the residual-deficient sets $L_R$ and $L_B$, which modify the locally fair propagation through their associated coefficients. As a result, the fluctuation dynamics include additional contributions arising from the residual redistribution mechanism.

Under the degree-degree uncorrelated mean-field approximation, and following the same degree-class averaging and variance decomposition procedure used in the LFPR$_N$ analysis, the resulting variance evolution equation for the Red-group degree class $k_R$ is given by

\begin{equation}
\begin{aligned}
\frac{\sigma_{(t)}^2(k_R)}{(1-\gamma)^2} &= (1-\phi)^2\, k_{R,\mathrm{in}}\sum_{k^{'}\subseteq L_R}\frac{P_{\mathrm{in}}(k^{'}\mid k_R)}{\mathrm{out}_B^2(k^{'})}\sigma^2_{(t-1)}(k^{'}) \\ & + (1-\phi)^2\, k_{R,\mathrm{in}}\sum_{k^{'}\subseteq L_R}\frac{P_{\mathrm{in}}(k^{'}\mid k_R)}{\mathrm{out}_B^2(k^{'})}(\overline{p}_U^{(t-1)})^2(k^{'}) \\ & + \phi^{2}\, k_{R,\mathrm{in}}\sum_{k^{'}\subseteq L_B} \frac{P_{\mathrm{in}}(k^{'}\mid k_R)}{\mathrm{out}_R^2(k^{'})}\sigma^2_{(t-1)}(k^{'}) \\ & + \phi^{2}\, k_{R,\mathrm{in}}\sum_{k^{'}\subseteq L_B} \frac{P_{\mathrm{in}}(k^{'}\mid k_R)}{\mathrm{out}_R^2(k^{'})}(\overline{p}_U^{(t-1)})^2(k^{'}) \\ & \, -k_{R,\mathrm{in}}\left[(1-\phi)\sum_{k^{'}\subseteq L_R}\frac{P_{\mathrm{in}}(k^{'}\mid k_R)}{\mathrm{out}_B(k^{'})}\overline{p}_U^{(t-1)}(k^{'}) \right. \\ & \left. +\,\phi\sum_{k^{'}\subseteq L_B}\frac{P_{\mathrm{in}}(k^{'}\mid k_R)}{\mathrm{out}_R(k^{'})}\overline{p}_U^{(t-1)}(k^{'}) \right]^2
\end{aligned}
\label{eq:eqFU1}
\end{equation}
Unlike LFPR$_N$, the variance dynamics of LFPR$_U$ include separate contributions from the locally deficient sets $L_R$ and $L_B$, each weighted by the corresponding conditional probabilities and propagation coefficients. In the stationary limit $t\rightarrow \infty$, and degree-degree uncorrelated networks assumption, Eq. (\ref{eq:eqFU1}) reduces to the following analytical expression for the degree-class variance:

\begin{equation*}
\begin{aligned}
\mathcal{A}_U &= \frac{(1-\phi)^2 P(L_R)} {(\beta_B^{L_R})^2\langle d_{\mathrm{in}}\rangle} \left\langle \frac{k_{\mathrm{in}}}{\mathrm{out}(k)} \right\rangle_{L_R} + \frac{\phi^2 P(L_B)} {(\beta_R^{L_B})^2\langle d_{\mathrm{in}}\rangle} \left\langle \frac{k_{\mathrm{in}}}{\mathrm{out}(k)} \right\rangle_{L_B},
\end{aligned}
\label{eq:AU}
\end{equation*}

\begin{equation*}
\begin{aligned}
\mathcal{B}_U &= \frac{(1-\phi)^2 P(L_R)} {(\beta_B^{L_R})^2\langle d_{\mathrm{in}}\rangle} \left\langle \frac{\overline{p}_U^{\,2}(k)} {\mathrm{out}(k)}\right\rangle_{L_R} \\ &\quad+ \frac{\phi^2 P(L_B)} {(\beta_R^{L_B})^2\langle d_{\mathrm{in}}\rangle} \left\langle \frac{\overline{p}_U^{\,2}(k)} {\mathrm{out}(k)} \right\rangle_{L_B}, \end{aligned} 
\label{eq:BU}
\end{equation*}

\begin{equation*}
\begin{aligned}
\mathcal{C}_U &= \frac{1} {n^2\langle d_{\mathrm{in}}\rangle^2} \left[ \frac{(1 -\phi)P(L_R)} {(\beta_B^{L_R})} + \frac{\phi P(L_B)} {(\beta_R^{L_B})} \right]^2.
\end{aligned}
\label{eq:CU}
\end{equation*}
\begin{equation}
\sigma^2(k_R) = \frac{(1-\gamma)^2 k_{R,\mathrm{in}} \left(\mathcal{B}_U -\mathcal{C}_U\right)}{1-(1-\gamma)^2\mathcal{A}_U}.
\label{eq:eqeq70}
\end{equation}
In large-scale networks, the $n^{-2}\, \langle d_{\mathrm{in}}\rangle^{-2}$ term in the numerator becomes negligible relative to the leading propagation contribution. Furthermore, under the heterogeneous mean-field approximation, the denominator correction is assumed to remain sufficiently small, so the denominator can be approximated by unity. Under these conditions, the LFPR$_U$ fluctuations are primarily determined by the residual-aware locally fair propagation contributions associated with the deficient sets $L_R$ and $L_B$. Accordingly, Eq. (\ref{eq:eqeq70}) simplifies to
\begin{equation}
\begin{aligned}
\sigma^2(k_R)  \approx & \frac{(1-\gamma)^2\, k_{R,\mathrm{in}}}{\langle d_{\mathrm{in}}\rangle} \left(\frac{(1-\phi)^2P(L_R)}{(\beta^{L_R}_{B})^2} \left\langle \frac{(\overline{p}_U(k))^2}{\mathrm{out}(k)} \right\rangle_{L_R} \right. \\ & \qquad \qquad \left. + \frac{\phi^2 P(L_B)}{(\beta^{L_B}_{R})^2}\left\langle \frac{(\overline{p}_U(k))^2}{\mathrm{out}(k)} \right\rangle_{L_B} \right)
\label{eq:eqFU3}
\end{aligned}
\end{equation}
For LFPR$_U$, only the degree-dependent residual-aware propagation component of the stationary approximation contributes to variations across degree classes. The teleportation and residual redistribution terms are uniform within the same degree class and therefore do not contribute to intra-class fluctuations. Accordingly, the degree-dependent part of the stationary LFPR$_U$ approximation can be written as
 \begin{equation}
\overline{p}_U(k_R) \approx (1-\gamma)\Theta_U \frac{k_{R,\mathrm{in}}}{n\, \langle d_{\mathrm{in}}\rangle}
\label{eq:eqFU4}
\end{equation}
 where $\Theta_U$ denotes the effective residual-aware propagation coefficient defined in Eq. (\ref{eq:ThetaU}). Substituting Eq. (\ref{eq:eqFU4}) into Eq. (\ref{eq:eqFU3}) yields.
 
 \begin{equation*}
 \begin{aligned}
\sigma^2(k_R) \approx & \frac{(1-\gamma)^4\Theta^2_{U} \, k_{R,\mathrm{in}}}{n^2\, \langle d_{\mathrm{in}} \rangle^3}  \left(\frac{(1-\phi)^2P(L_R)}{(\beta^{L_R}_{B})^2} \left\langle \frac{k^2_{\mathrm{in}}}{\mathrm{out}(k)} \right\rangle_{L_R} \right. \\ & \qquad \qquad \left. + \frac{\phi^2 P(L_B)}{(\beta^{L_B}_{R})^2}\left\langle \frac{k^2_{\mathrm{in}}}{\mathrm{out}(k)} \right\rangle_{L_B} \right)
\label{eq:eqFU5}
\end{aligned}
\end{equation*}
 Consequently, the coefficient of variation for the Red-group degree class $k_R$ is given by
\begin{equation}
\begin{aligned}
\frac{\sigma(k_R)}{\overline{p}_U(k_R)} &\simeq (1-\gamma)\Bigg[\Bigg(\frac{(1-\phi)^2P(L_R)} {(\beta^{L_R}_{B})^2} \left\langle \frac{k^2_{\mathrm{in}}} {\mathrm{out}(k)} \right\rangle_{L_R} \\ & + \frac{\phi^2P(L_B)} {(\beta^{L_B}_{R})^2} \left\langle \frac{k^2_{\mathrm{in}}} {\mathrm{out}(k)} \right\rangle_{L_B} \Bigg) \frac{1} {k_{R,\mathrm{in}} \,\langle d_{\mathrm{in}}\rangle} \Bigg]^{\frac12}.
\end{aligned}
\label{eq:eqFU6}
\end{equation}
Following an analogous derivation for the Blue-group degree classes, the corresponding expression is obtained as
\begin{equation}
\begin{aligned}
\frac{\sigma(k_B)}{\overline{p}_U(k_B)} &\simeq (1-\gamma)\Bigg[\Bigg(\frac{(1-\phi)^2P(L_R)} {(\beta^{L_R}_{B})^2} \left\langle \frac{k^2_{\mathrm{in}}} {\mathrm{out}(k)} \right\rangle_{L_R} \\ & + \frac{\phi^2P(L_B)} {(\beta^{L_B}_{R})^2} \left\langle \frac{k^2_{\mathrm{in}}} {\mathrm{out}(k)} \right\rangle_{L_B} \Bigg) \frac{1} {k_{B,\mathrm{in}} \,\langle d_{\mathrm{in}}\rangle} \Bigg]^{\frac12}.
\end{aligned}
\label{eq:eqFU7}
\end{equation}
For sufficiently large in-degree, Eqs. (\ref{eq:eqFU6}) and (\ref{eq:eqFU7}) exhibit the following asymptotic scaling behavior:
\begin{equation}
\frac{\sigma(k_R)}{\overline{p}_U(k_R)} \sim (k_{R,\mathrm{in}})^{-1/2} \qquad \frac{\sigma(k_B)}{\overline{p}_U(k_B)} \sim (k_{B,\mathrm{in}})^{-1/2}
\label{eq:eqFU8}
\end{equation}
Equation (\ref{eq:eqFU8}) indicates that the relative fluctuations of LFPR$_U$ decrease with increasing in-degree according to the same inverse-square-root scaling behavior. Consequently, despite the residual-aware locally fair propagation corrections, the mean-field LFPR$_U$ approximation remains asymptotically stable for structurally influential high-degree nodes. In contrast, lower-degree nodes are more strongly affected by residual-deficient neighborhood structure and therefore exhibit comparatively larger fluctuation effects.

\section{One-Step Local Refinement}
Although the mean-field approximations derived in Sections \ref{subsec4.1} and \ref{subsec4.2} capture the dominant stationary behavior of the locally fair propagation dynamics, they estimate node scores through degree-class averaging. As a result, nodes in the same degree class receive identical approximate scores, even though they may exhibit different local connectivity structures. To incorporate these local structural variations while preserving the computational efficiency of the mean-field approximation, we introduce a One-Step Refinement (ORF) mechanism that applies the corresponding locally fair propagation operator to the analytically obtained mean-field estimate. The proposed refinement uses the mean-field solution as an informed initialization and applies the original locally fair propagation operator exactly once. This operator-level correction incorporates node-specific neighborhood information into the class-level approximation without requiring repeated global power iterations over the full fairness-aware transition matrix.

The ORF mechanism applies uniformly to both LFPR$_N$  and LFPR$_U$. Let $\overline{p}_X(u)$denote the stationary mean-field estimate for node $u$, where $X\in \{N, U\}$ identifies the corresponding locally fair propagation model. Starting from $\overline{p}_X$, ORF evaluates the associated fairness-aware propagation operator once to obtain the refined node-level estimate
\begin{equation}
\tilde{p}_X(u) = \gamma v_X(u) + (1-\gamma)\sum_{j\in V}\overline{p}_X(j)P_X[j, u]
\label{eq:eqorf1}
\end{equation}
where $P_X$ denotes the fairness-aware transition matrix associated with model $X$, and $v_X$ denotes the corresponding teleportation distribution. For LFPR$_N$, substituting the transition matrix $P_N$ defined in Section \ref{subsec4.1} into Eq. (\ref{eq:eqorf1}):
\begin{equation*}
\begin{aligned}
\tilde{p}_N(u) = \gamma v_X(u) + & (1-\gamma) \sum_{j\in V}\overline{p}_N(j) \left[\phi P_R[j,u] + \right. \\ & \qquad \left. (1-\phi) P_B[j,u] \right] 
\end{aligned}
\end{equation*}
Similarly, for LFPR$_U$, Let $P_U = P_L + \delta_R x^T + \delta_By^T$ denote the corresponding residual-aware transition matrix. Substituting $P_U$ in Eq. (\ref{eq:eqorf1}), yields
\begin{equation*}
\begin{aligned}
\tilde{p}_U(u) & = \gamma v_X(u) + (1-\gamma) \sum_{j\in V}\overline{p}_U(j) \left[P_L[j,u] \right. \\ & \qquad \left. +\delta_R(j)x(u) + \delta_B(j)y(u) \right]
\end{aligned}
\end{equation*}
The proposed ORF mechanism performs a single operator-level correction of the mean-field estimate. Specifically, the mean-field stage captures the degree-class-level propagation structure, whereas ORF evaluates the original fairness-aware propagation dynamics using these analytical estimates as the input distribution. This introduces node-specific neighborhood information into the refined scores while preserving the computational advantage of avoiding convergence-level iteration.

From a computational perspective, ORF bridges the analytical degree-class approximation and the original node-level propagation dynamics through only one fairness-aware graph traversal. Unlike exact LFPR computation, which repeatedly applies the propagation operator until convergence, ORF evaluates this operator once from an informed mean-field initialization. Consequently, the framework combines closed-form mean-field estimation with a lightweight node-level correction, providing improved approximation fidelity without reintroducing the iterative cost of exact LFPR computation.

\section{Complexity Analysis}
\label{sec:complexity}
\subsection{Exact LFPR Methods}
Both LFPR$_N$ and LFPR$_U$ consist of a preprocessing stage followed by iterative locally fair propagation. During preprocessing, the graph is scanned once to construct the adjacency structures and compute the statistics required by the respective propagation mechanism. Since each node and edge is processed a constant number of times, this stage requires $(\mathcal{O}(m+n))$ time and $(\mathcal{O}(m+n))$ space.

After preprocessing, both methods perform iterative sparse propagation until convergence. In each iteration, all graph edges are traversed once, requiring $(\mathcal{O}(m))$ work, while node-level redistribution, normalization, and convergence computations require $(\mathcal{O}(n))$ work. Hence, the total computational work per iteration is $(\mathcal{O}(m+n))$. Using $(p)$ processing threads, the parallel runtime per iteration is reduced to $(\mathcal{O}((m+n)/p))$. Therefore, after $(T)$ iterations, both methods have a parallel runtime of $\mathcal{O}((T(m+n))/p)$ and space complexity of $(\mathcal{O}(m+n))$. LFPR$_N$ and LFPR$_U$ differ only in their redistribution mechanisms. LFPR$_N$ uses neighborhood-based group redistribution, whereas LFPR$_U$ employs residual redistribution coefficients. These differences affect only constant factors and do not change the asymptotic complexity
\subsection{Approximation Methods}
Similar to the exact methods, the approximation algorithms require a single preprocessing pass to construct the sparse graph representation and compute the statistics needed for subsequent score estimation. Consequently, the preprocessing complexity remains $(\mathcal{O}(m+n))$ in both time and space.

The Mean-Field LFPR$_N$ and LFPR$_U$ approximations eliminate iterative propagation by replacing local neighborhood interactions with aggregate graph statistics. Following preprocessing, both methods compute node scores through a constant number of parallel passes over the node set to evaluate the approximation model and perform normalization. Consequently, the total computational work is $(\mathcal{O}(n))$, yielding a parallel runtime of $(\mathcal{O}(n/p))$ using $(p)$ processing threads. Since the approximation stage requires only node-level statistics and score vectors, the additional memory requirement is $(\mathcal{O}(n))$ during the approximation stage.

The ORF refinement methods augment the mean-field approximations with a single locality-aware propagation step to recover neighborhood information lost during statistical averaging. Following the initial approximation stage, each method performs one propagation sweep over the graph together with node-level update and normalization operations. Since each edge and node is processed a constant number of times, the total computational work is $(\mathcal{O}(m+n))$. Using $(p)$ processing threads, the corresponding parallel runtime is $(\mathcal{O}((m+n)/p))$. The refinement stage requires only a constant number of additional node-level vectors, resulting in an additional memory requirement of $(\mathcal{O}(n))$. Consequently, including the graph representation constructed during preprocessing, the overall space complexity remains $(\mathcal{O}(m+n))$ for both refinement methods. Table \ref{tab:complexity_comparison} summarizes the complexity bounds.

\begin{table}[!ht]
\centering
\caption{Computational costs of the exact, mean-field (MF), and One-Step Refinement (ORF) LFPR methods. Here, n, m, T, and p denote the node count, edge count, iterations to convergence, and processing threads, respectively.}
\label{tab:complexity_comparison}
\centering
\setlength{\tabcolsep}{2.8pt}
\renewcommand{\arraystretch}{1.05}

\begin{tabular}{lcccc}
\hline
\textbf{Method} &
\textbf{Prep.} &
\textbf{Seq.} &
\textbf{Parallel} &
\textbf{Space} \\
\hline

Exact LFPR$_N$ & $O(m+n)$ & $O(T(m+n))$ & $O\!\left(\frac{T(m+n)}{p}\right)$ & $O(m+n)$ \\

Exact LFPR$_U$ & $O(m+n)$ & $O(T(m+n))$ & $O\!\left(\frac{T(m+n)}{p}\right)$ & $O(m+n)$ \\

MF LFPR$_N$ & $O(m+n)$ & $O(n)$ & $O\!\left(\frac{n}{p}\right)$ & $O(m+n)$ \\

MF LFPR$_U$ & $O(m+n)$ & $O(n)$ & $O\!\left(\frac{n}{p}\right)$ & $O(m+n)$ \\

ORF LFPR$_N$ & $O(m+n)$ & $O(m+n)$ & $O\!\left(\frac{m+n}{p}\right)$ & $O(m+n)$ \\

ORF LFPR$_U$ & $O(m+n)$ & $O(m+n)$ & $O\!\left(\frac{m+n}{p}\right)$ & $O(m+n)$ \\

\hline
\end{tabular}
\end{table}

\section{Experimental Evaluation}
\subsection{Experimental Setup}
We evaluate the proposed LFPR approximation framework on multiple real-world networks. The experiments examine the approximation accuracy of the mean-field and one-step refined LFPR methods relative to the corresponding exact LFPR stationary distributions, their ability to preserve the fairness properties of the original LFPR models, and their computational scalability on large networks. We further assess the effectiveness of the one-step refinement in improving approximation fidelity while retaining computational efficiency. Finally, the theoretical predictions of the mean-field analysis are validated through degree-based LFPR statistics and fluctuation behavior.

Table \ref{tab:dataset_statistics} reports the structural characteristics of the real-world networks, including node and edge counts, the average in-degree $\langle k_{\mathrm{in}} \rangle$, the average Red and Blue group-specific out-degrees $\langle \mathrm{out}_R \rangle$ and $\langle \mathrm{out}_B \rangle$, and the fraction of Red nodes used as the target fairness ratio $\phi$. The selected networks exhibit diverse structural characteristics in terms of network size, group-specific connectivity, and group imbalance, enabling evaluation of the proposed LFPR frameworks under heterogeneous network structures. Group memberships are defined according to the corresponding sensitive attributes or community labels available for each dataset.

\begin{table}[!t]
\caption{Structural characteristics of the real-world networks used in the experiments.}
\label{tab:dataset_statistics}
\centering
\footnotesize
\setlength{\tabcolsep}{2.6pt}
\renewcommand{\arraystretch}{1.05}
\begin{tabular}{lcccccc}
\hline
\textbf{Dataset} &
\textbf{Nodes} &
\textbf{Edges} &
$\langle k_{\mathrm{in}}\rangle$ &
$\langle \mathrm{out}_R\rangle$ &
$\langle \mathrm{out}_B\rangle$ &
\textbf{Red Frac.} \\
\hline
Twitch-Gamers  & 168114 & 6797557  & 80.87 & 0.54  & 80.33 & 0.03  \\
Pokec-G       & 1632803 & 30622564 & 18.75 & 8.30  & 10.44 & 0.49  \\
DBLP-Gender   & 2328560 & 35360648 & 15.18 & 11.26 & 3.92  & 0.71  \\
Pokec-P       & 1632803 & 30622564 & 18.75 & 12.80 & 5.95  & 0.66  \\
Pokec-R       & 1632803 & 30622564 & 18.75 & 16.96 & 1.80  & 0.90  \\
Web-Google   & 875713  & 5105039  & 5.83  & 0.12  & 5.71  & 0.019 \\
\hline
\end{tabular}
\end{table}
The evaluation is conducted on real-world collaboration, social, and web networks with diverse sensitive-attribute configurations. The Twitch Gamers \cite{rozemberczki2021twitch} network is an undirected social network of Twitch users collected from the public API in Spring 2018. Nodes represent Twitch users, and edges denote mutual follower relationships. The Dblp-Gender \cite{dblp2026xml} network is an undirected co-authorship network constructed from the official DBLP XML bibliography using publications from 2020 to 2026, where nodes represent authors and edges denote co-authorship relationships. The Pokec \cite{takac2012data} network is a directed social network derived from the Slovak online platform Pokec, in which directed edges represent relationships between users. The Web-Google \cite{leskovec2009community} network is a directed web graph, where nodes correspond to web pages, and directed edges represent hyperlinks between them. Collectively, these datasets span collaboration, social, and web networks with diverse structural characteristics, connectivity patterns, and levels of group imbalance, enabling evaluation of the proposed LFPR approximation framework.

For each dataset, nodes are assigned to one of two groups based on the corresponding sensitive attribute or community label. For the Twitch Gamers dataset, users are partitioned according to account status, where active and inactive (dead) accounts constitute the two groups. For the DBLP-Gender dataset, author gender is inferred from first names using the gender-guesser \cite{menendez2020damegender} Python package. For the Pokec network, we evaluate the proposed LFPR approximation framework under three sensitive-attribute configurations. Specifically, Pokec-G partitions users based on gender, Pokec-R defines groups according to regional affiliation by distinguishing users associated with major administrative regions of Slovakia from those with external, foreign, or unspecified regional affiliations, and Pokec-P partitions users according to profile visibility, assigning public and non-public accounts to different groups.

Since the Web-Google dataset does not contain sensitive-attribute information, binary group labels are derived from the network structure using the Louvain community detection \cite{blondel2008fast} algorithm. The resulting community partition is subsequently converted into two groups for fairness evaluation. Throughout all experiments, the damping factor is fixed at $(\gamma = 0.15)$ \cite{brin1998anatomy, tsioutsiouliklis2021fairness}, following standard practice in the PageRank and fairness-aware ranking literature.
\subsection{Approximation Performance Evaluation}
We first evaluate the approximation accuracy of the proposed mean-field and ORF-refined LFPR methods with respect to the exact LFPR stationary distributions. We use Pearson correlation ($r$) to measure score-level agreement between the approximate and exact LFPR vectors, while Spearman correlation ($\rho_s$) evaluates preservation of the ranking order. To examine approximation performance across groups, both correlations are computed for the entire network as well as for the Red and Blue node groups separately. This enables assessment of the agreement between the approximate and exact LFPR scores and rankings at both the global and group levels. The resulting correlation statistics for the mean-field and ORF-refined LFPR approximations are summarized in Table \ref{tab:correlation_results}.

\begin{table}[!t]
\centering
\setlength{\tabcolsep}{2.6pt}
\renewcommand{\arraystretch}{1.05}
\caption{Correlation analysis of proposed LFPR approximation methods with respect to exact LFPR stationary distributions.} \label{tab:correlation_results}
\begin{tabular}{llcccccc}
\hline
\multirow{2}{*}{\textbf{Dataset}} &
\multirow{2}{*}{\textbf{Method}} &
\multicolumn{3}{c}{\textbf{Pearson Corr. ($r$)}} &
\multicolumn{3}{c}{\textbf{Spearman Corr. ($\rho_s$)}} \\

\cmidrule(lr){3-5}
\cmidrule(lr){6-8}

&
& \textbf{Global} & \textbf{Red} & \textbf{Blue}
& \textbf{Global} & \textbf{Red} & \textbf{Blue} \\

\hline
\rule{0pt}{2.2ex}
\multirow{4}{*}{Twitch-Gamers}
& MF-LFPR$_N$  & 0.99 & 0.96 & 0.99 & 0.97 & 0.96 & 0.99 \\
& ORF-LFPR$_N$ & 0.99 & 0.99 & 0.99 & 0.99 & 0.99 & 0.99 \\
& MF-LFPR$_U$  & 0.99 & 0.99 & 0.99 & 0.98 & 0.99 & 0.98 \\
& ORF-LFPR$_U$ & 0.99 & 0.99 & 0.99 & 0.99 & 0.99 & 0.99 \\

\hline
\rule{0pt}{2.2ex}
\multirow{4}{*}{Pokec-G}
& MF-LFPR$_N$  & 0.96 & 0.97 & 0.95 & 0.95 & 0.96 & 0.95 \\
& ORF-LFPR$_N$ & 0.99 & 0.99 & 0.99 & 0.99 & 0.99 & 0.99 \\
& MF-LFPR$_U$  & 0.96 & 0.97 & 0.95 & 0.95 & 0.96 & 0.95 \\
& ORF-LFPR$_U$ & 0.99 & 0.99 & 0.99 & 0.99 & 0.99 & 0.99 \\

\hline
\rule{0pt}{2.2ex}
\multirow{4}{*}{Dblp-Gender}
& MF-LFPR$_N$  & 0.94 & 0.94 & 0.95 & 0.83 & 0.82 & 0.86 \\
& ORF-LFPR$_N$ & 0.96 & 0.96 & 0.97 & 0.93 & 0.93 & 0.92 \\
& MF-LFPR$_U$  & 0.94 & 0.94 & 0.95 & 0.83 & 0.83 & 0.83 \\
& ORF-LFPR$_U$ & 0.97 & 0.97 & 0.97 & 0.95 & 0.95 & 0.95 \\

\hline
\rule{0pt}{2.2ex}
\multirow{4}{*}{Pokec-P}
& MF-LFPR$_N$  & 0.96 & 0.96 & 0.96 & 0.96 & 0.96 & 0.96 \\
& ORF-LFPR$_N$ & 0.99 & 0.99 & 0.99 & 0.99 & 0.99 & 0.99 \\
& MF-LFPR$_U$  & 0.97 & 0.96 & 0.97 & 0.96 & 0.96 & 0.97 \\
& ORF-LFPR$_U$ & 0.99 & 0.99 & 0.99 & 0.99 & 0.99 & 0.99 \\

\hline
\rule{0pt}{2.2ex}
\multirow{4}{*}{Pokec-R}
& MF-LFPR$_N$  & 0.94 & 0.95 & 0.83 & 0.94 & 0.95 & 0.91 \\
& ORF-LFPR$_N$ & 0.99 & 0.98 & 0.99 & 0.99 & 0.99 & 0.99 \\
& MF-LFPR$_U$  & 0.94 & 0.94 & 0.98 & 0.95 & 0.95 & 0.97 \\
& ORF-LFPR$_U$ & 0.99 & 0.99 & 0.99 & 0.99 & 0.99 & 0.99 \\

\hline
\rule{0pt}{2.2ex}
\multirow{4}{*}{web-Google}
& MF-LFPR$_N$  & 0.90 & 0.96 & 0.91 & 0.83 & 0.86 & 0.85 \\
& ORF-LFPR$_N$ & 0.96 & 0.94 & 0.96 & 0.95 & 0.95 & 0.95 \\
& MF-LFPR$_U$  & 0.90 & 0.97 & 0.91 & 0.85 & 0.86 & 0.85 \\
& ORF-LFPR$_U$ & 0.96 & 0.98 & 0.96 & 0.96 & 0.94 & 0.96 \\

\hline
\end{tabular}
\end{table}
Table \ref{tab:correlation_results} demonstrates that the mean-field approximations achieve strong agreement with the exact LFPR stationary distributions across all evaluated datasets. The proposed ORF mechanism consistently improves the score- and rank-level agreement of the mean-field estimates, with particularly notable improvements on datasets where the class-level approximation exhibits lower initial correlation. The largest improvements are observed on DBLP-Gender and Web-Google, particularly in Spearman correlation, where the refinement substantially increases the correlation values relative to the corresponding mean-field approximation. Across the Twitch-Gamers and Pokec networks, the refined methods achieve near-perfect agreement with the exact LFPR solutions, with most Pearson ($r$) and Spearman ($\rho_s$) correlations approaching unity. Similar trends are observed for both the Red and Blue node groups, indicating that the refinement improves approximation fidelity while preserving group-specific ranking behavior. Overall, these results demonstrate that the proposed mean-field approximation captures the dominant stationary behavior of the LFPR models, while the one-step refinement enhances approximation accuracy by incorporating local structural information.

We next assess the ability of the proposed mean-field and refined LFPR approximations to preserve the group-level fairness targets of the original LFPR models. To quantify the deviation between the group-level PageRank mass produced by an approximation and the corresponding target fairness mass, we use the absolute FairnessGap metric:
\begin{equation*}
\text{FairnessGap}(g) = \left|\sum_{i\in g}\hat{p}(i) - \phi_{g}\right|
\end{equation*}
where $g\in\{R,B\}$ denotes the corresponding demographic group, $\hat{p}(i)$ is the approximate LFPR score of node $i$, and $\phi_g$ denotes the target fairness mass associated with the corresponding group. Smaller FairnessGap values indicate closer agreement with the desired group-level fairness constraint and are reported in Table \ref{tab:fairness_gap_results}.

\begin{table}[!t]
\centering
\caption{FairnessGap values of the proposed LFPR approximation methods across the networks.}
\label{tab:fairness_gap_results}
\renewcommand{\arraystretch}{1.05}
\begin{tabular}{llcc}
\hline

{\textbf{Dataset}} &{\textbf{Method}} &

\textbf{Red} & \textbf{Blue} \\

\hline
\multirow{4}{*}{Twitch-Gamers}
& MF-LFPR$_N$  & $3.12\times10^{-17}$ & $9.99\times10^{-16}$  \\
& ORF-LFPR$_N$ & $9.82\times10^{-7}$ & $9.82\times10^{-7}$  \\
& MF-LFPR$_U$  & $1.90\times10^{-5}$ & $1.09\times10^{-5}$  \\
& ORF-LFPR$_U$ & $5.59\times10^{-16}$ & $1.22\times10^{-15}$  \\

\hline
\rule{0pt}{2.2ex}
\multirow{4}{*}{Pokec-G}
& MF-LFPR$_N$  & $9.49\times10^{-15}$ & $9.44\times10^{-15}$  \\
& ORF-LFPR$_N$ & $5.37\times10^{-9}$ & $5.37\times10^{-9}$  \\
& MF-LFPR$_U$  & $3.99\times10^{-3}$ & $3.99\times10^{-3}$  \\
& ORF-LFPR$_U$ & $1.75\times10^{-14}$ & $1.67\times10^{-15}$  \\

\hline
\rule{0pt}{2.2ex}
\multirow{4}{*}{Dblp-Gender}
& MF-LFPR$_N$  & $1.25\times10^{-14}$ & $5.27\times10^{-15}$  \\
& ORF-LFPR$_N$ & $9.63\times10^{-8}$ & $9.63\times10^{-8}$  \\
& MF-LFPR$_U$  & $8.72\times10^{-3}$ & $8.72\times10^{-3}$  \\
& ORF-LFPR$_U$ & $2.33\times10^{-15}$ & $1.67\times10^{-15}$  \\

\hline
\rule{0pt}{2.2ex}
\multirow{4}{*}{Pokec-P}
& MF-LFPR$_N$  & $3.55\times10^{-15}$ & $1.78\times10^{-15}$  \\
& ORF-LFPR$_N$ & $7.09\times10^{-8}$ & $7.09\times10^{-8}$  \\
& MF-LFPR$_U$  & $1.12\times10^{-3}$ & $1.12\times10^{-3}$  \\
& ORF-LFPR$_U$ & $4.00\times10^{-15}$ & $1.07\times10^{-14}$  \\

\hline
\rule{0pt}{2.2ex}
\multirow{4}{*}{Pokec-R}
& MF-LFPR$_N$  & $7.22\times10^{-15}$ & $8.19\times10^{-16}$  \\
& ORF-LFPR$_N$ & $4.21\times10^{-8}$ & $4.21\times10^{-8}$  \\
& MF-LFPR$_U$  & $1.04\times10^{-2}$ & $1.04\times10^{-2}$   \\
& ORF-LFPR$_U$ & $6.94\times10^{-14}$ & $1.30\times10^{-14}$  \\

\hline
\rule{0pt}{2.2ex}
\multirow{4}{*}{Web-Google}
& MF-LFPR$_N$  & $1.10\times10^{-15}$ & $5.60\times10^{-14}$  \\
& ORF-LFPR$_N$ & $2.22\times10^{-7}$ & $2.22\times10^{-7}$  \\
& MF-LFPR$_U$  & $1.34\times10^{-2}$ & $1.34\times10^{-2}$  \\
& ORF-LFPR$_U$ & $1.39\times10^{-17}$ & $2.05\times10^{-13}$  \\

\hline
\end{tabular}

\end{table}

Across all datasets, the proposed one-step refined approximations preserve the target group-level PageRank mass with very small deviations, typically below $10^{-7}$. Similarly, the mean-field LFPR$_N$ approximation maintains near-zero FairnessGap values across all evaluated datasets. In contrast, the mean-field LFPR$_U$ approximation exhibits small but non-zero fairness deviations, ranging from approximately $10^{-5}$ to $10^{-2}$. The ORF step substantially reduces these deviations while improving the score and rank-level agreement reported in Table \ref{tab:correlation_results}. Overall, the proposed approximation framework achieves high approximation accuracy while closely preserving the fairness properties of the original LFPR models.

In addition to fairness preservation, we evaluate the approximation performance of the proposed methods relative to the exact LFPR stationary distributions. We quantify this error using the UtilityLoss metric, which measures the average absolute deviation between the exact and approximate LFPR scores. Formally, UtilityLoss is defined as
\begin{equation*}
\text{UtilityLoss}(p,\hat{p}) = \frac{1}{|V|}\sum_{i\in V} \left|p(i) - \hat{p}(i)\right|
\end{equation*}
where $p(i)$ and $\hat{p}(i)$ denote the exact and approximate LFPR scores of node $i$, respectively, and $|V|$ is the number of nodes in the network.  To examine the approximation behavior across groups, we also compute the corresponding UtilityLoss errors separately for the Red and Blue node subsets. Table \ref{tab:l1_results} summarizes the resulting errors for the proposed LFPR approximation frameworks.
\begin{table}[!t]
\caption{UtilityLoss values of proposed LFPR approximation methods.}
\label{tab:l1_results}
\centering
\setlength{\tabcolsep}{2.6pt}
\renewcommand{\arraystretch}{1.05}

\begin{tabular}{llccc}
\hline
{\textbf{Dataset}} &{\textbf{Method}} 
& \textbf{Global} & \textbf{Red} & \textbf{Blue} \\
\hline
\rule{0pt}{2.2ex}
\multirow{4}{*}{Twitch-Gamers}
& MF-LFPR$_N$  & $7.36\times10^{-7}$ & $2.66\times10^{-6}$ & $6.75\times10^{-7}$ \\
& ORF-LFPR$_N$ & $3.76\times10^{-7}$ & $1.89\times10^{-7}$ & $3.82\times10^{-7}$ \\
& MF-LFPR$_U$  & $6.82\times10^{-7}$ & $6.93\times10^{-7}$ & $6.99\times10^{-7}$ \\
& ORF-LFPR$_U$ & $3.61\times10^{-7}$ & $6.93\times10^{-8}$ & $3.70\times10^{-7}$ \\

\hline
\rule{0pt}{2.2ex}
\multirow{4}{*}{Pokec-G}
& MF-LFPR$_N$  & $1.34\times10^{-7}$ & $1.35\times10^{-7}$ & $1.32\times10^{-7}$ \\
& ORF-LFPR$_N$ & $5.86\times10^{-8}$ & $5.79\times10^{-8}$ & $5.92\times10^{-8}$ \\
& MF-LFPR$_U$  & $8.73\times10^{-8}$ & $7.89\times10^{-8}$ & $9.55\times10^{-8}$ \\
& ORF-LFPR$_U$ & $3.02\times10^{-8}$ & $2.86\times10^{-8}$ & $3.17\times10^{-8}$ \\

\hline
\rule{0pt}{2.2ex}
\multirow{4}{*}{Dblp-Gender}
& MF-LFPR$_N$  & $1.45\times10^{-7}$ & $1.50\times10^{-7}$ & $1.32\times10^{-7}$ \\
& ORF-LFPR$_N$ & $1.04\times10^{-7}$ & $1.05\times10^{-7}$ & $1.01\times10^{-7}$ \\
& MF-LFPR$_U$  & $1.11\times10^{-7}$ & $1.14\times10^{-7}$ & $1.03\times10^{-7}$ \\
& ORF-LFPR$_U$ & $6.96\times10^{-8}$ & $7.37\times10^{-8}$ & $5.99\times10^{-8}$ \\

\hline
\rule{0pt}{2.2ex}
\multirow{4}{*}{Pokec-P}
& MF-LFPR$_N$  & $1.24\times10^{-7}$ & $1.24\times10^{-7}$ & $1.25\times10^{-7}$ \\
& ORF-LFPR$_N$ & $6.30\times10^{-8}$ & $6.36\times10^{-8}$ & $6.20\times10^{-8}$ \\
& MF-LFPR$_U$  & $8.98\times10^{-8}$ & $9.56\times10^{-8}$ & $7.87\times10^{-8}$ \\
& ORF-LFPR$_U$ & $3.70\times10^{-8}$ & $3.83\times10^{-8}$ & $3.43\times10^{-8}$ \\

\hline
\rule{0pt}{2.2ex}
\multirow{4}{*}{Pokec-R}
& MF-LFPR$_N$  & $1.45\times10^{-7}$ & $1.34\times10^{-7}$ & $2.41\times10^{-7}$ \\
& ORF-LFPR$_N$ & $6.84\times10^{-8}$ & $6.99\times10^{-8}$ & $5.51\times10^{-8}$ \\
& MF-LFPR$_U$  & $1.07\times10^{-7}$ & $1.08\times10^{-7}$ & $9.15\times10^{-8}$ \\
& ORF-LFPR$_U$ & $4.44\times10^{-8}$ & $4.52\times10^{-8}$ & $3.78\times10^{-8}$ \\

\hline
\rule{0pt}{2.2ex}
\multirow{4}{*}{Web-Google}
& MF-LFPR$_N$  & $6.82\times10^{-7}$ & $1.10\times10^{-6}$ & $6.19\times10^{-7}$ \\
& ORF-LFPR$_N$ & $3.62\times10^{-7}$ & $1.00\times10^{-8}$ & $3.69\times10^{-7}$ \\
& MF-LFPR$_U$  & $5.82\times10^{-7}$ & $8.07\times10^{-7}$ & $5.77\times10^{-7}$ \\
& ORF-LFPR$_U$ & $3.46\times10^{-7}$ & $1.56\times10^{-8}$ & $3.53\times10^{-7}$ \\

\hline
\end{tabular}
\end{table}

Across all datasets, the proposed approximation methods achieve UtilityLoss values on the order of $10^{-8}$ to $10^{-6}$, indicating close agreement between the approximate and exact LFPR scores. ORF mechanism consistently reduces the global UtilityLoss for both LFPR$_N$ and LFPR$_U$ across all datasets. Similar reductions are observed for the protected and unprotected groups, indicating that the refinement improves performance for both groups. This observation is also consistent with the correlation analysis and further highlights the effectiveness of combining the mean-field approximation with a one-step local refinement

\subsection{Degree-Based Approximation Analysis}
\begin{figure}[!t]
\centering

\subfloat{%
\includegraphics[width=0.48\columnwidth]
{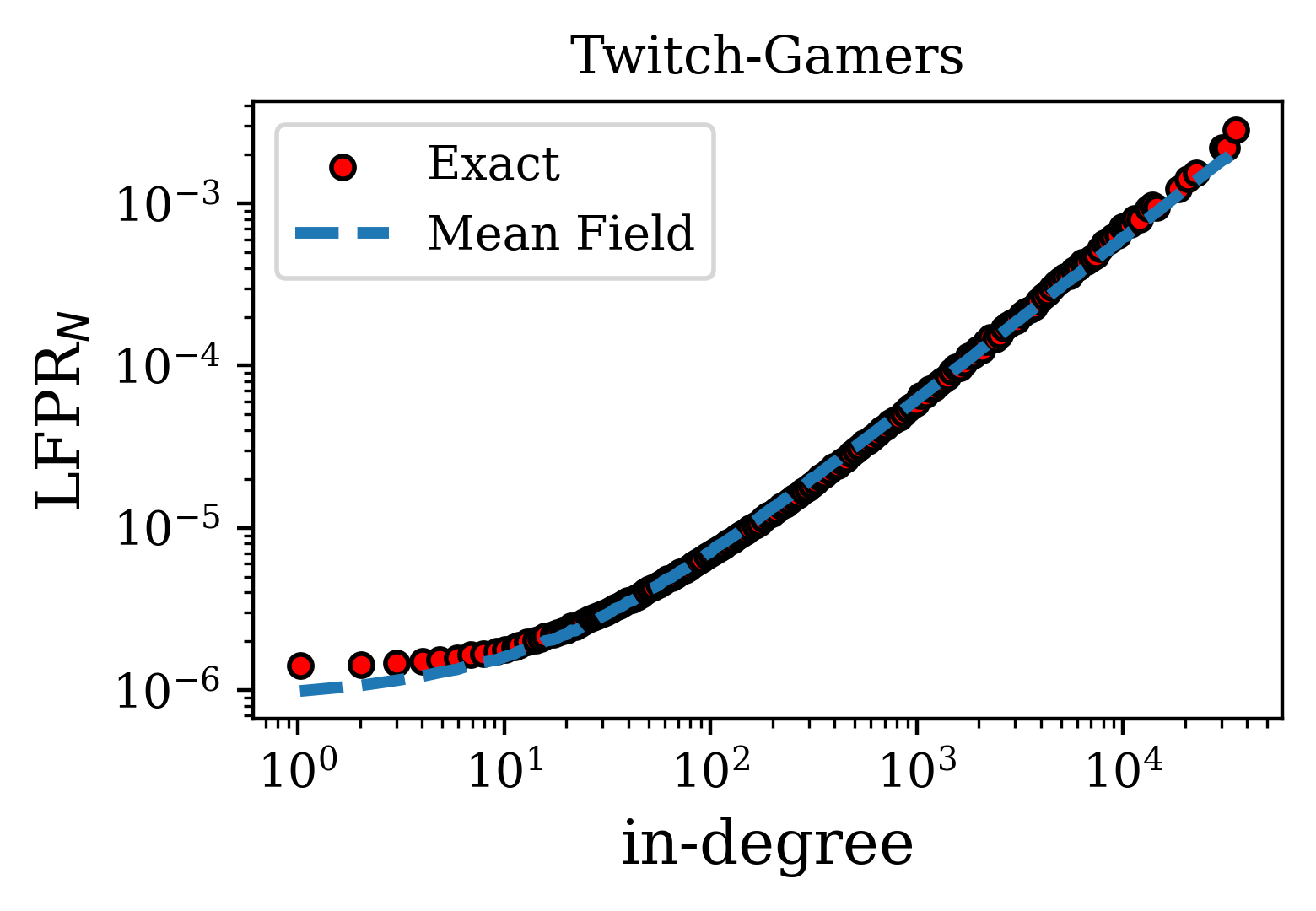}}
\hfill
\subfloat{%
\includegraphics[width=0.48\columnwidth]
{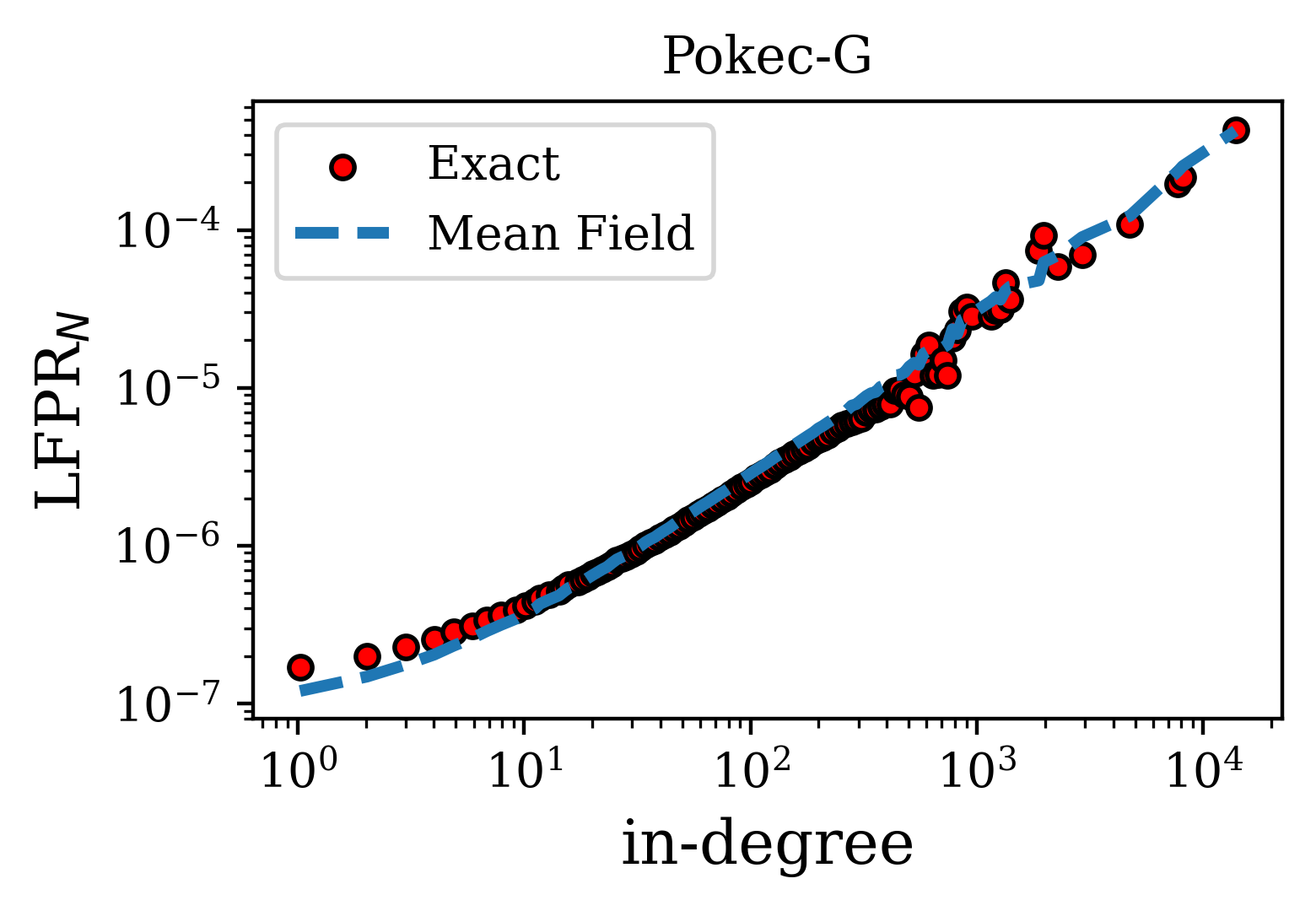}}

\vspace{-1mm}

\subfloat{%
\includegraphics[width=0.48\columnwidth]
{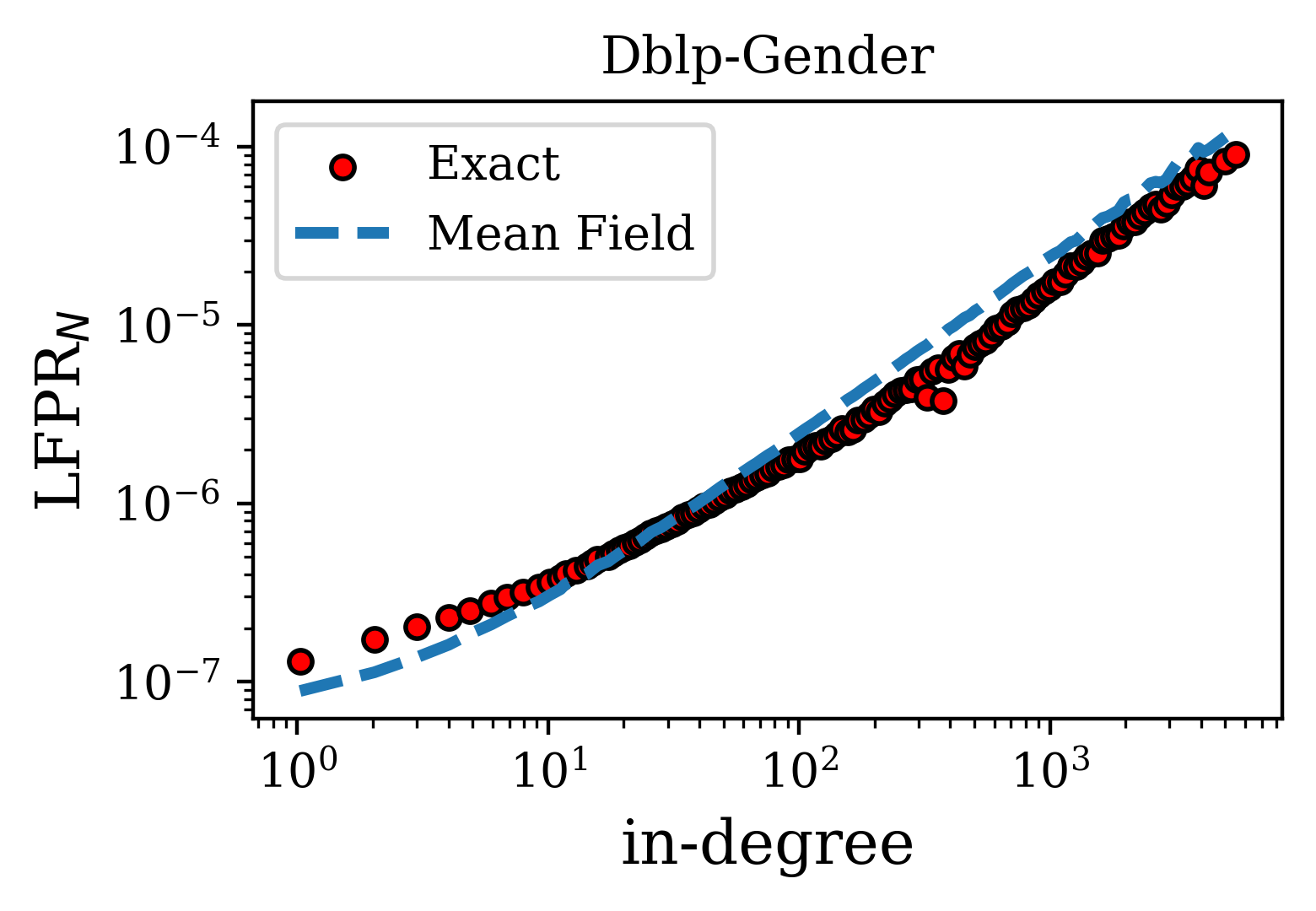}}
\hfill
\subfloat{%
\includegraphics[width=0.48\columnwidth]
{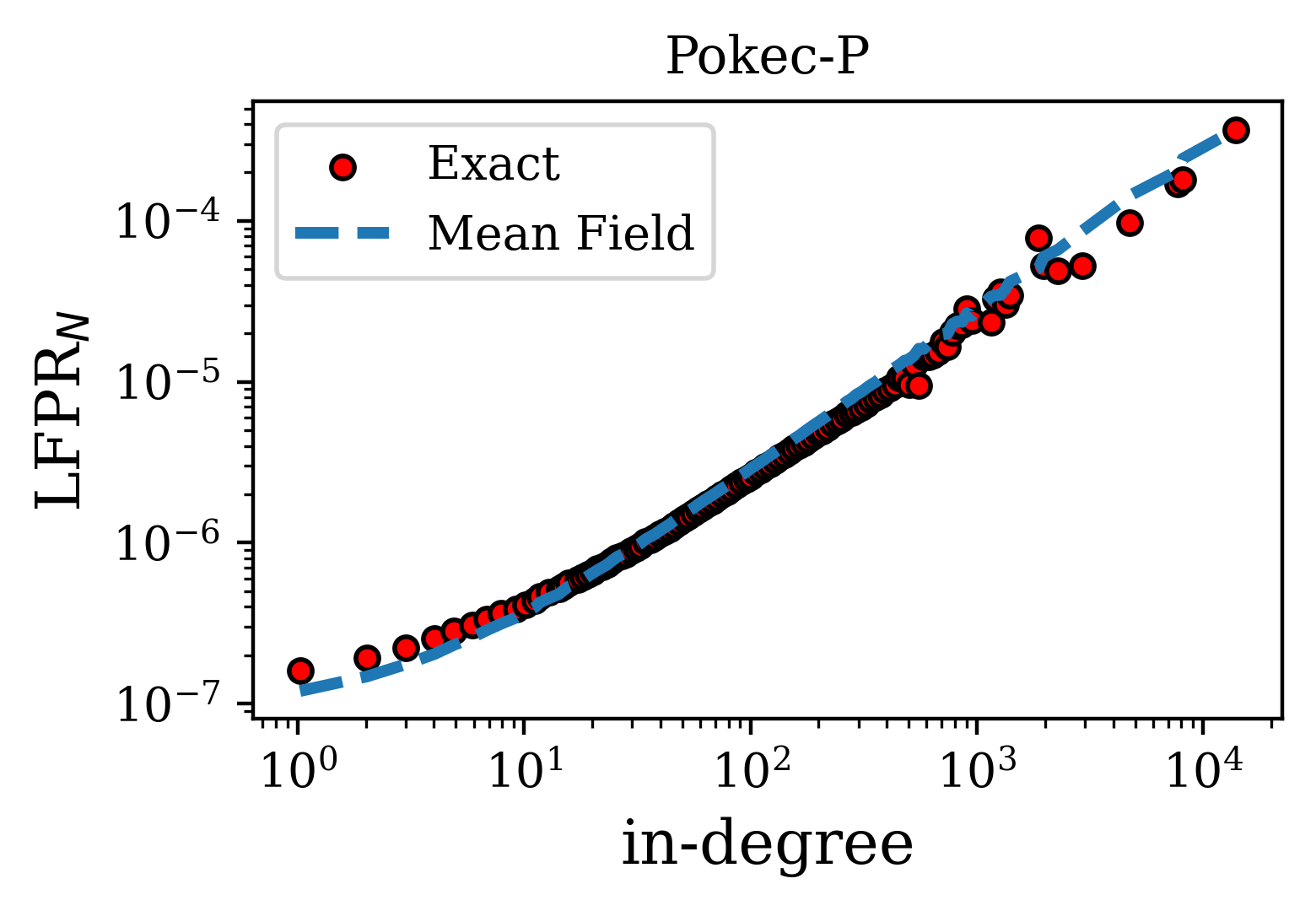}}

\vspace{-1mm}

\subfloat{%
\includegraphics[width=0.48\columnwidth]
{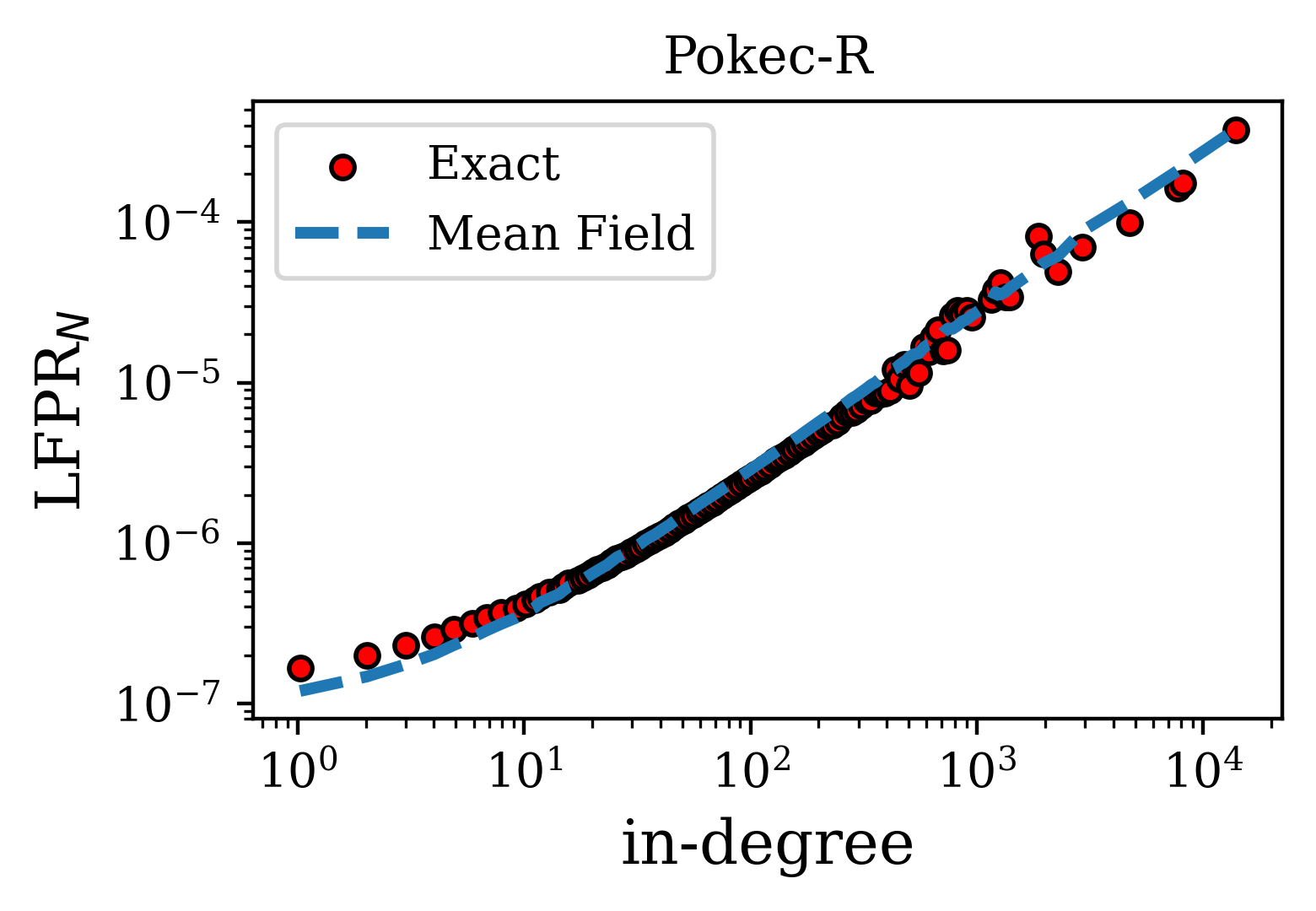}}
\hfill
\subfloat{%
\includegraphics[width=0.48\columnwidth]
{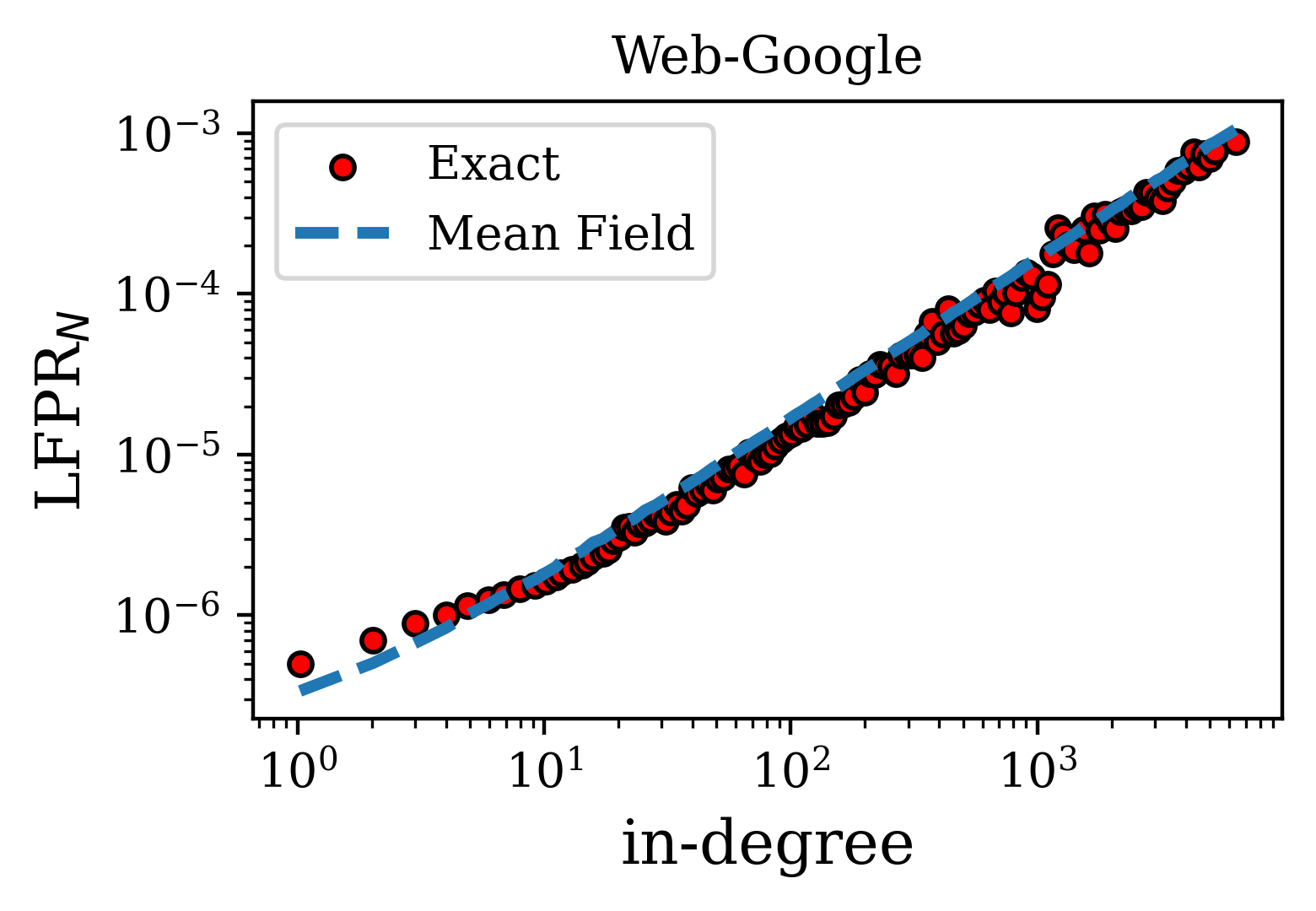}}

\caption{Comparison of the average exact and mean-field LFPR$_N$ scores across logarithmically spaced in-degree bins for six real-world networks. The dashed line denotes the mean-field approximation.}
\label{fig:mf_lfprn}
\end{figure}

\begin{figure}[!t]
\centering

\subfloat{%
\includegraphics[width=0.48\columnwidth]
{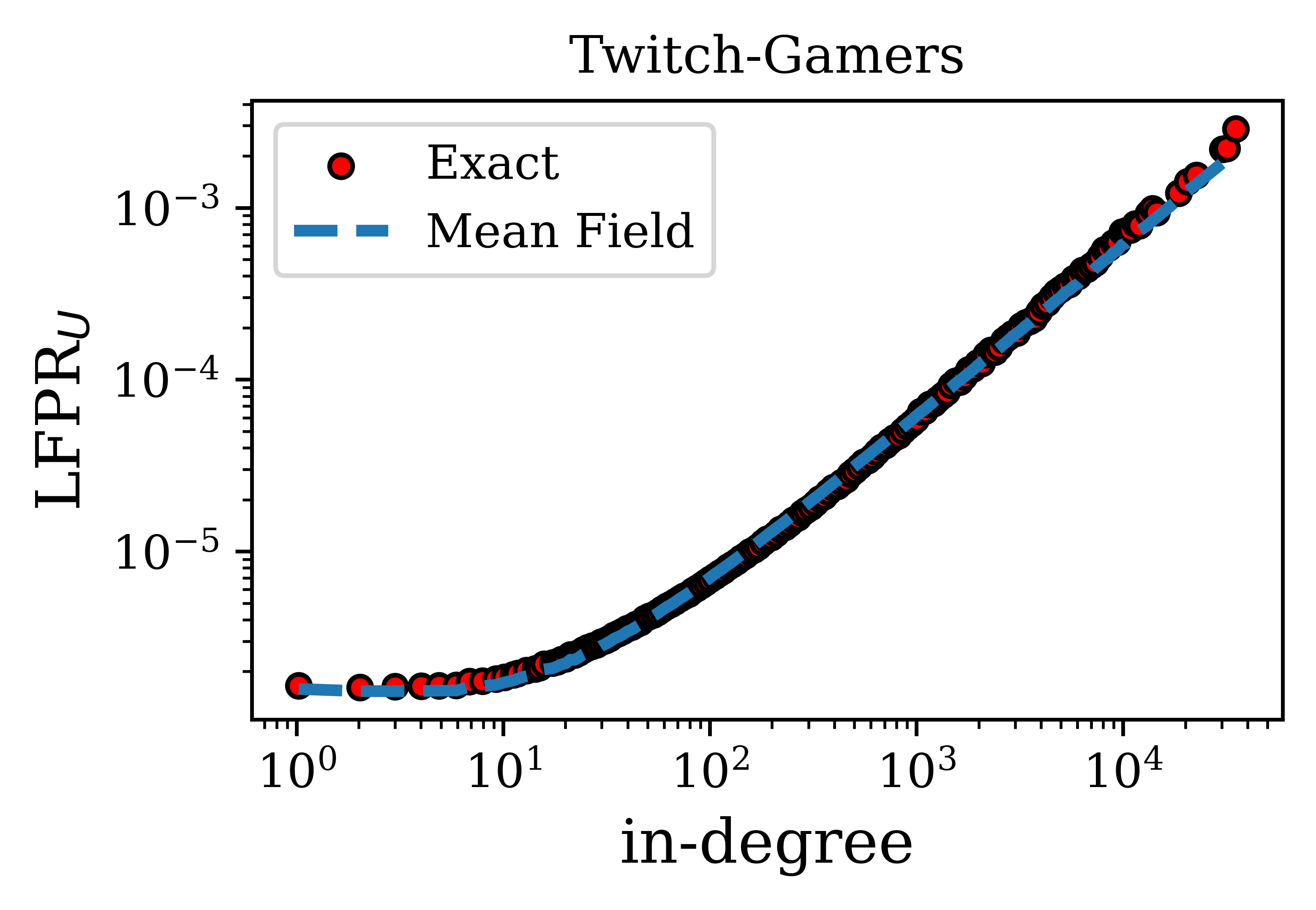}}
\hfill
\subfloat{%
\includegraphics[width=0.48\columnwidth]
{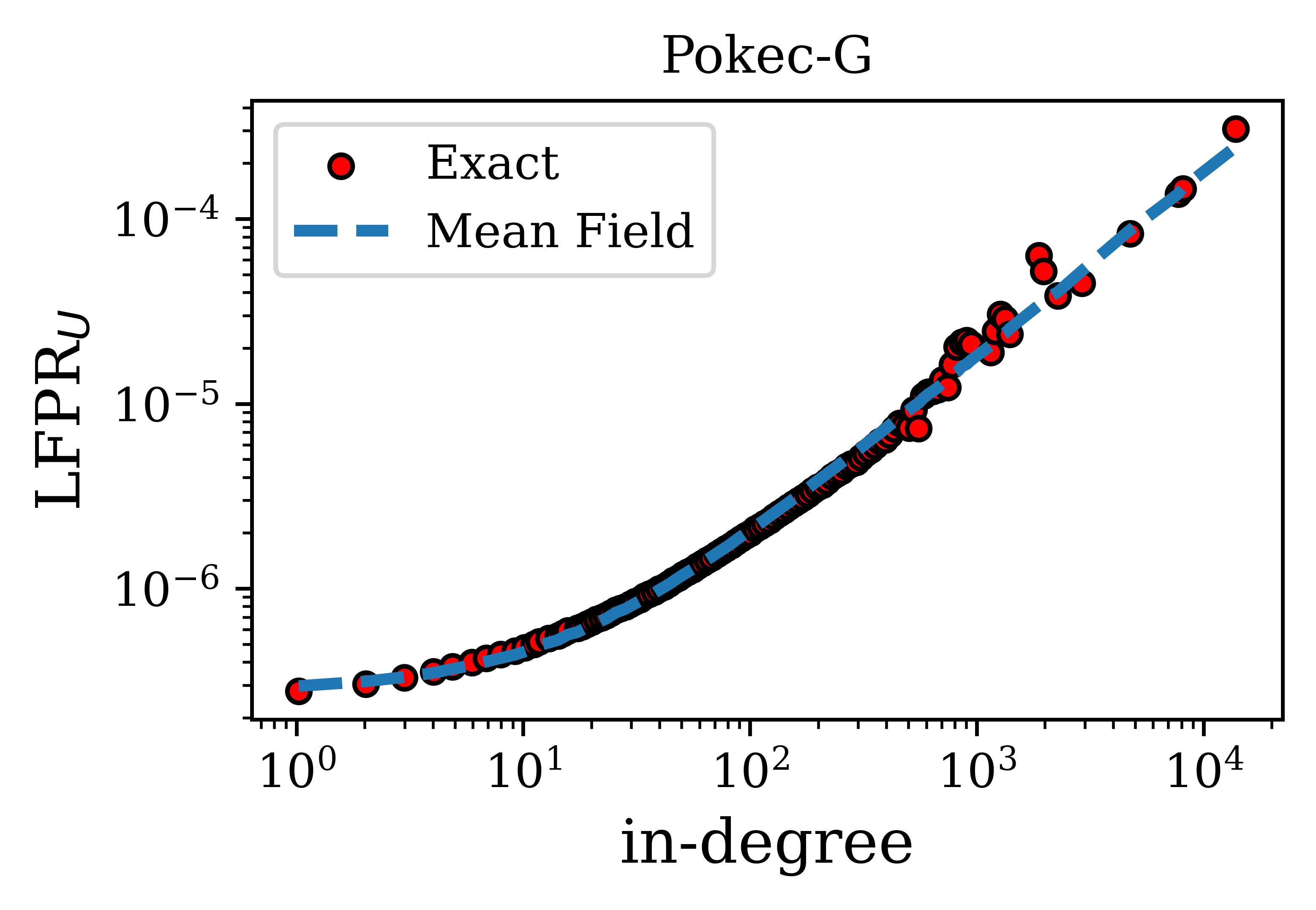}}

\vspace{-1mm}

\subfloat{%
\includegraphics[width=0.48\columnwidth]
{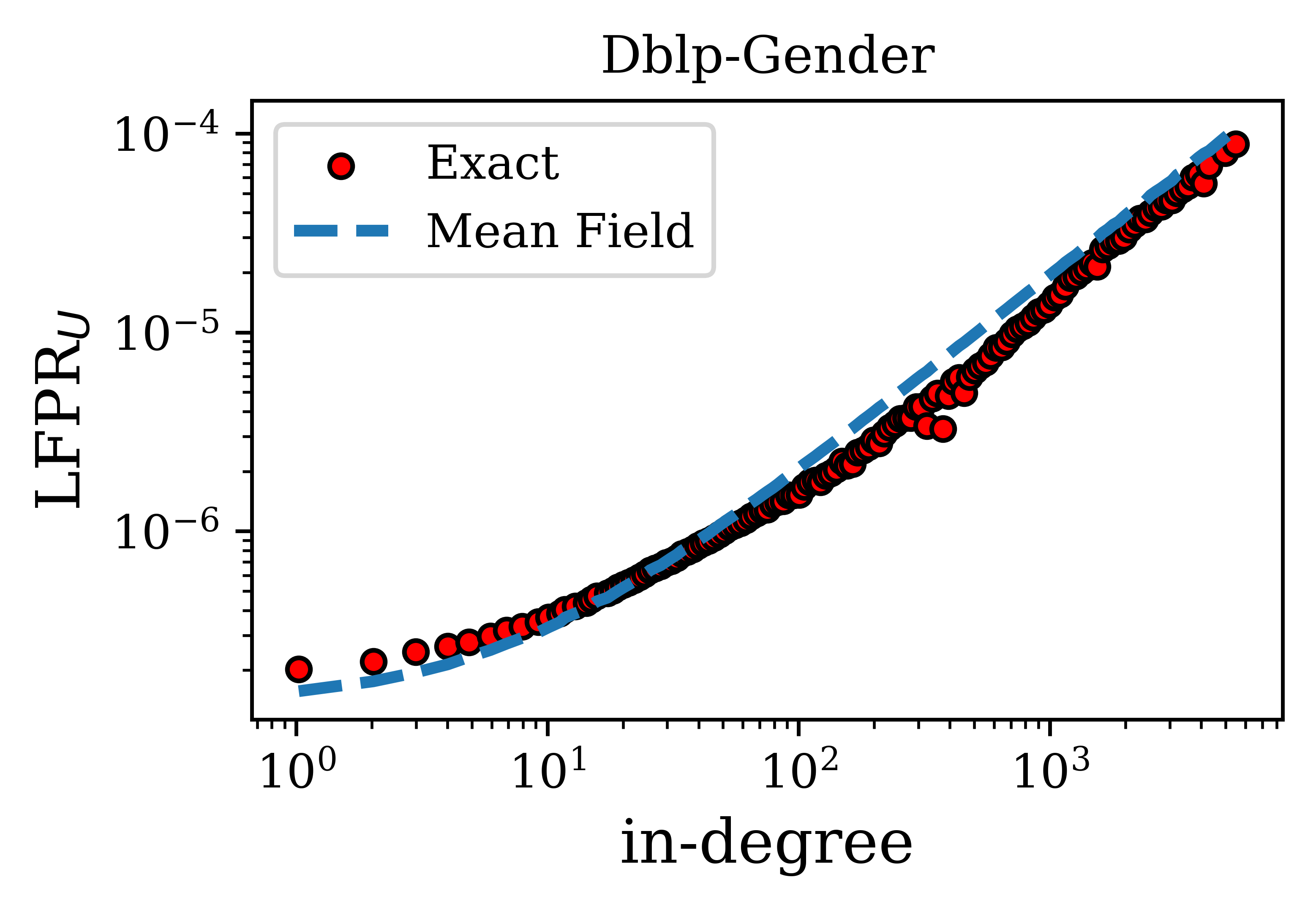}}
\hfill
\subfloat{%
\includegraphics[width=0.48\columnwidth]
{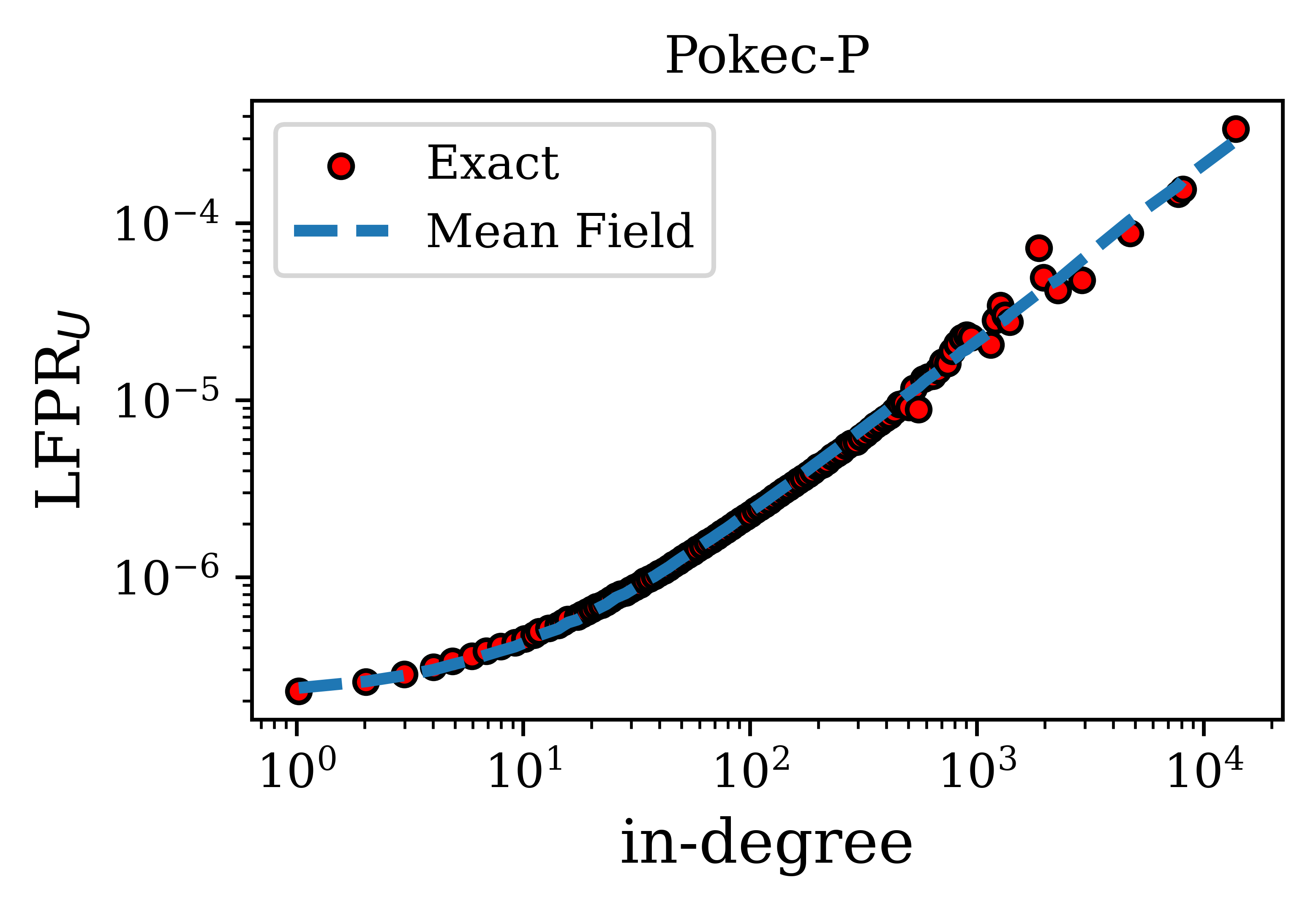}}

\vspace{-1mm}

\subfloat{%
\includegraphics[width=0.48\columnwidth]
{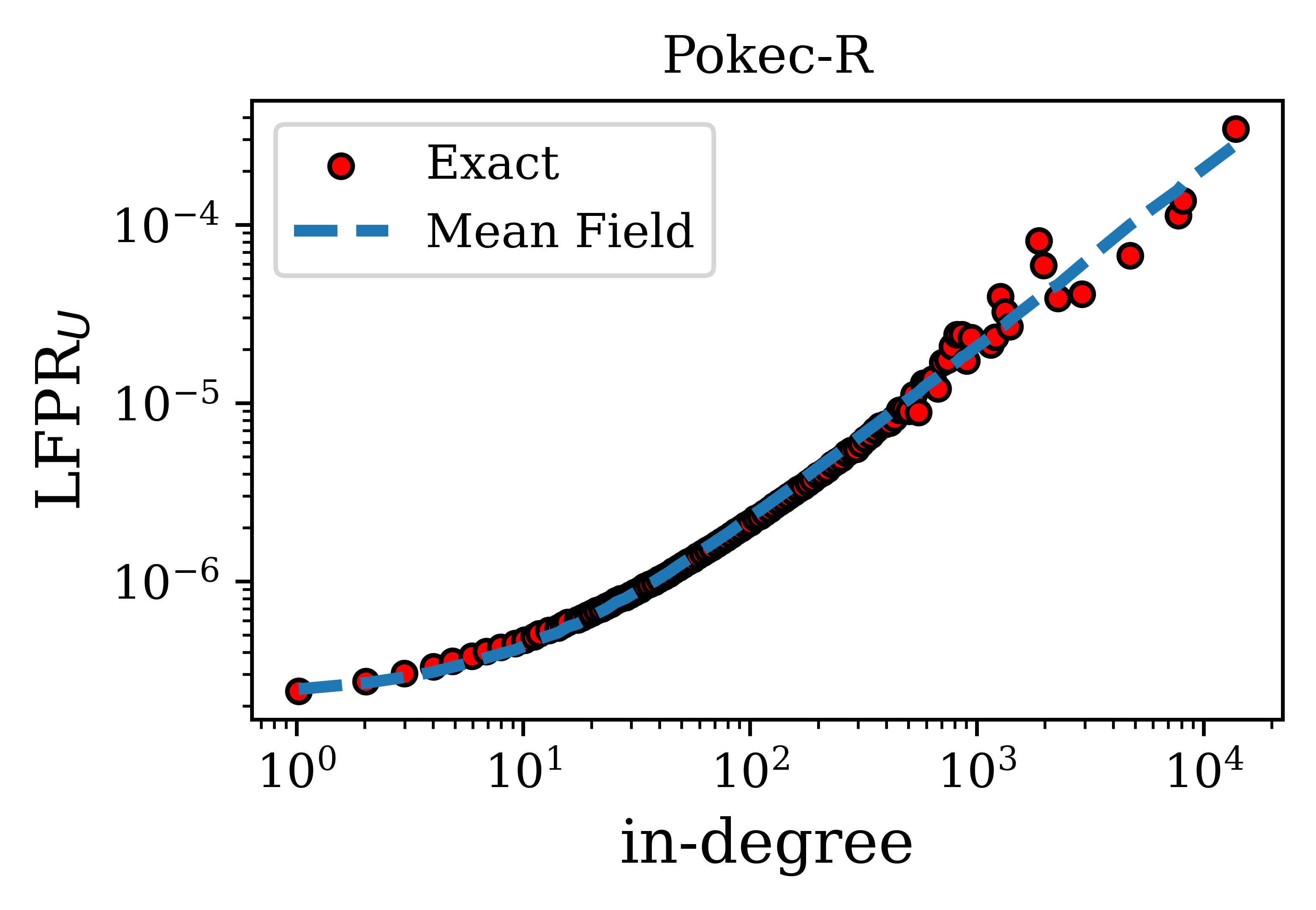}}
\hfill
\subfloat{%
\includegraphics[width=0.48\columnwidth]
{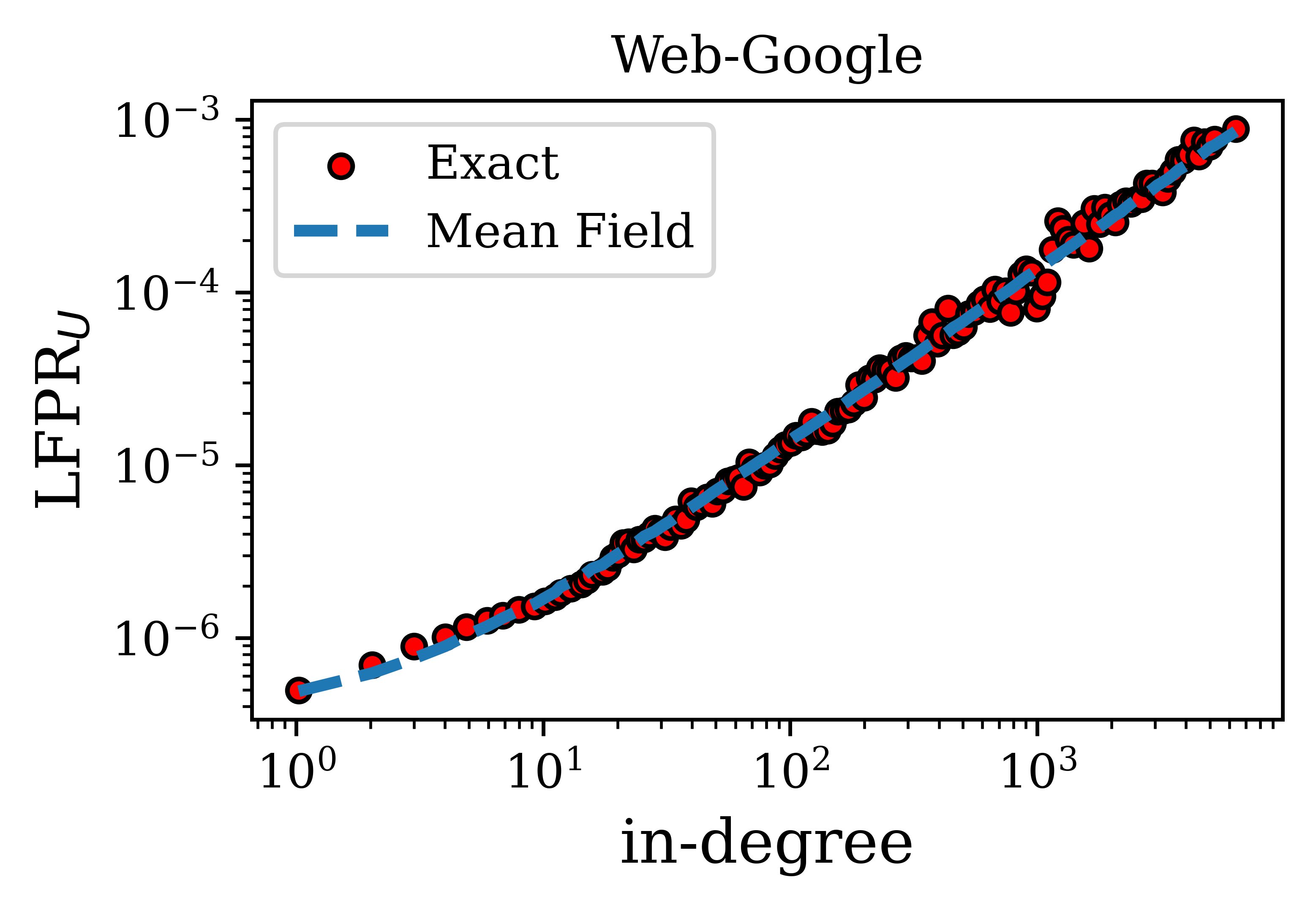}}

\caption{Comparison of the average exact and mean-field LFPR$_U$ scores across logarithmically spaced in-degree bins for six real-world networks. The dashed line denotes the mean-field approximation.}
\label{fig:mf_lfpru}

\end{figure}

We next evaluate the quality of the proposed LFPR approximations by comparing the average LFPR scores produced by the exact and approximate LFPR methods. Due to local structural heterogeneity, nodes within the same degree class may exhibit variations in their LFPR scores. To obtain a clearer view of the approximation quality, nodes are grouped into logarithmically spaced in-degree bins, and the average LFPR score within each bin is computed. Figures \ref{fig:mf_lfprn} and \ref{fig:mf_lfpru} compare the exact LFPR scores with the corresponding mean-field LFPR$_N$ and LFPR$_U$ approximations, respectively, across all datasets. Overall, both approximation methods exhibit strong agreement with the exact LFPR values across most degree classes, indicating that the proposed analytical formulations closely characterize the stationary behavior of the LFPR models. Across all six datasets, the approximate and exact curves closely overlap across most degree classes, supporting the effectiveness of the degree-class mean-field representation in estimating the exact LFPR scores.

\begin{figure}[!t]
\centering

\subfloat{%
\includegraphics[width=0.48\columnwidth]
{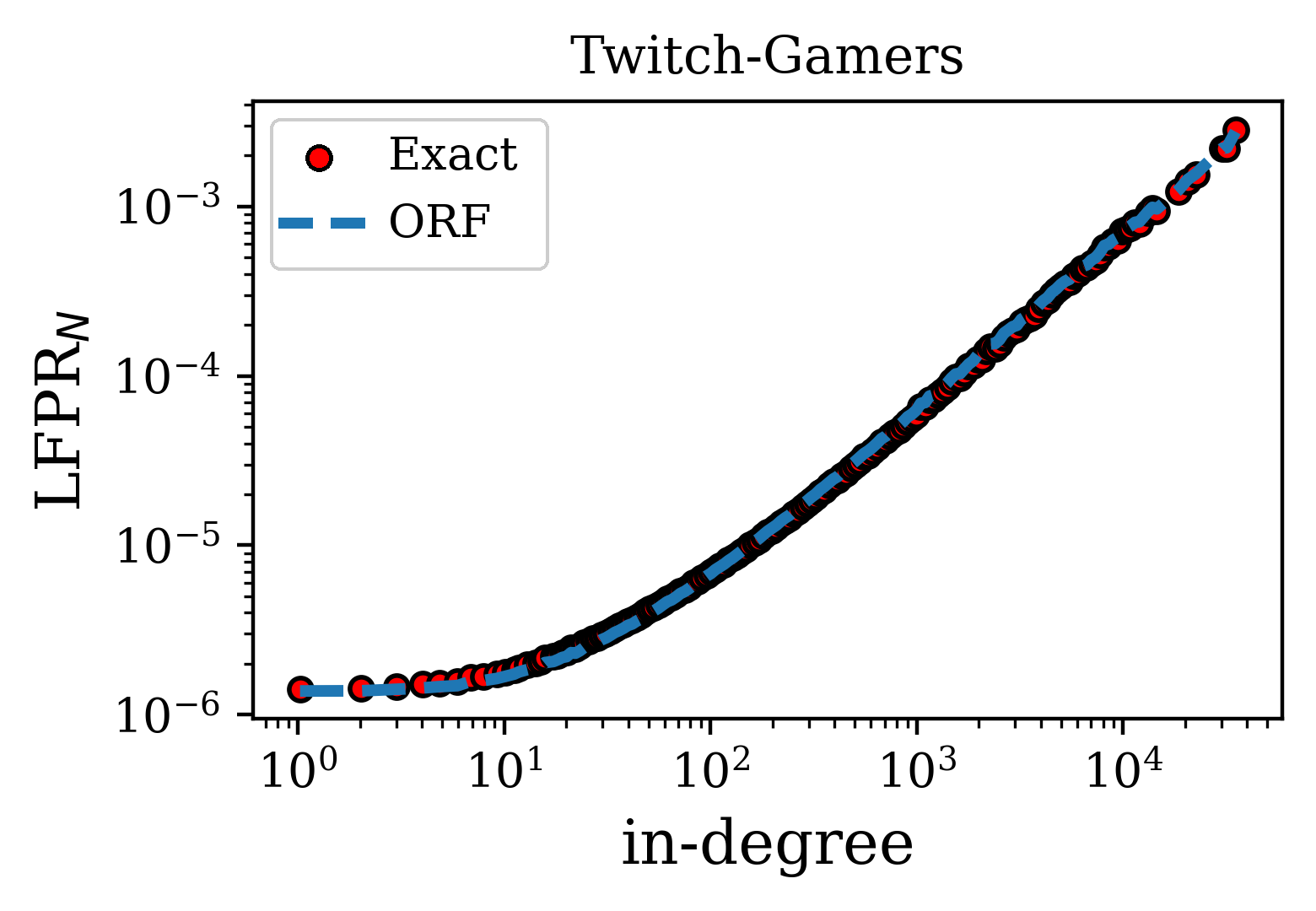}}
\hfill
\subfloat{%
\includegraphics[width=0.48\columnwidth]
{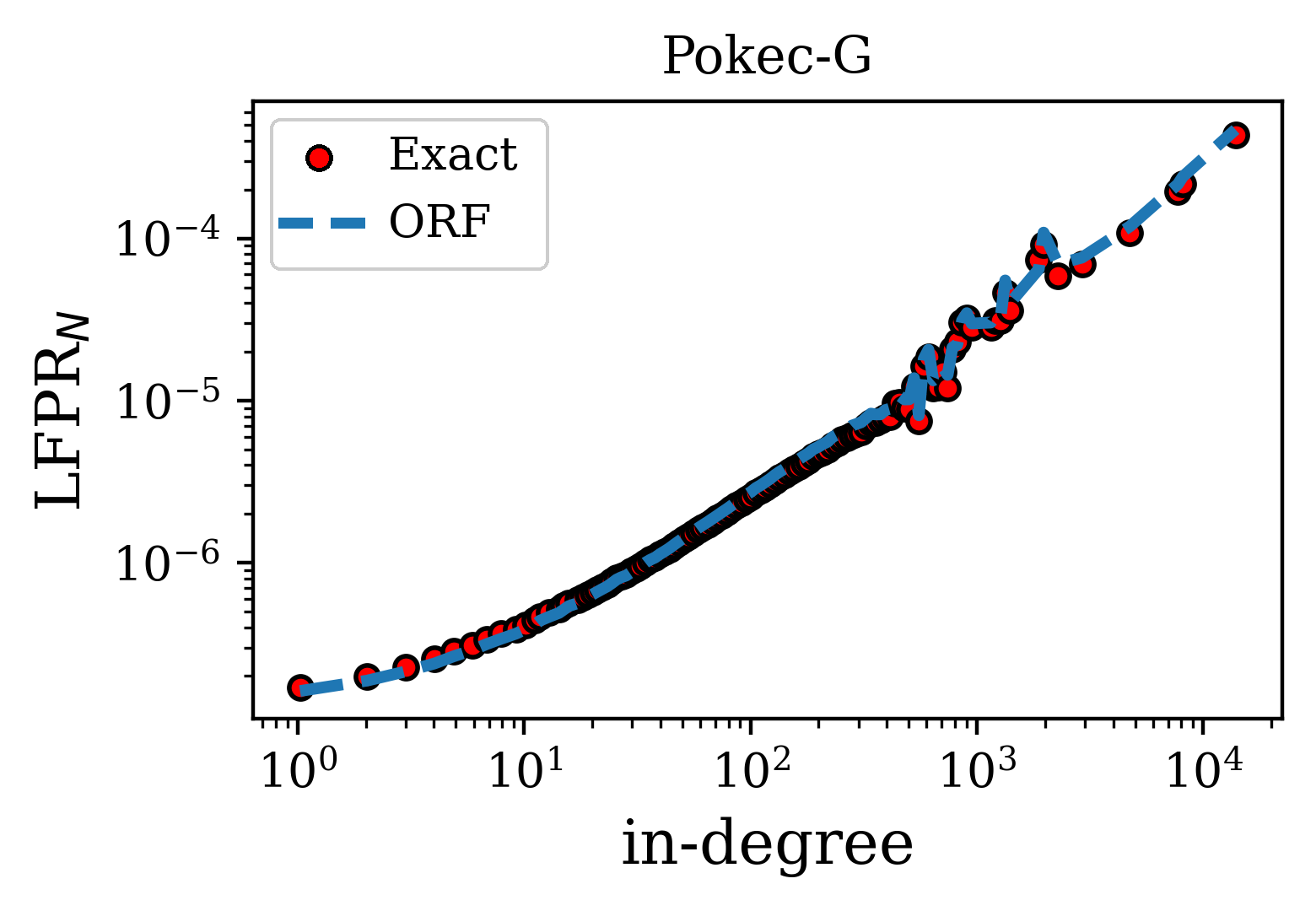}}

\vspace{-1mm}

\subfloat{%
\includegraphics[width=0.48\columnwidth]
{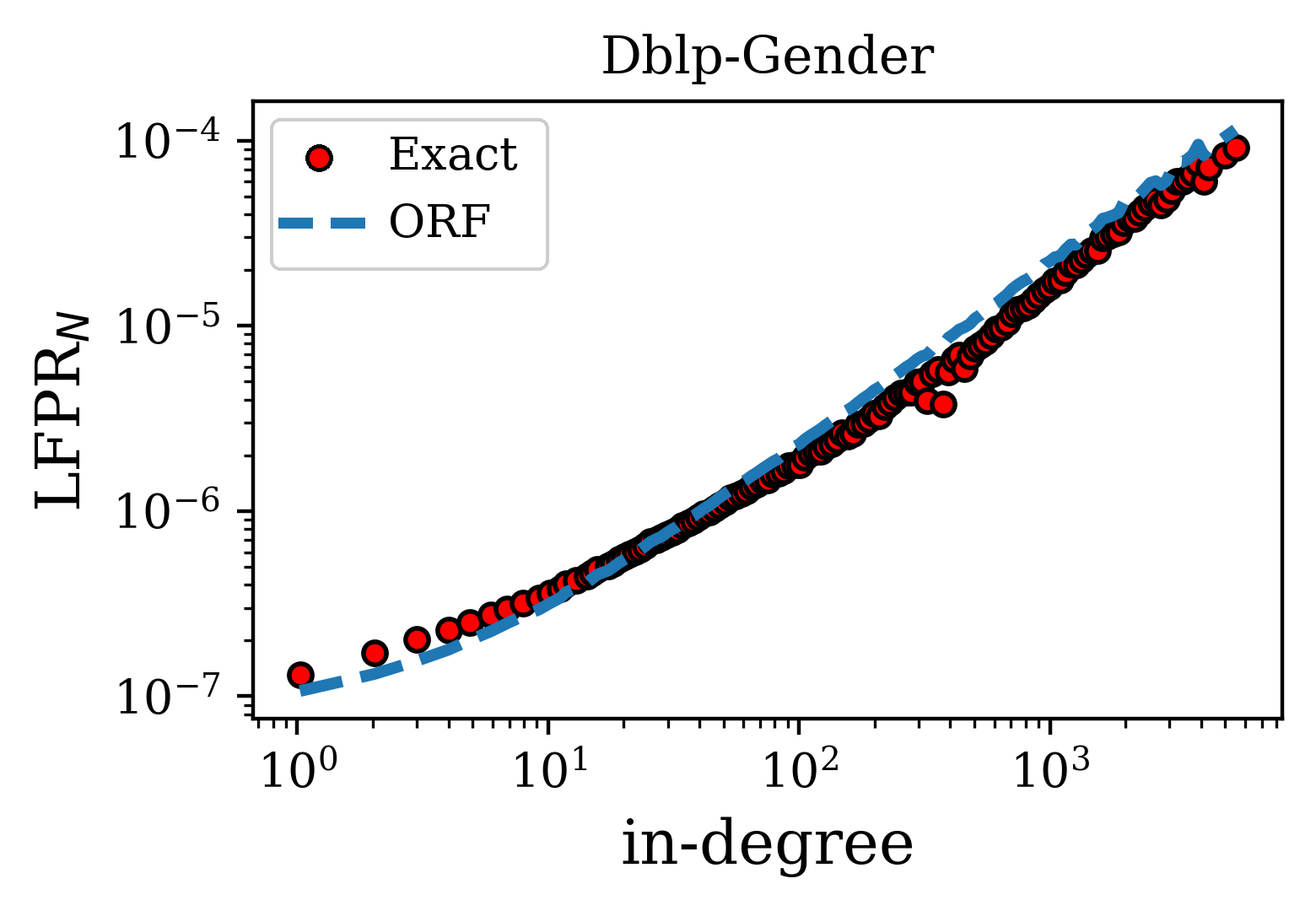}}
\hfill
\subfloat{%
\includegraphics[width=0.48\columnwidth]
{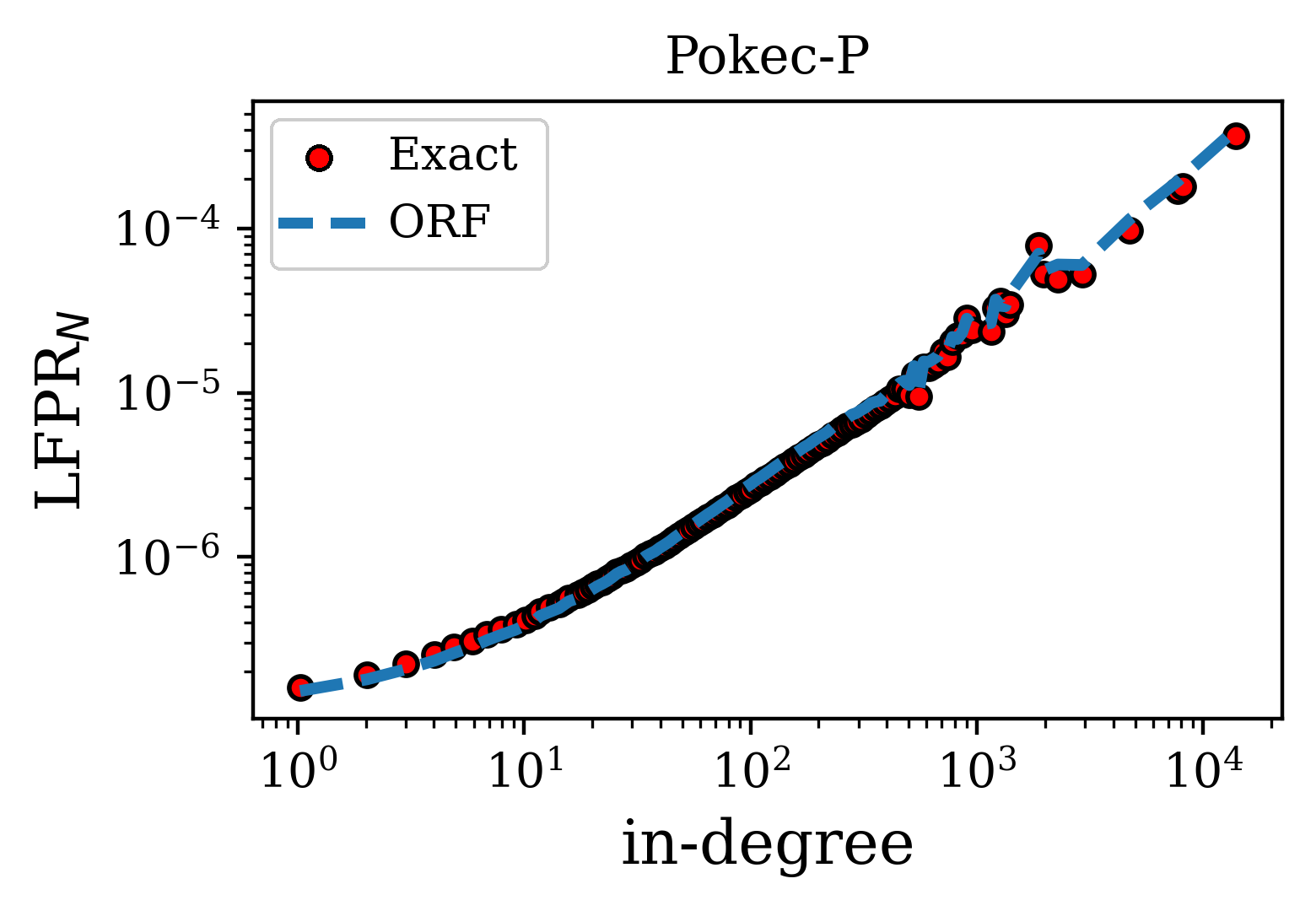}}

\vspace{-1mm}

\subfloat{%
\includegraphics[width=0.48\columnwidth]
{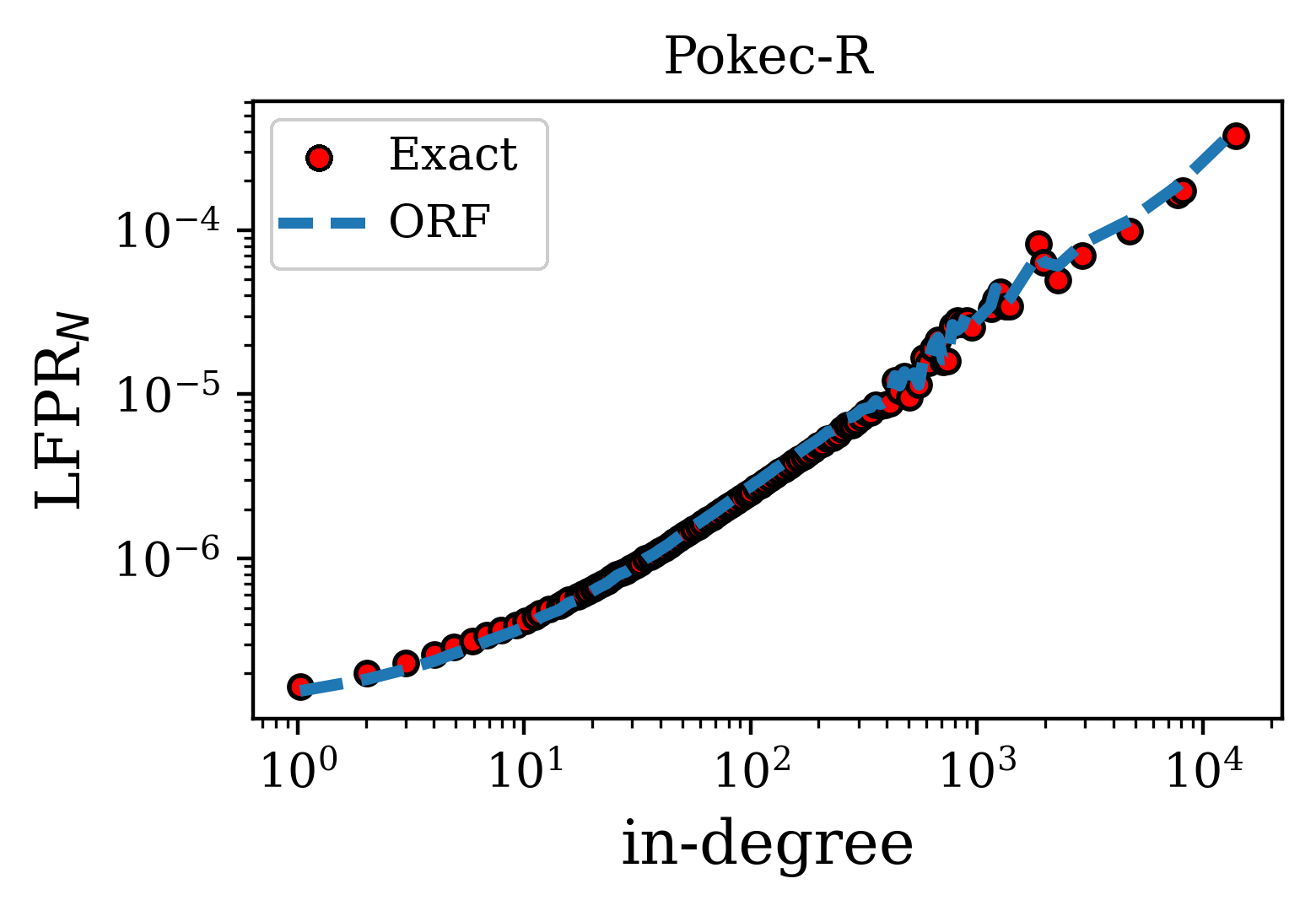}}
\hfill
\subfloat{%
\includegraphics[width=0.48\columnwidth]
{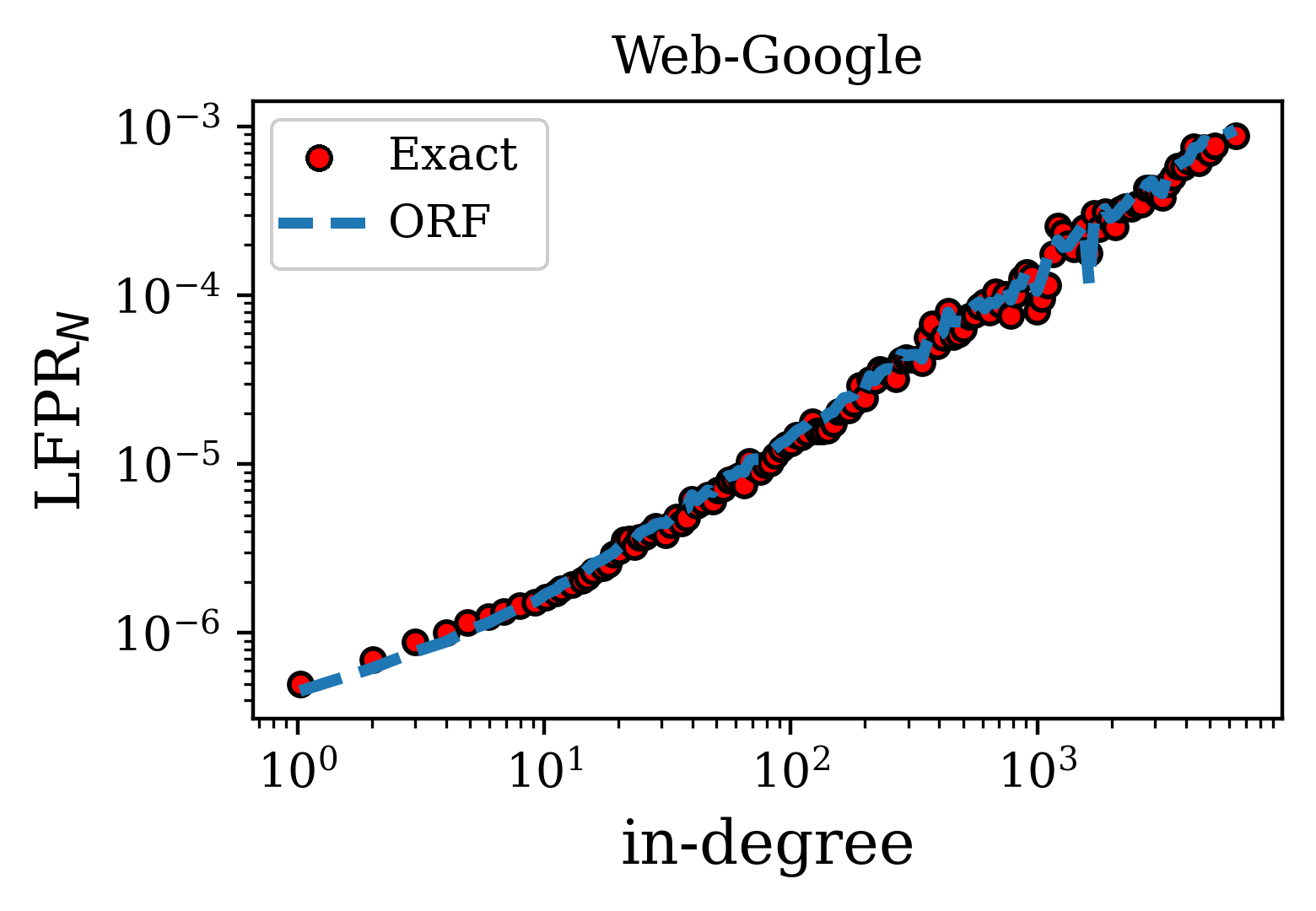}}

\caption{Comparison of average exact and one-step refined LFPR$_N$ scores across logarithmically spaced in-degree bins for six real-world networks.}
\label{fig:ofr_lfprn}

\end{figure}

\begin{figure}[!t]
\centering

\subfloat{%
\includegraphics[width=0.48\columnwidth]
{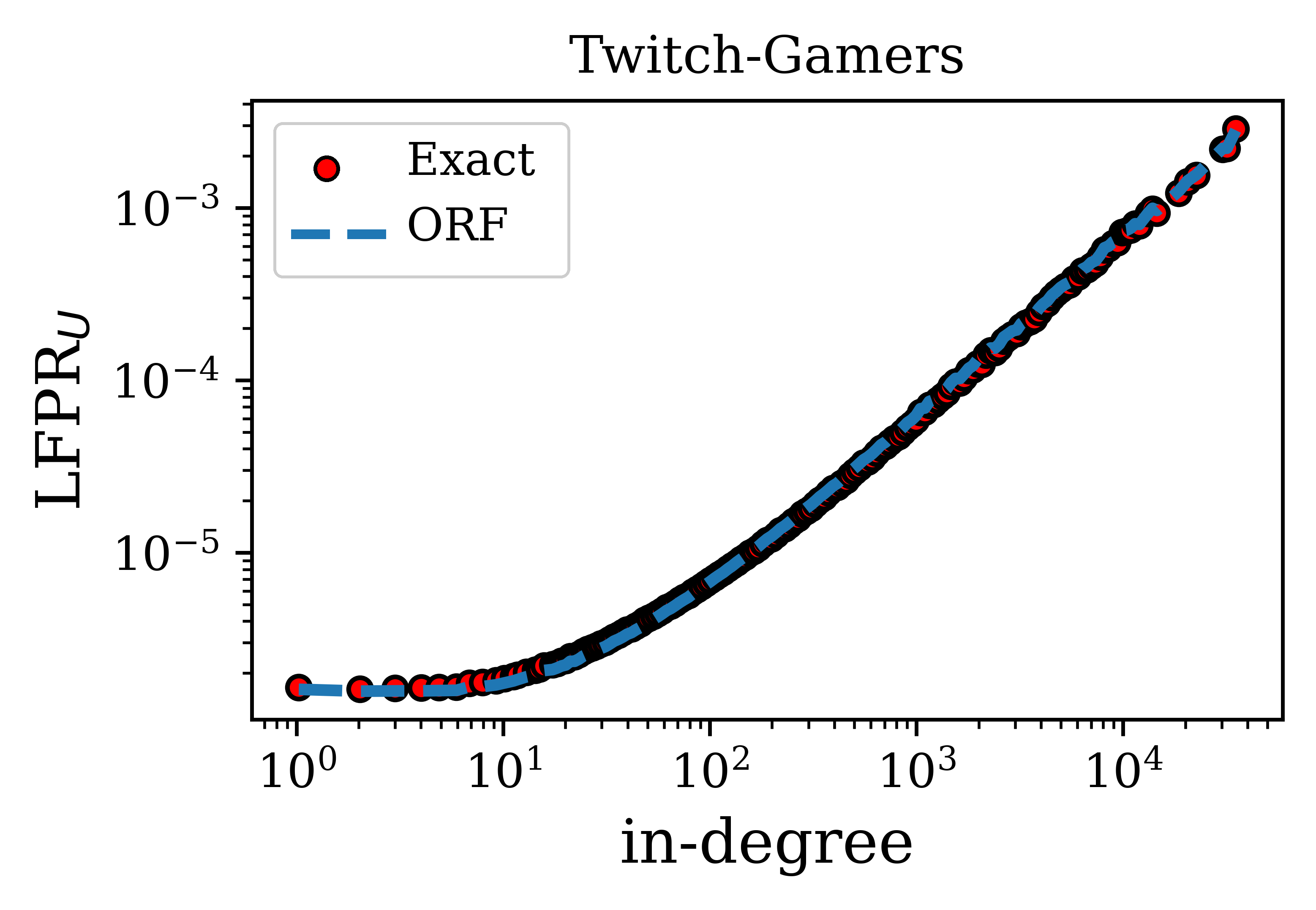}}
\hfill
\subfloat{%
\includegraphics[width=0.48\columnwidth]
{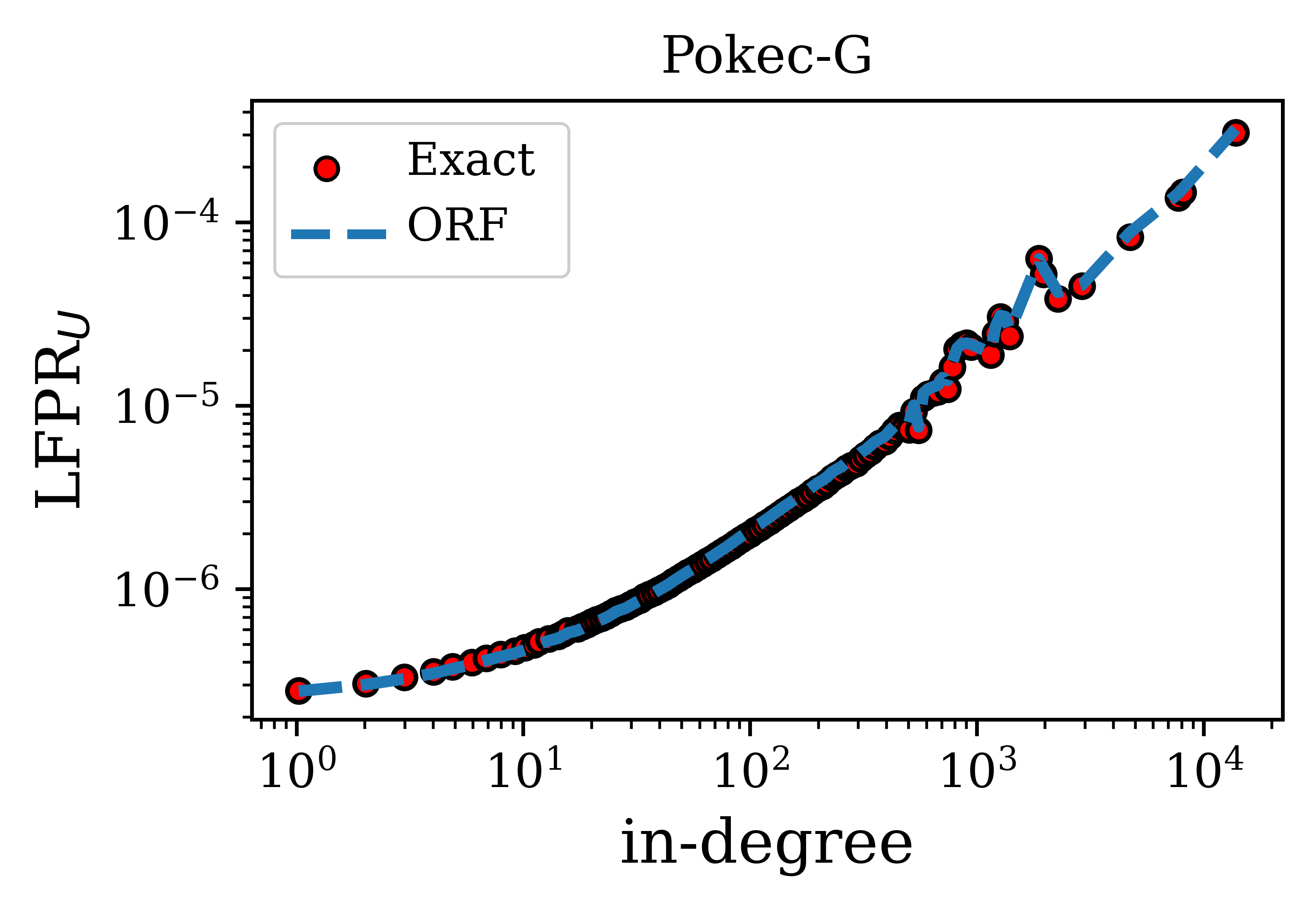}}

\vspace{-1mm}

\subfloat{%
\includegraphics[width=0.48\columnwidth]
{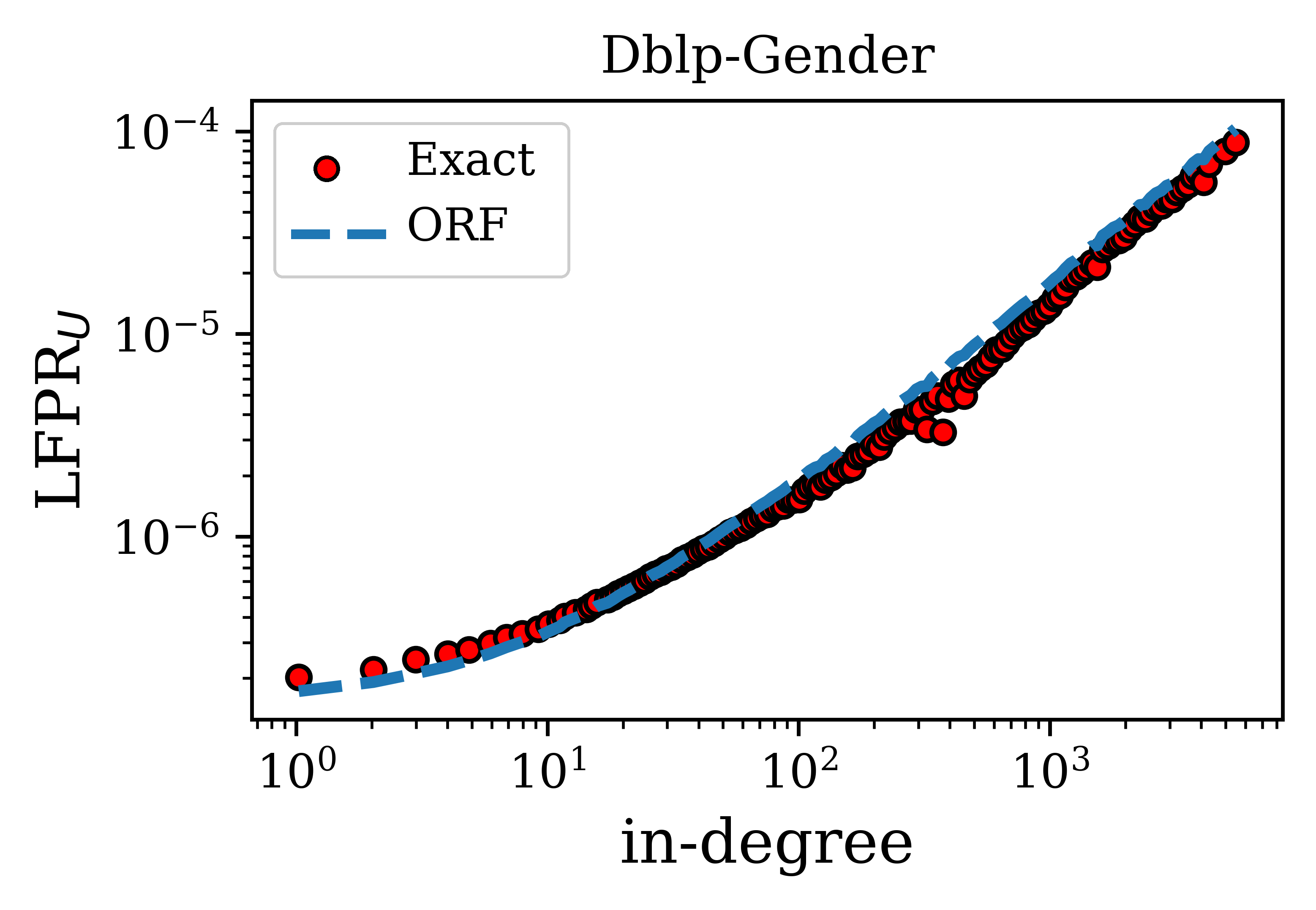}}
\hfill
\subfloat{%
\includegraphics[width=0.48\columnwidth]
{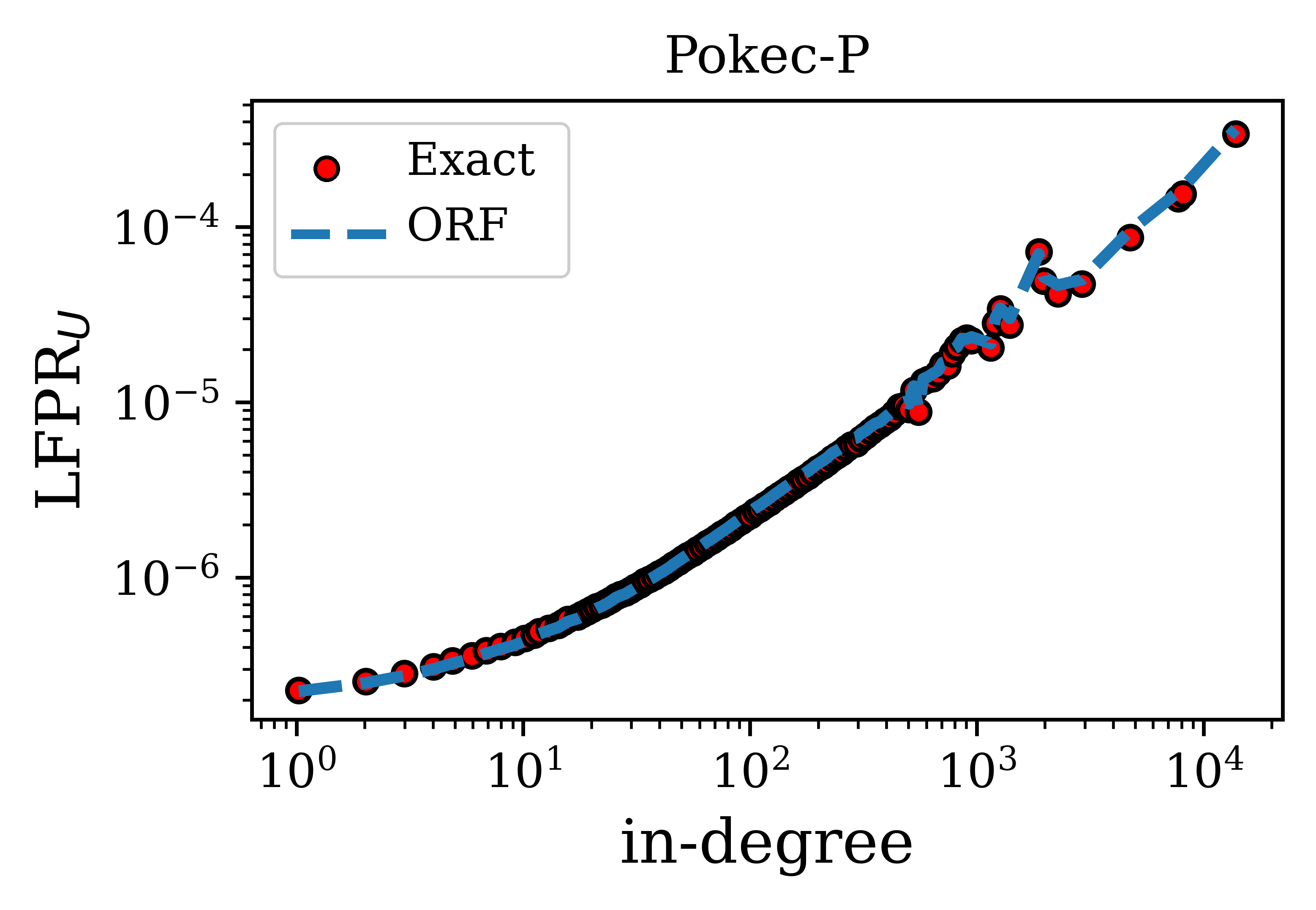}}

\vspace{-1mm}

\subfloat{%
\includegraphics[width=0.48\columnwidth]
{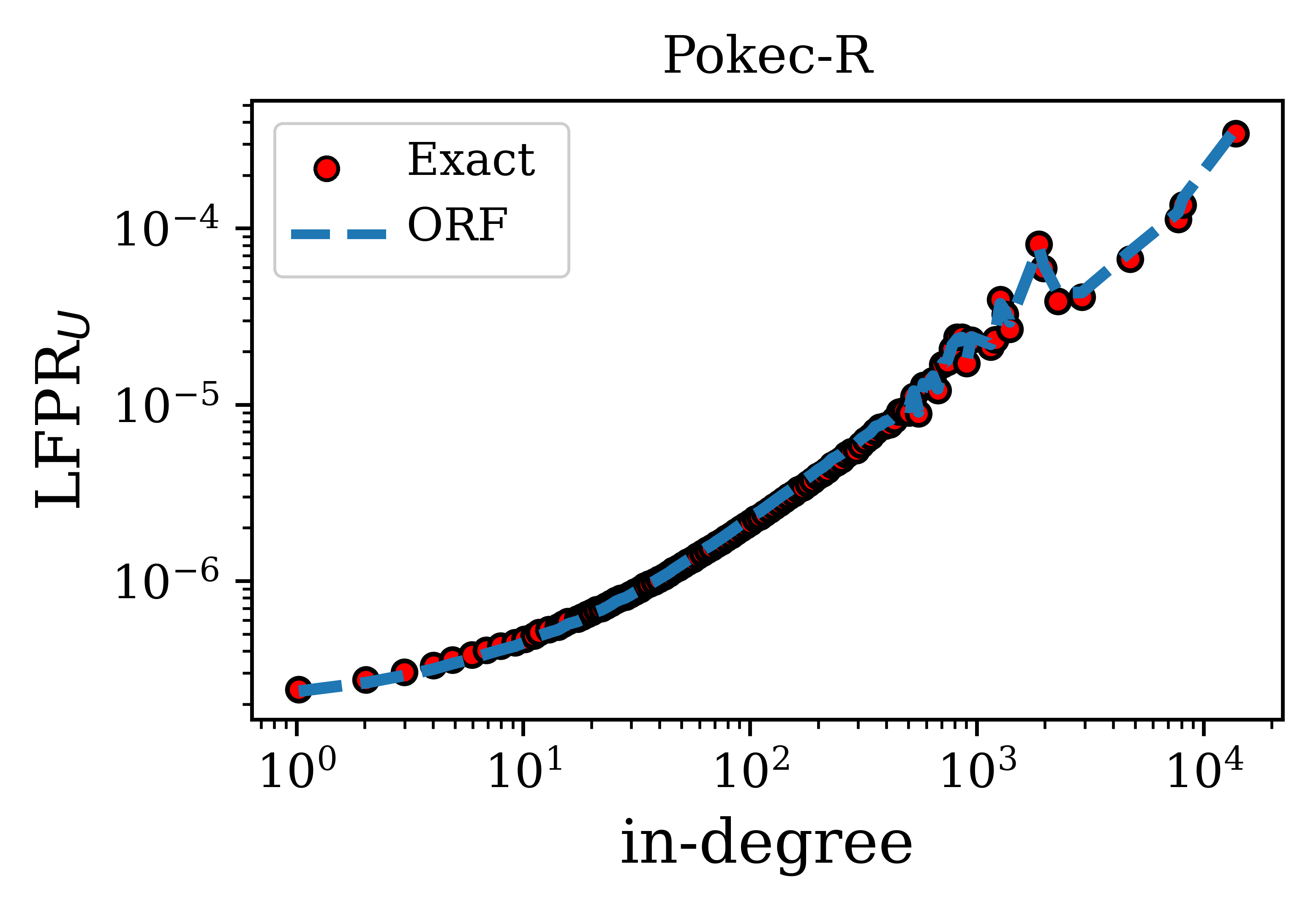}}
\hfill
\subfloat{%
\includegraphics[width=0.48\columnwidth]
{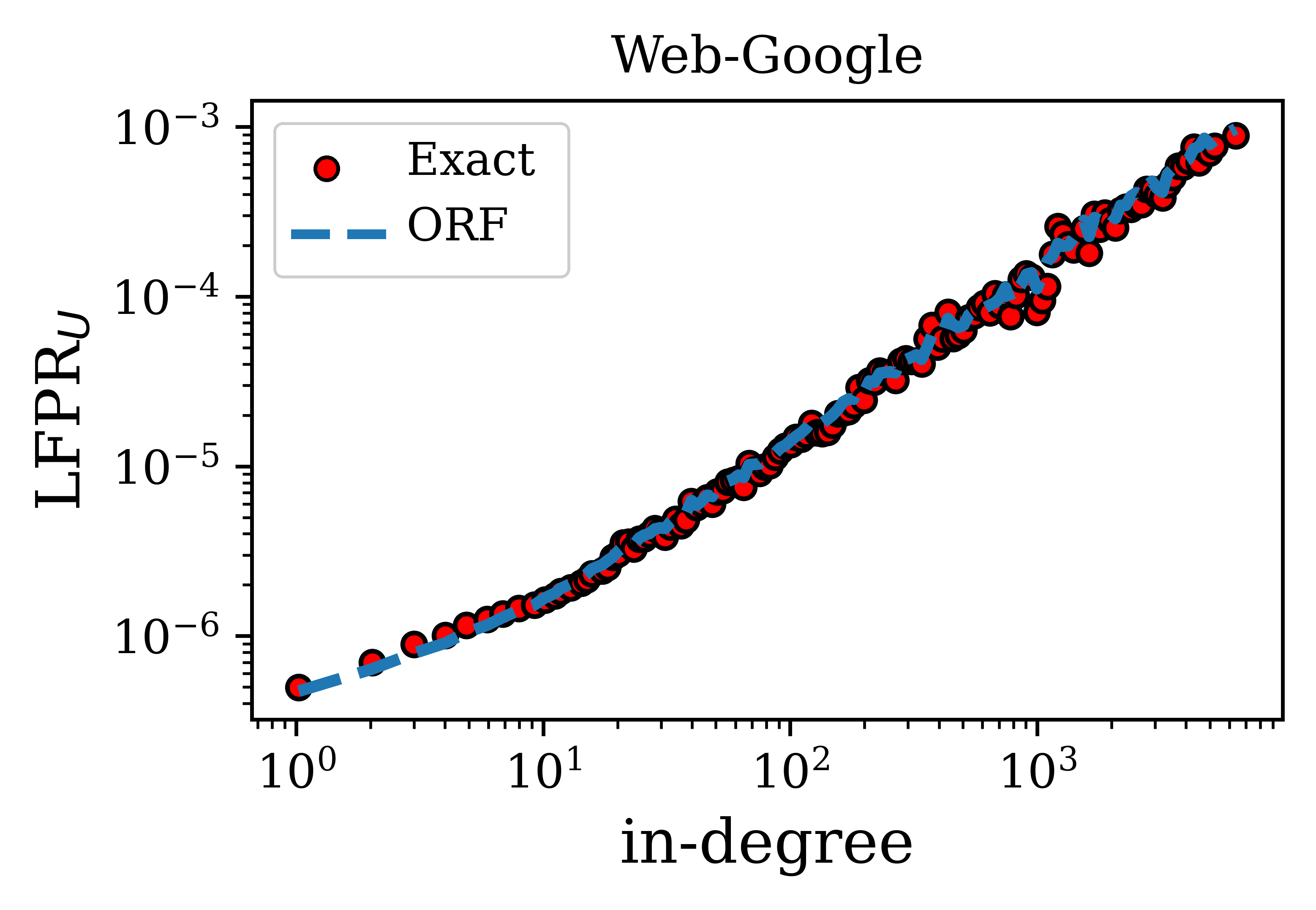}}

\caption{Comparison of average exact and one-step refined LFPR$_U$ scores across logarithmically spaced in-degree bins for six real-world networks.}
\label{fig:ofr_lfpru}

\end{figure}

Figures \ref{fig:ofr_lfprn} and \ref{fig:ofr_lfpru} illustrate the effect of the proposed One-Step Refinement (ORF) on the LFPR$_N$ and LFPR$_U$ approximations, respectively. Compared with the corresponding mean-field approximations, ORF improves the agreement between the approximate and exact LFPR scores across all datasets. Although the mean-field formulations already provide accurate estimates of the LFPR scores, applying the original fairness-aware propagation operator once to the mean-field estimate incorporates node-specific neighborhood information into the refined scores. Consequently, the refined curves exhibit closer agreement with the exact LFPR values across the observed degree classes, with the largest improvements occurring in degree ranges where the mean-field curves exhibit visible deviations from the exact values. The consistent improvements observed across networks with diverse structural characteristics support the effectiveness of the proposed refinement strategy. Overall, the one-step refinement improves approximation fidelity while retaining the efficiency of the analytical framework.

\begin{figure}[!t]
\centering

\subfloat{%
\includegraphics[width=0.48\columnwidth]
{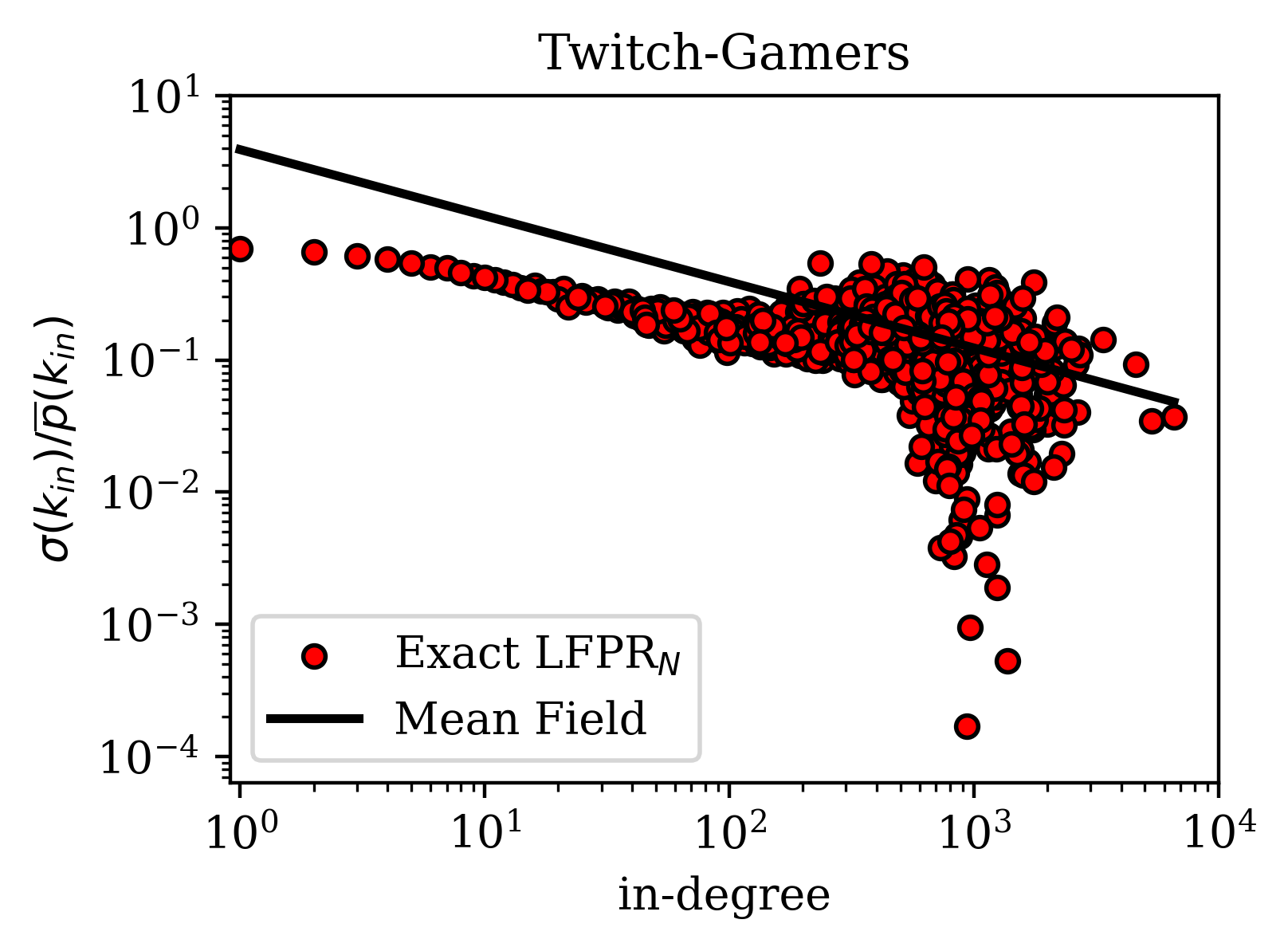}}
\hfill
\subfloat{%
\includegraphics[width=0.48\columnwidth]
{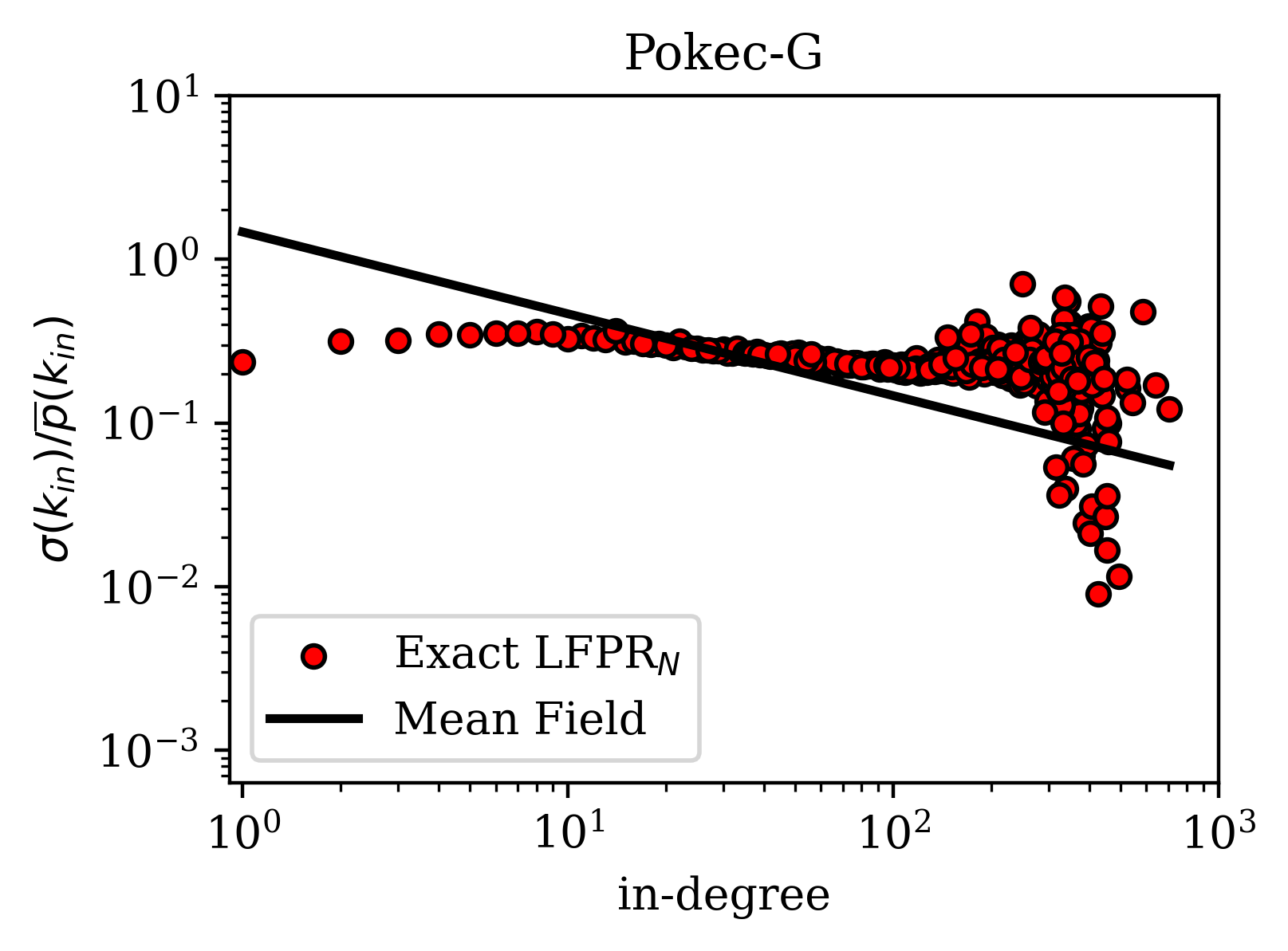}}

\vspace{-1mm}

\subfloat{%
\includegraphics[width=0.48\columnwidth]
{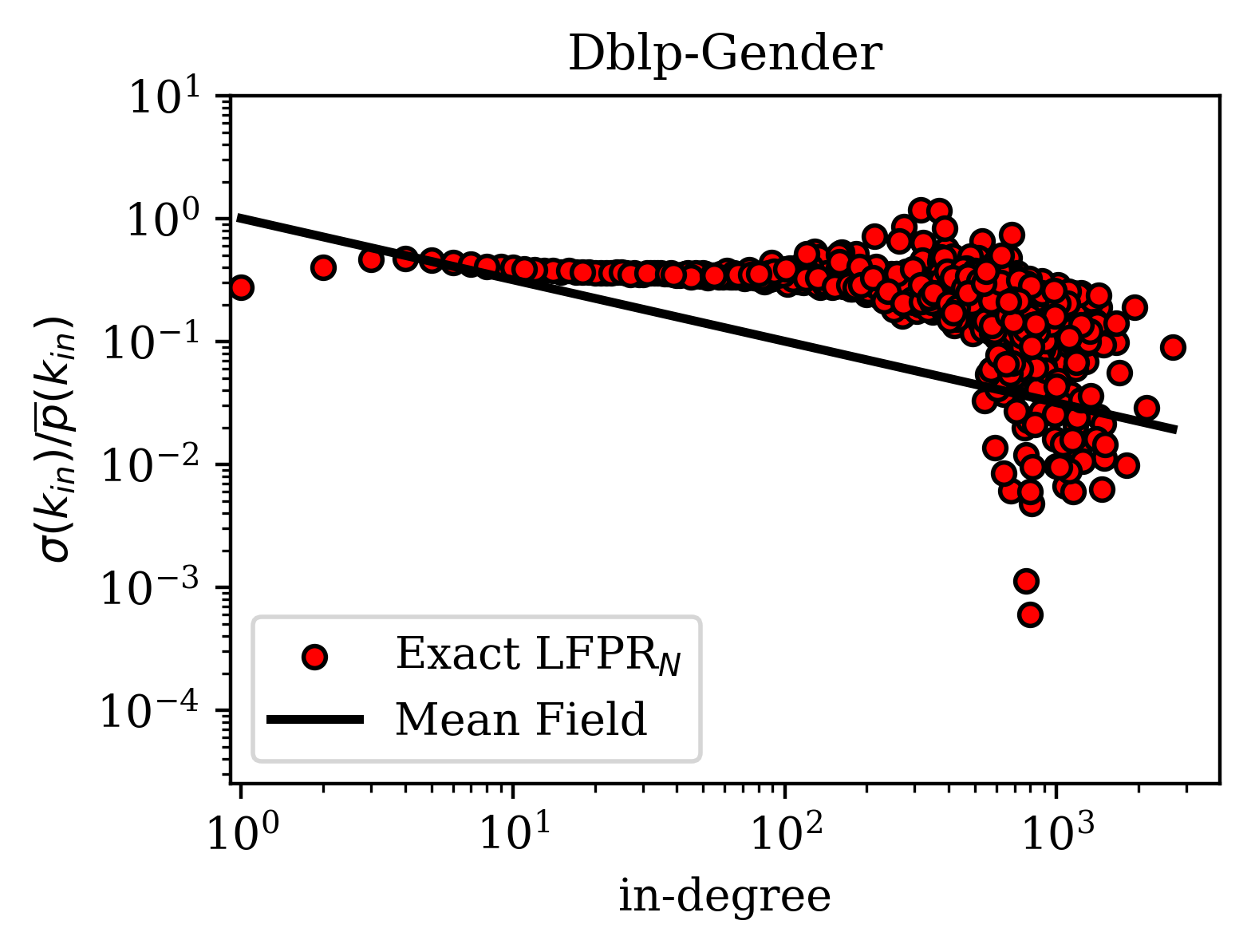}}
\hfill
\subfloat{%
\includegraphics[width=0.48\columnwidth]
{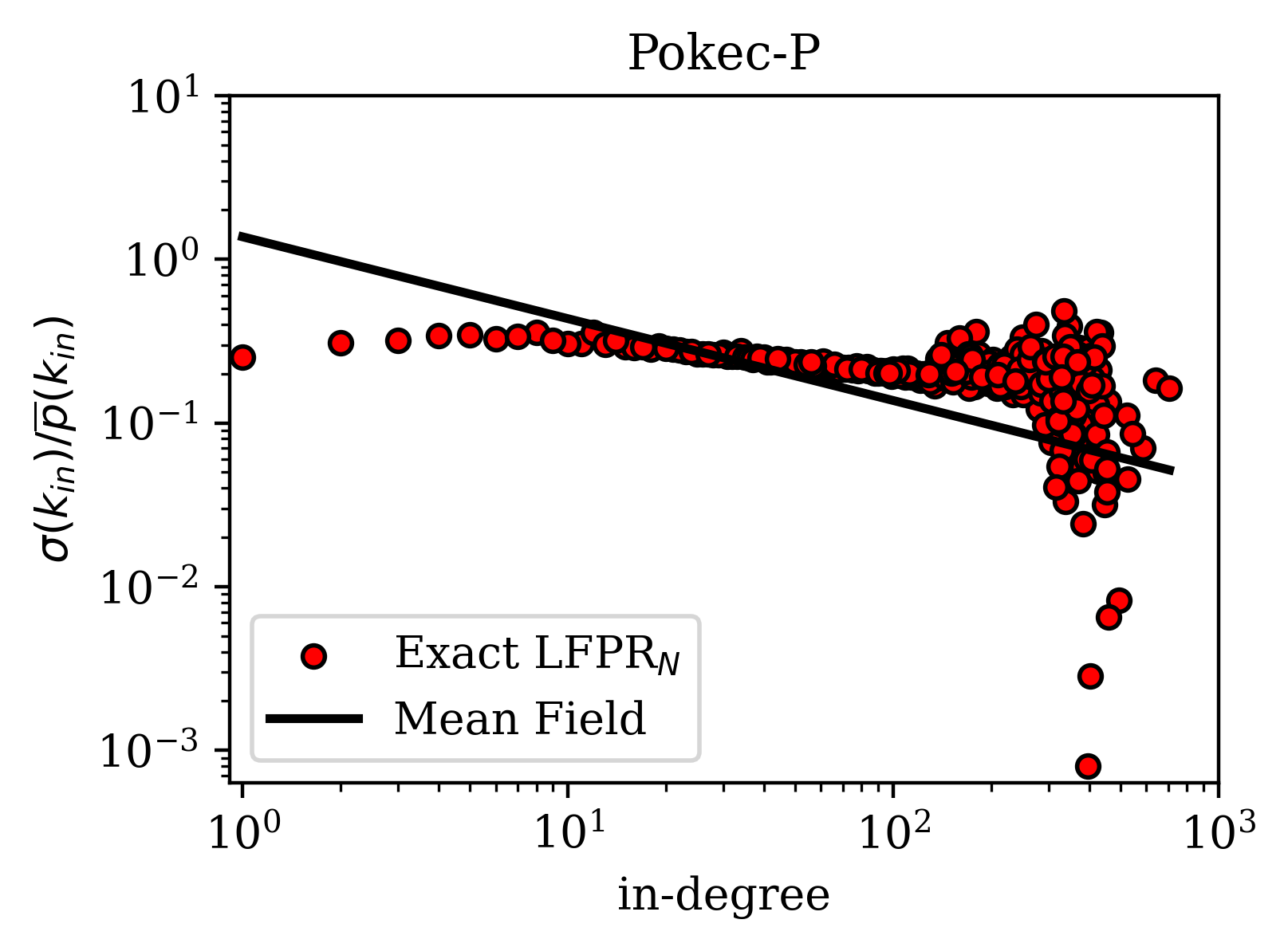}}

\vspace{-1mm}

\subfloat{%
\includegraphics[width=0.48\columnwidth]
{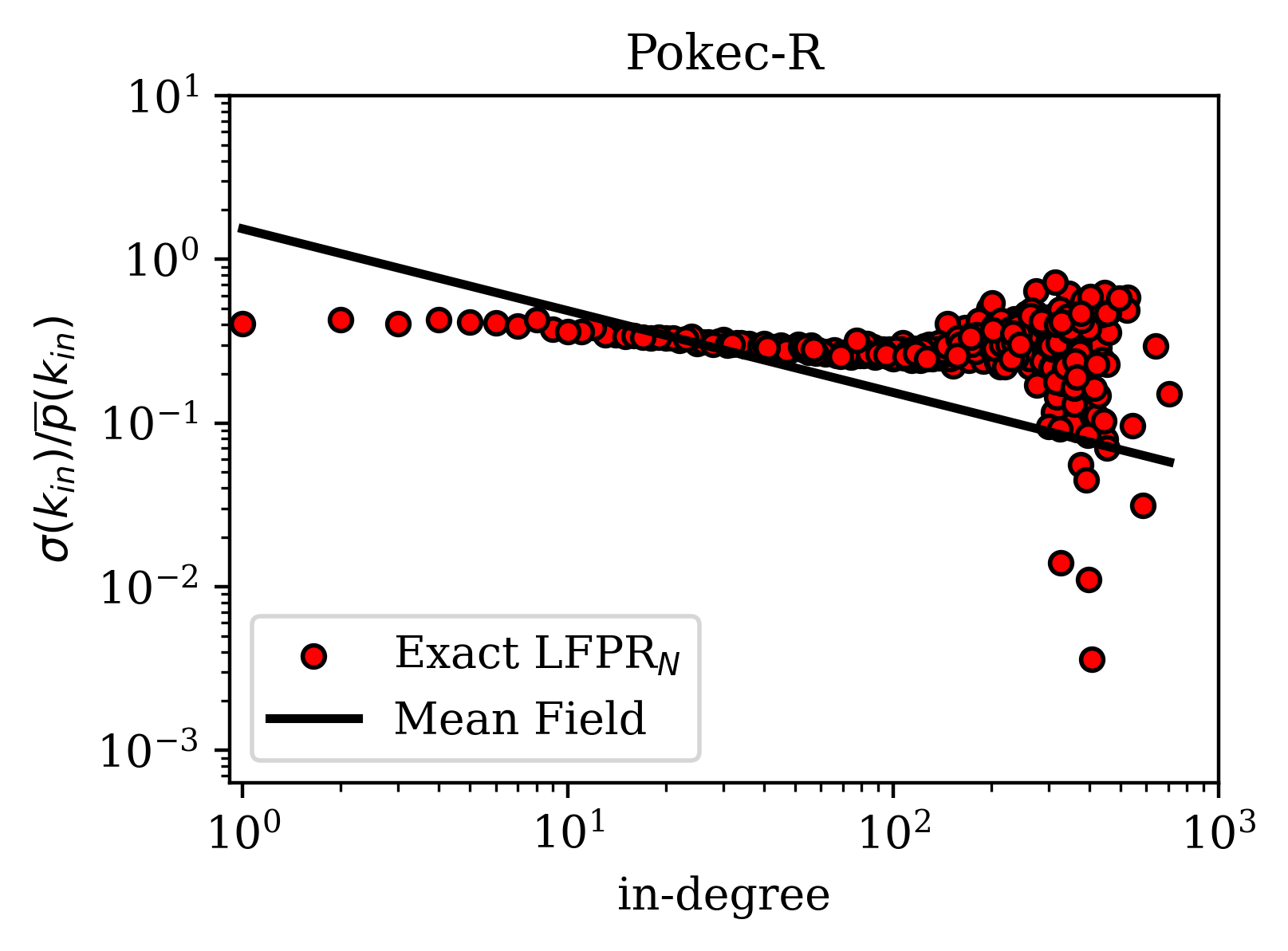}}
\hfill
\subfloat{%
\includegraphics[width=0.48\columnwidth]
{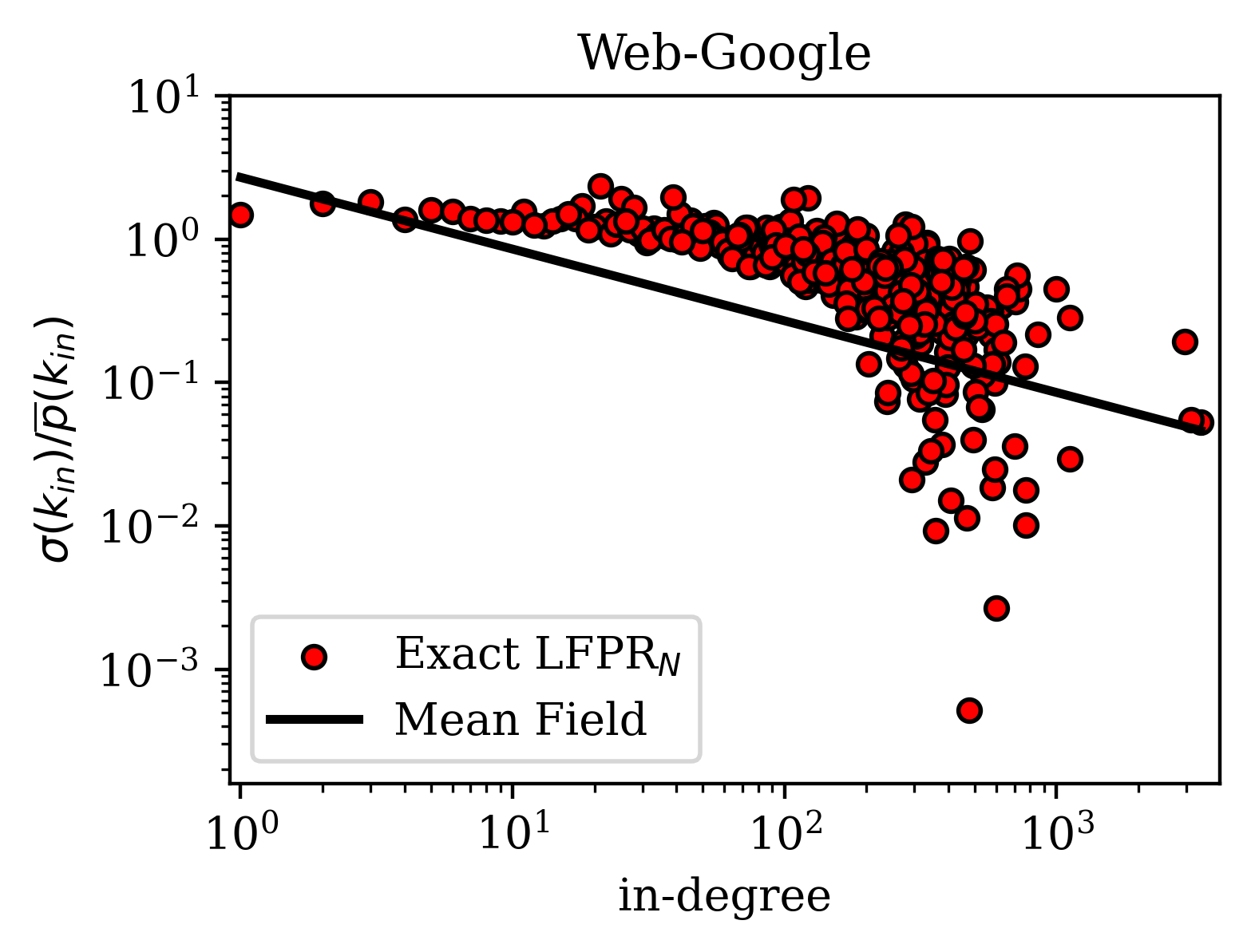}}

\caption{Coefficient of variation of LFPR$_N$ with respect to in-degree.}
\label{fig:fluct_lfprn}

\end{figure}

\begin{figure}[!t]
\centering

\subfloat{%
\includegraphics[width=0.48\columnwidth]
{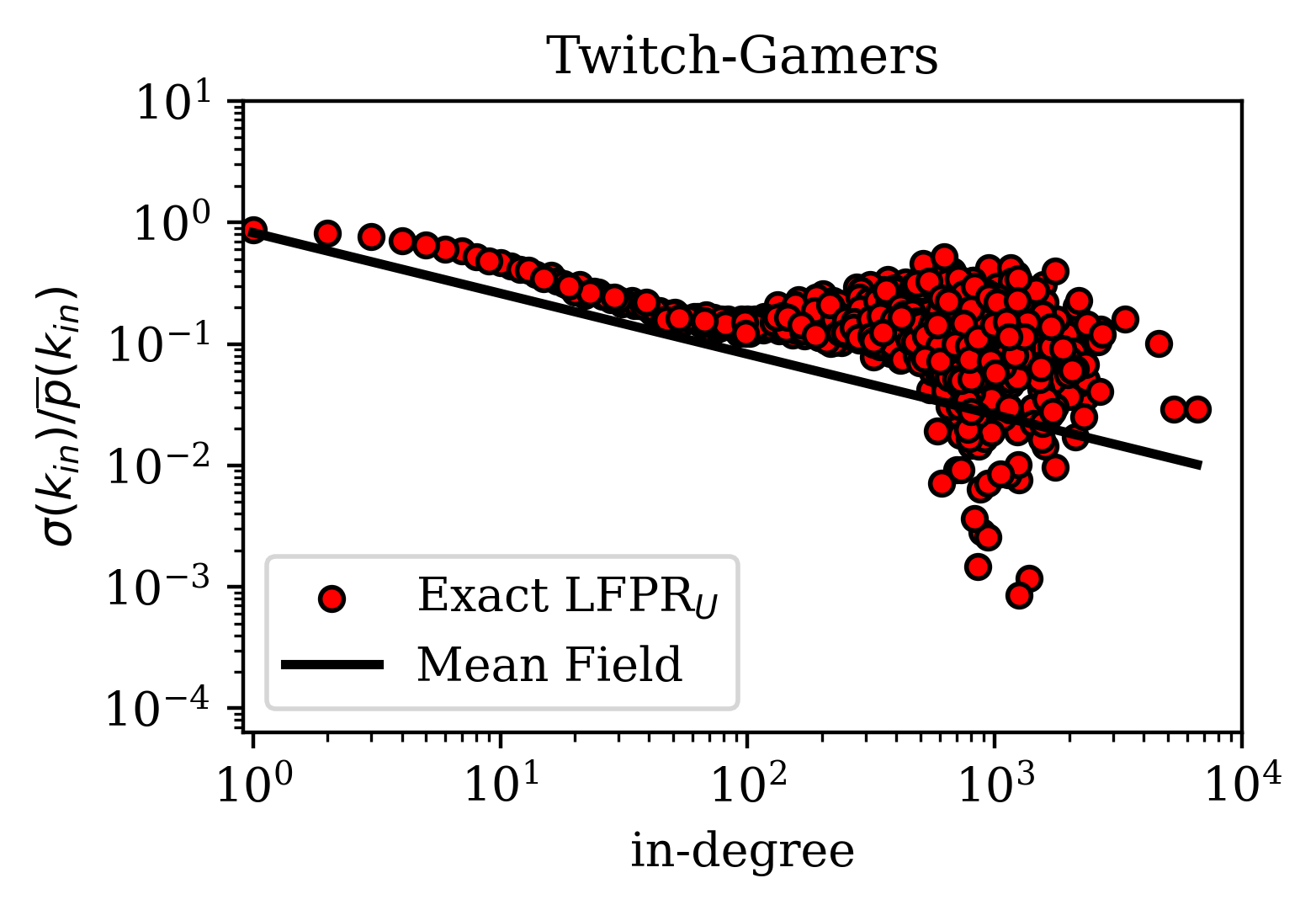}}
\hfill
\subfloat{%
\includegraphics[width=0.48\columnwidth]
{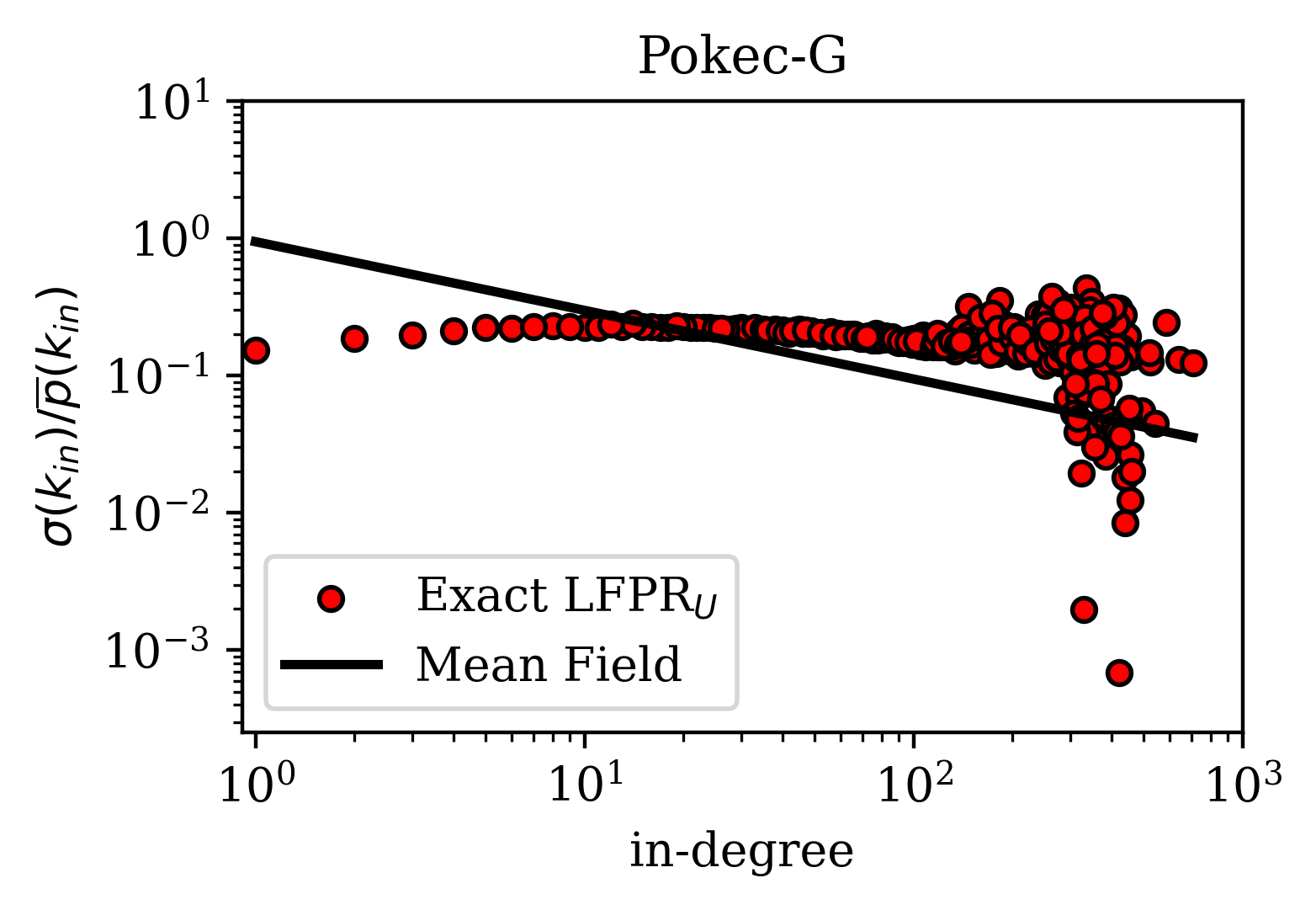}}

\vspace{-1mm}

\subfloat{%
\includegraphics[width=0.48\columnwidth]
{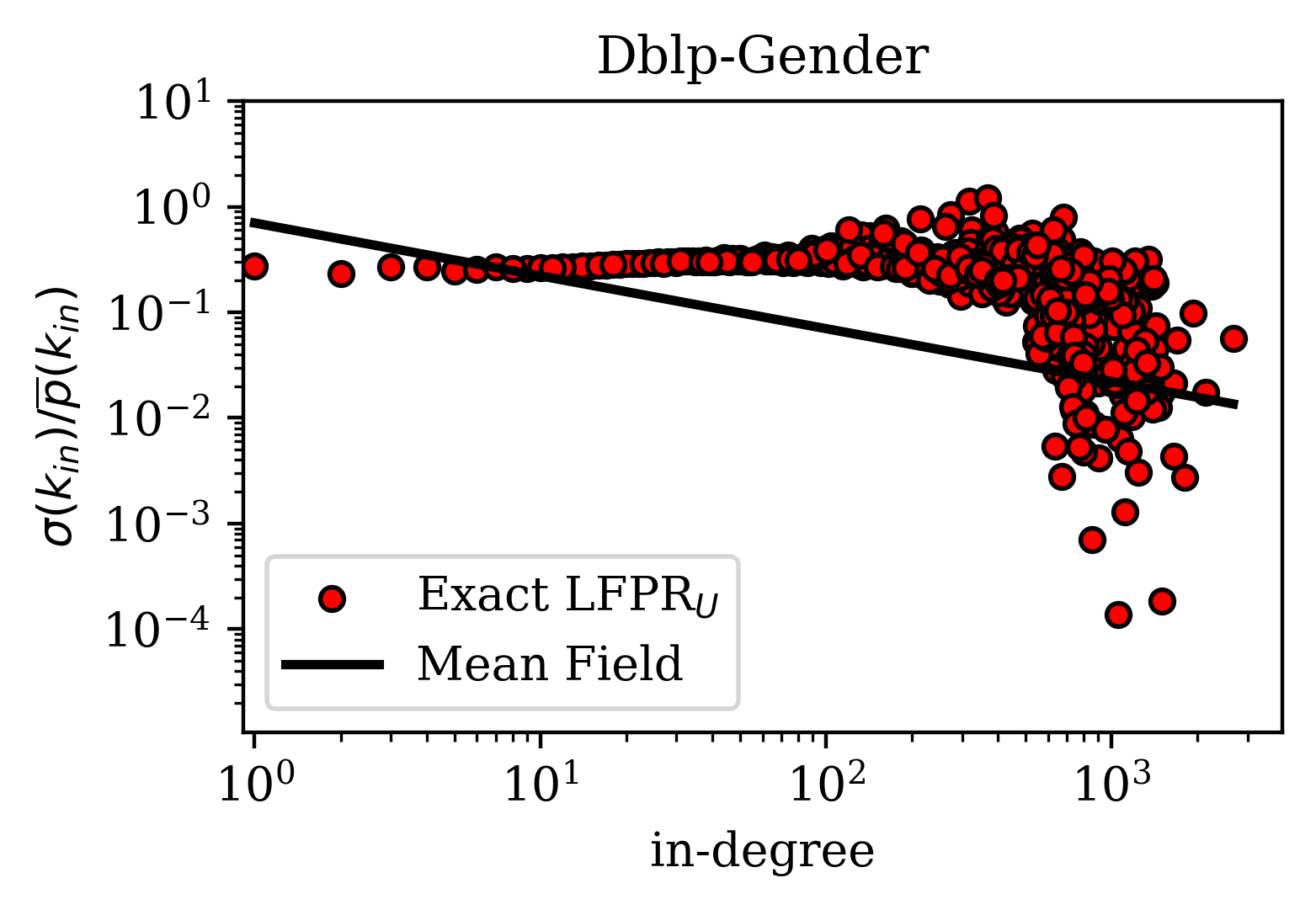}}
\hfill
\subfloat{%
\includegraphics[width=0.48\columnwidth]
{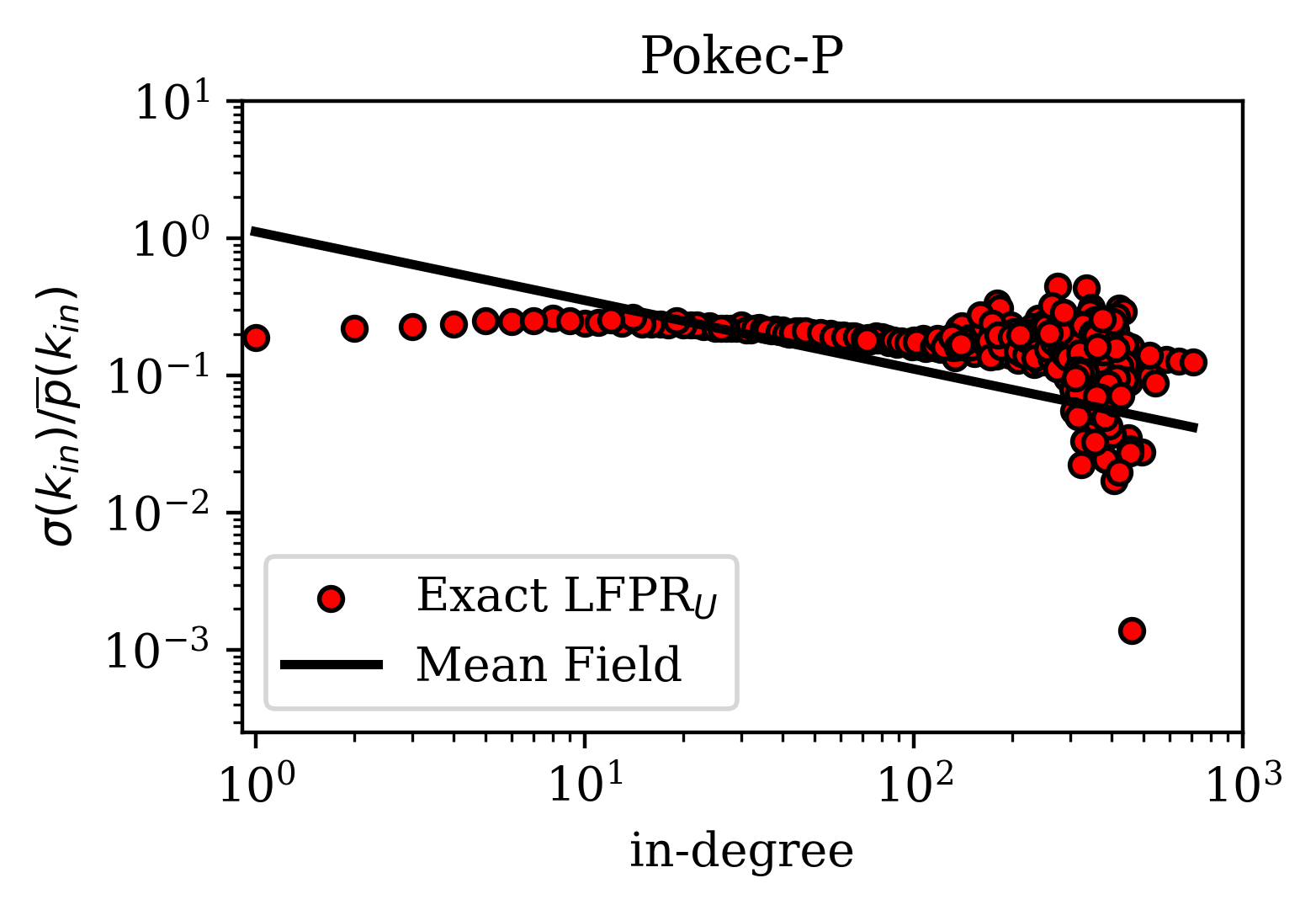}}

\vspace{-1mm}

\subfloat{%
\includegraphics[width=0.48\columnwidth]
{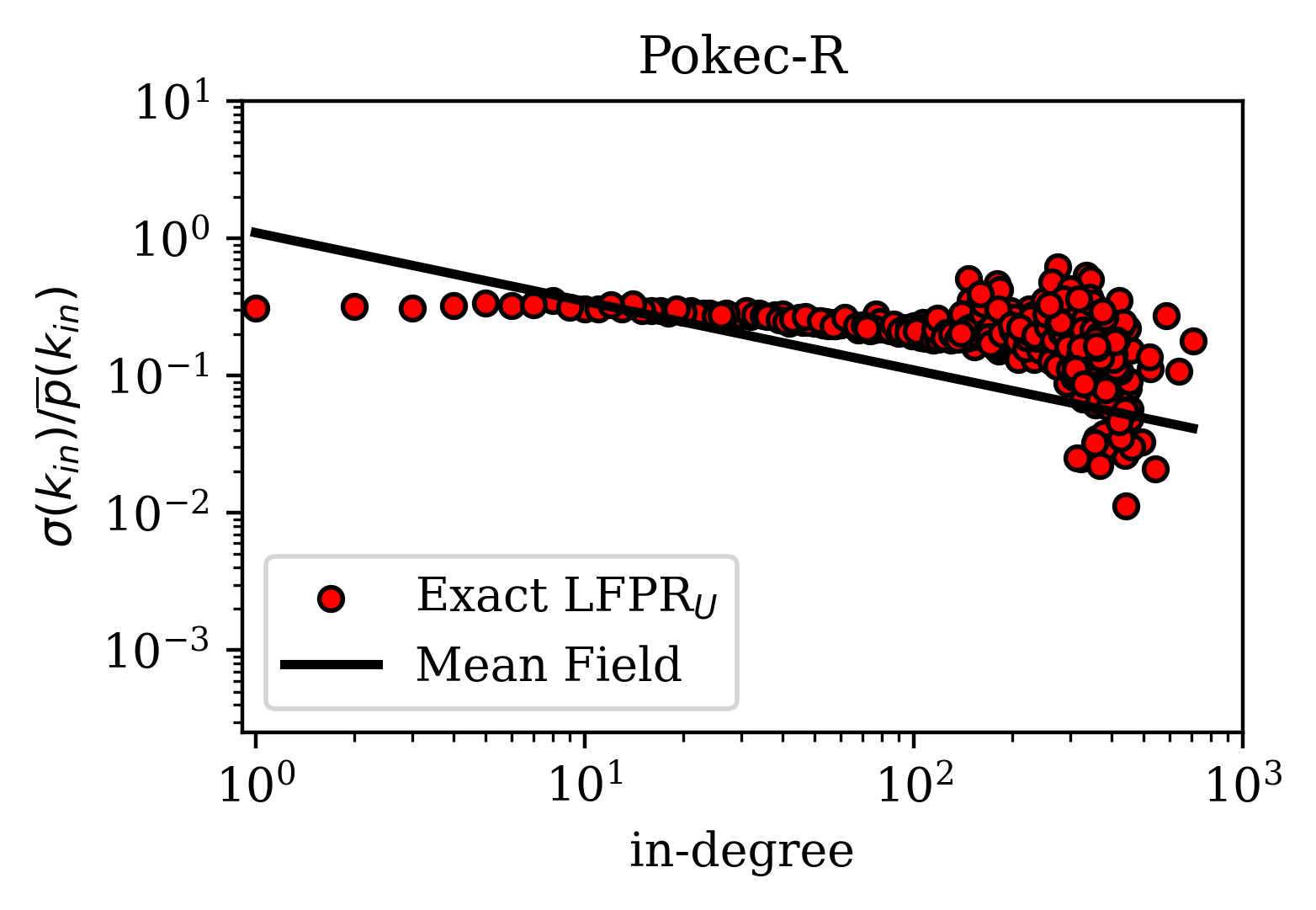}}
\hfill
\subfloat{%
\includegraphics[width=0.48\columnwidth]
{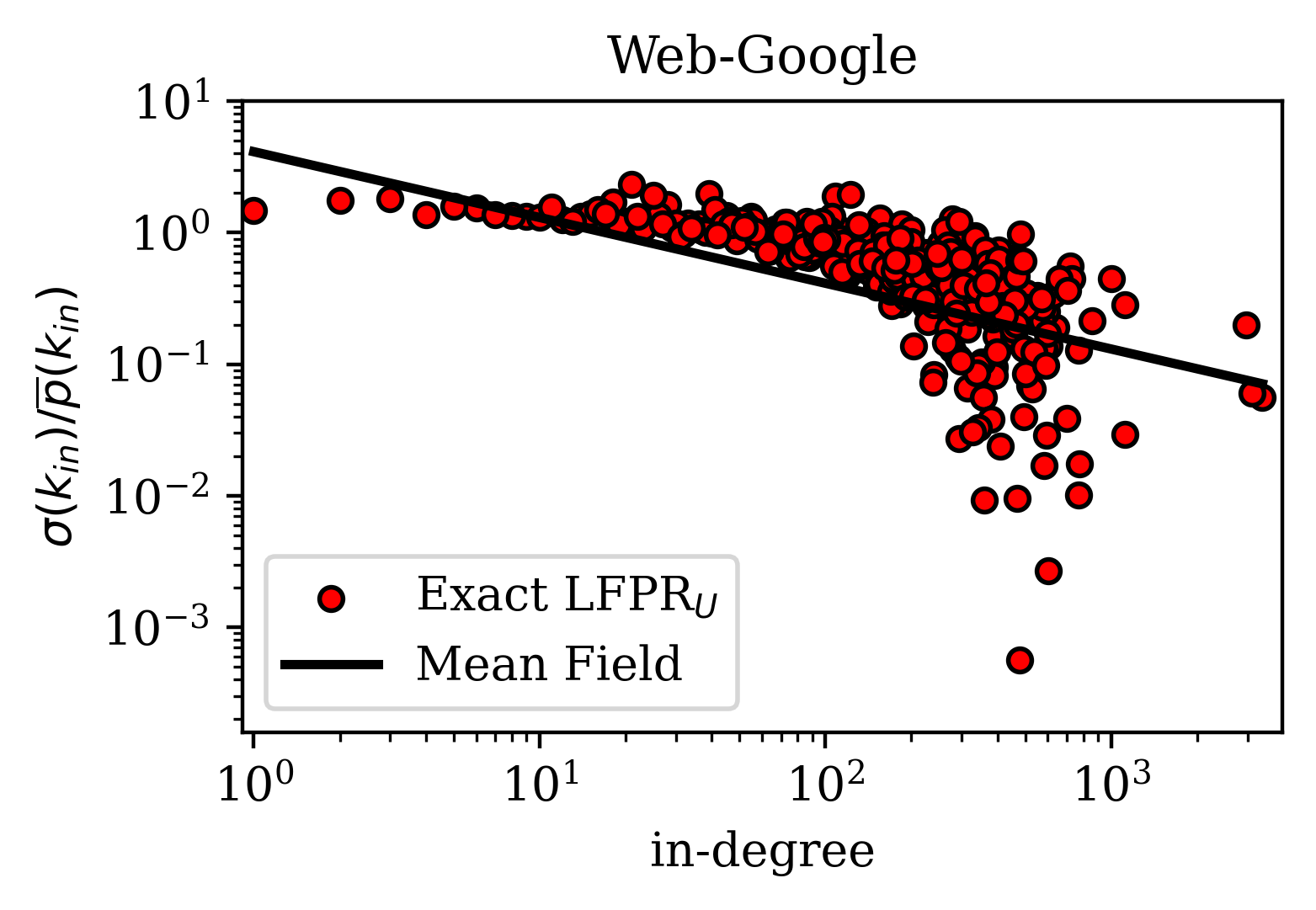}}

\caption{Coefficient of variation of LFPR$_U$ with respect to in-degree.}
\label{fig:fluct_lfpru}

\end{figure}

As a final validation of the mean-field framework, we investigate the fluctuations of LFPR scores around the corresponding degree-class mean-field estimates. As discussed in Section \ref{sec:Fluct}, the theoretical analysis predicts that relative fluctuations decrease with increasing in-degree. Figures \ref{fig:fluct_lfprn} and \ref{fig:fluct_lfpru} report the coefficient of variation of LFPR$_N$ and LFPR$_U$, respectively, as a function of node in-degree. In both cases, the empirical fluctuations exhibit a clear decreasing trend with increasing in-degree, indicating that LFPR scores become increasingly concentrated around their corresponding degree-class means for higher-degree nodes. This behavior is consistent with the theoretical analysis, which predicts a corresponding reduction in relative fluctuations. Overall, the agreement between the empirical trends and the theoretical predictions provides supporting evidence for the degree-dependent concentration behavior underlying the proposed mean-field approximations.

\subsection{Runtime and Scalability Evaluation}
To evaluate the computational scalability of our LFPR approximation frameworks, we assess runtime execution performance of the exact, mean-field, and refined LFPR methods across multiple real-world networks of varying size. All methods are evaluated using the same implementation environment and hardware configuration to maintain consistency across experiments. The experiments are performed on a Windows~11 Pro workstation with an Intel Core i7-13700 processor (2.1 GHz, 16 cores and 24 threads) and 32 GB of RAM. Table \ref{tab:runtime_results} reports the execution times of the exact LFPR algorithms and the corresponding mean-field and ORF-refined LFPR approximations across the considered networks.

\begin{table}[!t]
\caption{Runtime (seconds) of the exact, mean-field (MF), and one-step refined (ORF) LFPR methods across the evaluated networks.}
\label{tab:runtime_results}
\centering
\renewcommand{\arraystretch}{1.05}

\begin{tabular}{lcccccc}
\hline
\textbf{Dataset} &
\multicolumn{3}{c}{\textbf{LFPR$_N$}} &
\multicolumn{3}{c}{\textbf{LFPR$_U$}} \\
\cmidrule(lr){2-4}
\cmidrule(lr){5-7}
&
\textbf{Exact} &
\textbf{MF} &
\textbf{ORF} &
\textbf{Exact} &
\textbf{MF} &
\textbf{ORF} \\
\hline

Twitch-Gamers & 19.09 & 1.28 & 4.80 & 9.52 & 1.96 & 4.02 \\
Pokec-G       & 93.31 & 1.38 & 5.39 & 71.31 & 2.28 & 4.13 \\
DBLP-Gender   & 130.34 & 1.30 & 4.95 & 99.50 & 2.28 & 4.28 \\
Pokec-P       & 97.14 & 1.35 & 5.32 & 78.17 & 2.04 & 4.22 \\
Pokec-R       & 117.91 & 1.40 & 5.05 & 77.60 & 1.97 & 4.06 \\
Web-Google    & 47.26 & 1.30 & 5.05 & 37.52 & 2.02 & 4.05 \\
\hline
\end{tabular}
\end{table}
The runtime results in Table \ref{tab:runtime_results} are consistent with the complexity analysis presented in Section \ref{sec:complexity}. Across all networks, the mean-field approximations achieve substantial reductions in execution time compared with the corresponding exact LFPR methods. The refined methods incur a modest additional computational cost relative to the mean-field approximations due to the single fairness-aware propagation step, yet remain considerably faster than the exact algorithms. This behavior is consistently observed for both LFPR$_N$ and LFPR$_U$ across all evaluated networks. Overall, the results demonstrate that the proposed mean-field and ORF approximations substantially reduce runtime while preserving the scalability advantage predicted by the theoretical complexity analysis.

\section{Conclusion}
We developed a scalable analytical framework for Neighborhood Locally Fair PageRank (LFPR$_N$) and Uniform Locally Fair PageRank (LFPR$_U$). By reformulating the fairness-aware propagation dynamics using a group-aware mean-field approximation, the framework estimates stationary LFPR scores without repeated iterative propagation over fairness-aware transition matrices. A lightweight One-Step Refinement (ORF) mechanism further improves approximation accuracy while retaining computational efficiency. Theoretical fluctuation analysis shows that relative fluctuations decrease with increasing in-degree, indicating that the mean-field estimate becomes more representative for higher-degree nodes. Experimental results on multiple real-world networks demonstrate that the proposed methods closely approximate exact LFPR stationary scores, preserve the fairness characteristics of the corresponding LFPR models, and substantially reduce computational cost. Overall, the framework provides a scalable and theoretically grounded approach to fairness-aware graph ranking.

\ifCLASSOPTIONcompsoc
  \section*{Acknowledgments}
\else
  \section*{Acknowledgment}
\fi

The authors acknowledge financial support from the Scheme for Promotion of Academic and Research Collaboration (SPARC), funded by the Ministry of Education, Government of India [Sanction No. SPARC/2025-2026/P4340].

\ifCLASSOPTIONcaptionsoff
  \newpage
\fi

\bibliographystyle{IEEEtran}
\bibliography{sn-bibliography}
\end{document}